\documentclass[acmsmall]{acmart}

\AtBeginDocument{%
  }

\setcopyright{acmlicensed}
\copyrightyear{2025}
\acmYear{2025}
\acmDOI{XXXXXXX.XXXXXXX}

\usepackage[frozencache=true,cachedir=.]{minted}
\usepackage{xspace}
\usepackage{tikz}
\usepackage[framemethod=tikz]{mdframed}

\usepackage[ruled,vlined,linesnumbered]{algorithm2e}
\SetAlFnt{\footnotesize}  
\SetAlCapFnt{\footnotesize}  
\SetAlCapNameFnt{\footnotesize}  
\usepackage{comment}
\usepackage{graphicx}
\usepackage{subcaption}  
\usepackage{caption}     
\usepackage{float}       
\usepackage[utf8]{inputenc}
\usepackage{float}       
\usepackage{pifont}     
\usepackage{multirow}    

\usepackage{xcolor}
\usepackage{listings}
\usepackage{orcidlink}
\usepackage{pifont}
\usepackage{booktabs}
\usepackage{multirow}

\definecolor{functioncolor}{HTML}{4672C4}
\definecolor{classcolor}{HTML}{D79655}
\definecolor{keywordcolor}{HTML}{C678DD}
\definecolor{commentcolor}{HTML}{7F7F7F}
\definecolor{stringcolor}{HTML}{A2D27E}

\newcommand{\tool}{\textsc{PCREQ}\xspace}
\newcommand{\benchmark}{\textsc{REQBench}\xspace}
\acmJournal{JACM}
\acmVolume{37}
\acmNumber{4}
\acmArticle{111}
\acmMonth{8}

\begin{document}

\title{\tool: Automated Inference of Compatible Requirements for Python Third-party Library Upgrades}


\author{Huashan Lei}
\affiliation{%
  \institution{College of Computer Science and Technology and Key Laboratory for Safety-critical Software Development and Verification, Nanjing University of Aeronautics and Astronautics}
  \city{Nanjing}
  \country{China}}
\email{leihuashan@nuaa.edu.cn}

\author{Guanping Xiao}
\authornote{Guanping Xiao is the corresponding author.}
\affiliation{%
  \institution{College of Computer Science and Technology and Key Laboratory for Safety-critical Software Development and Verification, Nanjing University of Aeronautics and Astronautics; State Key Laboratory for Novel Software Technology, Nanjing University}
  \city{Nanjing}
  \country{China}}  
\email{gpxiao@nuaa.edu.cn}

\author{Yepang Liu}
\affiliation{%
  \institution{Department of Computer Science and Engineering, Southern University of Science and Technology}
  \city{Shenzhen}
  \country{China}}
\email{liuyp1@sustech.edu.cn}

\author{Zheng Zheng}
\affiliation{%
  \institution{School of Automation Science and Electrical Engineering, Beihang University}
  \city{Beijing}
  \country{China}}
\email{zhengz@buaa.edu.cn}





\renewcommand{\shortauthors}{Lei et al.}

\begin{abstract}
Python third-party libraries (TPLs) are essential in modern software development, but upgrades often cause compatibility issues, leading to system failures. These issues fall into two categories: version compatibility issues (VCIs) and code compatibility issues (CCIs). Existing tools mainly detect dependency conflicts but overlook code-level incompatibilities, with no solution fully automating the inference of compatible versions for both VCIs and CCIs. To fill this gap, we propose \tool, the first approach to automatically infer compatible requirements by combining version and code compatibility analysis. \tool integrates six modules: knowledge acquisition, version compatibility assessment, invoked APIs and modules extraction, code compatibility assessment, version change, and missing TPL completion. \tool collects candidate versions, checks for conflicts, identifies API usage, evaluates code compatibility, and iteratively adjusts versions to generate a compatible \mintinline{python}{requirements.txt} with a detailed repair report. To evaluate \tool, we construct \benchmark, a real-world benchmark with 2,095 upgrade scenarios derived from 34 real-world scientific/ML Python projects (including 406 scenarios unsolvable by pip). Results show \tool achieves a 94.03\% inference success rate, outperforming PyEGo (37.02\%), ReadPyE (37.16\%), and LLM-based approaches (GPT-4o, DeepSeek V3/R1) by 18--22\%. \tool processes each scenario from \benchmark in 60.79 s on average, demonstrating practical efficiency. \tool reduces manual effort in troubleshooting upgrades, advancing Python dependency maintenance automation. 
\end{abstract}



\received{20 February 2007}
\received[revised]{12 March 2009}
\received[accepted]{5 June 2009}

\maketitle

\section{Introduction}
Python has emerged as one of the most widely used programming languages, consistently ranking at the top of the TIOBE index
~\cite{TIOBE}. As of December 2025, it accounts for 23.64\% of the overall popularity across programming languages. 
A key factor contributing to Python's success is its rich ecosystem of third-party libraries (TPLs), with over 700,000 packages hosted on the Python Package Index (PyPI)~\cite{PyPI}. These TPLs significantly accelerate development by providing ready-to-use implementations of complex functionality.

Python TPLs undergo frequent updates to address security vulnerabilities, optimize performance, and fix bugs in prior versions  ~\cite{ye2022knowledge}. For example, according to PyPI, TensorFlow ~\cite{abadi2016tensorflow} and PyTorch ~\cite{paszke2019pytorch}, two of the most popular DL frameworks, release an average of 13 and 5 versions per year, respectively~\cite{lei2023deep}. 
Users often proactively upgrade TPLs to leverage these improvements, ensuring enhanced security, efficiency, and software stability. Staying up-to-date helps mitigate risks associated with deprecated features and performance bottlenecks in older versions.  

\begin{figure}[!t]
\centering
\includegraphics[width=\linewidth]{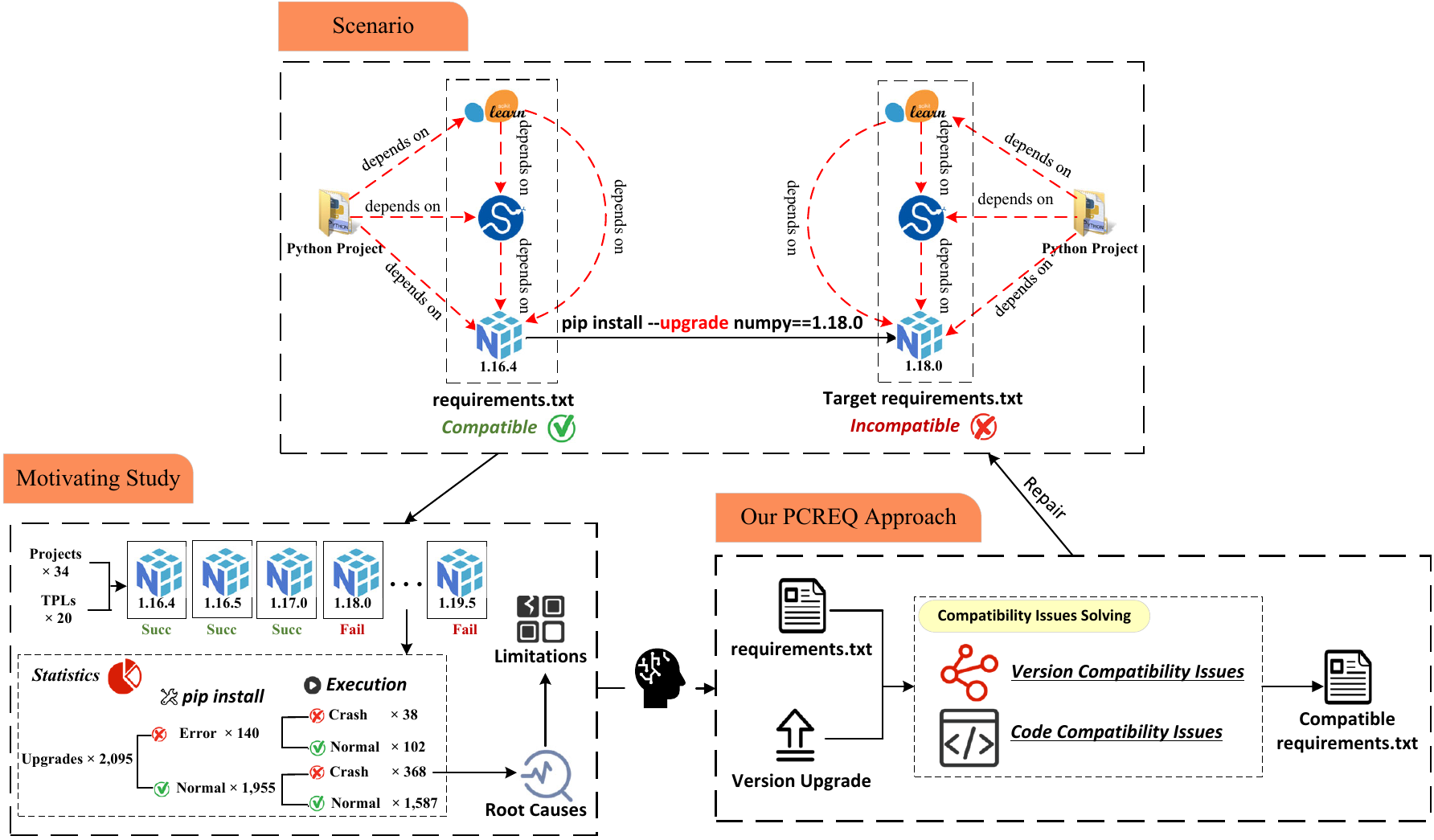}%
\Description{A high-level overview diagram that connects our motivating study to the proposed approach. It summarizes the workflow from the study setup and key observations to the main components of the approach and the resulting outcomes, with arrows indicating the progression between stages.}
\vspace{-4mm}
\caption{Overview of our motivating study and approach.}
\vspace{-4mm}
\label{all}
\end{figure}

As shown in Figure~\ref{all}, 
users usually upgrade TPLs in the local environment by running the command \mintinline{bash}{pip install --upgrade <library>==<version>}. 
However, the upgrade process can introduce compatibility issues that may result in system failures. These problems can be broadly categorized into \textbf{version compatibility issues (VCIs)} and \textbf{code compatibility issues (CCIs)}. Inferring compatible requirements (i.e., \mintinline{python}{requirements.txt}) after TPL upgrades is critical to the proper functioning of Python projects. 

On the one hand, version constraints between TPLs may be violated. When constraints are not satisfied, dependency conflicts can cause environment setup failures (i.e., version compatibility issues, VCIs). 
For example, the Python project \texttt{svoice} has two direct dependencies: torchvision 0.7.0 and torch 1.6.0. When upgrading torch from version 1.6.0 to 1.9.0, the version requirement of torchvision 0.7.0 for torch is strictly limited to 1.6.0 (i.e., ``== 1.6.0''), which will lead to a version dependency conflict between torchvision 0.7.0 and torch 1.9.0.

On the other hand, due to the complexity of the interactions between TPLs and between projects and TPLs, even if the version constraints are met and the build is successful, the project may still crash due to code compatibility issues (CCIs). For example, the Python project \texttt{deep-belief-network} has two direct dependencies: scipy 0.18.1 and scikit-learn 0.18.1. When upgrading scipy from version 0.18.1 to 1.3.0, CCI occurs between scipy and scikit-learn. 
Specifically, scikit-learn 0.18.1 relies on scipy for certain functionality. In the module \mintinline{python}{sklearn.model_selection._split.py}, scikit-learn imports the  \mintinline{python}{comb} function using the import statement \mintinline{python}{from scipy.misc import comb}. While the \mintinline{python}{comb} function exists in scipy 0.18.1, it was removed in version 1.3.0. As a result, when scikit-learn attempts to import \mintinline{python}{comb} after the upgrade, it raises the following error: \textit{ImportError: cannot import name `comb'}.

To better understand the impact of TPL upgrades, 
we conduct 2,095 TPL upgrade experiments covering 34 real-world scientific/ML Python projects and 20 TPLs (Figure~\ref{all}). Our empirical study shows that 140 (6.7\%) upgrades resulted in pip installation errors due to VCIs, of which 38 (27.1\% of the installation error cases) crashed in subsequent program runs, while the remaining 102 (72.9\%) maintained normal program operation despite installation failures. Interestingly, of the 1,955 (93.3\%) installations that completed, 368 (18.8\%) upgrades crashed the program, 
although dependency resolution was passed and there were no version conflicts. Through an in-depth analysis of the 406 project runtime crash cases (368+38), we find that these CCIs exist at two levels, i.e., the interaction between the project and TPLs (Project-TPL), and the interaction between different TPLs (TPL-TPL). Specifically, these issues can be categorized into the following four types: (1) module, (2) API name, (3) API parameter, and (4) API body.

Inferring compatible runtime environments for Python projects has garnered significant research attention, with tools like PyEGo~\cite{ye2022knowledge} and ReadPyE~\cite{cheng2023revisiting} offering solutions in this domain. However, existing approaches predominantly focus on managing version constraints between dependencies, often neglecting deeper code-level compatibility issues. 
While version constraints ensure that dependency requirements are satisfied during installation, they do not inherently guarantee code compatibility across different versions. Even when version specifications are met, subtle API changes, deprecated features, or behavioral modifications in TPLs can introduce runtime errors. This critical gap highlights the need for more advanced tooling that extends beyond version conflict resolution to actively detect and verify code compatibility across TPL versions.

To fill this gap, we propose \tool, a novel approach that automates the inference of compatible requirements for Python TPL upgrades. Unlike existing solutions focusing solely on version constraints, \tool integrates static analysis techniques to assess both version and code compatibility before upgrading a TPL. 
As depicted in Figure~\ref{overview}, \tool comprises the following six components: \ding{182} knowledge acquisition, \ding{183} version compatibility assessment, \ding{184} invoked APIs and modules extraction, \ding{185} code compatibility assessment, \ding{186} version change, and \ding{187} missing TPL completion. Given a Python project, the requirements of the project, the Python version, and the target TPL version to be upgraded, \tool precisely detects version and code compatibility issues and infers the compatible requirements. After automated inference, \tool generates a compatible requirements file and a report, encompassing the detection and repair process.

To evaluate \tool, we construct a real-world benchmark, i.e., \benchmark, including a total of 2,095 upgrade scenarios concerning diverse code compatibility issues. The benchmark covers 406 pip unsolved TPL upgrade scenarios. 
We conduct a comprehensive comparative analysis of \tool against state-of-the-art (SOTA) tools, i.e.,  PyEGo and ReadPyE, in terms of inference performance. Furthermore, we compare \tool with DeepSeek (V3 and R1)~\cite{DeepSeek} and ChatGPT (GPT-4o)~\cite{ChatGPT}, representing popular open-source and closed-source large language models (LLMs), respectively. 
Finally, we assess and discuss the efficiency of \tool.


In summary, we make the following key contributions: 

\begin{itemize}
    \item \textbf{Empirical Study.} We conduct a large-scale empirical study on 2,095 TPL upgrade experiments in real-world Python projects to investigate the frequency and causes of compatibility issues during TPL upgrades. 
    
    \item \textbf{Automated Inference Approach.} We propose and implement a novel approach, \tool, that combines both version and code compatibility analysis to infer the compatible requirements for TPL upgrades in Python projects. 

     \item 
     \textbf{Benchmark.} We construct \benchmark, a real-world benchmark, containing 2,095 upgrade scenarios and their associated labels, for evaluating compatible requirements inference approaches of TPL upgrades in Python projects. 
     
      \item \textbf{Evaluation and Analysis.} 
      We evaluate \tool on \benchmark. In the TPL upgrade scenario, \tool achieves a success rate of 94.03\% in inferring compatible requirements. Compared with SOTA tools, i.e., PyEGo (37.02\%) and ReadPyE (37.16\%), \tool outperforms them by 61.24\% and 58.02\%, respectively. Additionally, when compared with GPT-4o (75.56\%), DeepSeek V3 (71.98\%), and DeepSeek R1 (76.32\%), \tool achieves improvements of 18.47\%, 22.05\%, and 17.71\%, respectively. Finally, we assess the efficiency of \tool, which processes each upgrade scenario in \benchmark in 60.79 s on average. These results demonstrate the effectiveness of \tool in inferring compatible requirements in Python TPL upgrade scenarios.

\end{itemize}


\sloppy
\section{Motivating Study}
\label{sec:motivatingstudy}

Upgrading TPLs in Python projects is essential but often problematic. To better understand the impact of TPL upgrades, we conducted an empirical study on real-world Python projects. The goal is to investigate how often upgrade failures occur due to version or code compatibility issues and to identify their causes. This study also surveys existing tools to highlight gaps and challenges that motivate our approach.

\subsection{Study Design and Methodology}


\textbf{Project Selection.} To examine real-world upgrade issues, we selected a subset of open-source Python projects from a public dataset~\cite{lei2023deep}. 
The dataset consists of real-world deep-learning projects hosted on GitHub, including 90 PyTorch-based and 50 TensorFlow-based projects. The dataset was originally created to support analyses of framework version evolution and compatibility issues, providing a high-quality and reproducible basis for investigating version and code compatibility issues in Python ecosystems. 
To ensure reproducible and analyzable dependency-upgrade experiments, we applied two criteria: (1) the repository must contain a \mintinline{python}{requirements.txt} file that explicitly declares its TPL dependencies; and (2) this file must include version constraints (e.g., ``scipy==0.18.1'', ``numpy>=1.16.0'', ``torch<1.9''). Projects that listed only library names without version information were excluded, as such specifications neither reflect common engineering practice nor enable controlled dependency-upgrade experiments. 
By applying these criteria, we selected 34 projects with complete, well-structured, and version-specific dependency information.

\begin{figure}[!t]
\centering
\includegraphics[width=\linewidth]{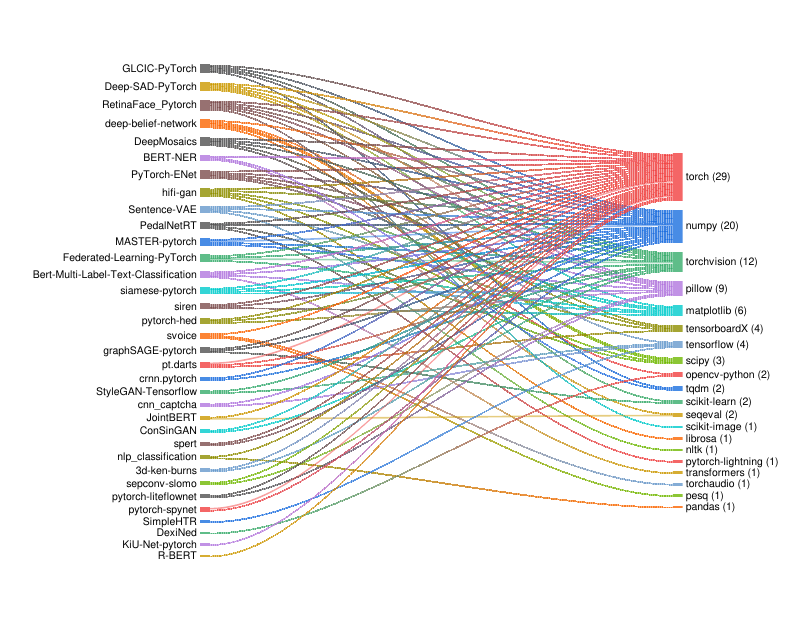}
\Description{A Sankey diagram showing the 103 project-library pairs across 34 projects and 20 third-party libraries.}
\vspace{-4mm}
\caption{Overview of the 103 project-library pairs across 34 projects and 20 TPLs.}
\vspace{-4mm}
\label{dis:protpl}
\end{figure}


\textbf{TPL Selection.} Next, we further performed a systematic filtering and validation process on the TPLs that each project depends on. Specifically, we first identified the directly declared dependencies from the project's \mintinline{python}{requirements.txt} file. These dependencies constitute the potential upgrade targets and define the core scope of subsequent compatibility analyses. However, relying solely on declarations in \mintinline{python}{requirements.txt} is insufficient to accurately reflect the dependencies that a project actually uses. To ensure the validity and representativeness of the selected libraries, we further validated each dependency.

In particular, we employed the open-source tool PCART~\cite{zhang2024pcart} to analyze whether the project truly invokes the APIs of a given TPL. PCART is an automated tool for detecting and repairing API parameter compatibility issues in Python programs. Given the project source code, a target TPL, and the execution command of the project, PCART performs a combined static and dynamic analysis and generates a report that details, for each library, the number of API calls and their runtime coverage. Using this report, we can accurately determine whether a project actually uses the APIs of a given library during execution. If a dependency appears in \mintinline{python}{requirements.txt} but PCART shows that the project never invokes any of its APIs, the dependency is not considered a valid ``upgrade target'' for our study and is excluded from further analysis.

Through this filtering process, we obtained a high-quality project-library mapping. As shown in Figure~\ref{dis:protpl}, we identified 20 actually used TPLs and formed 103 valid project-library pairs across 34 projects. Each pair consists of a project and a library whose APIs are actually invoked, directly indicating the concrete target library that needs to be upgraded and checked for compatibility within the project. As presented in Table~\ref{tab-domain}, the TPLs involved in this study span several important categories, including machine learning (ML), scientific computing, data analysis, and visualization. These libraries not only have high usage frequency in the ecosystem but also exhibit strong representativeness, enabling our study to provide broadly applicable insights into dependency upgrade and compatibility verification problems.

Note that based on the statistical results, each project's \mintinline{python}{requirements.txt} contains 17 TPLs on average. This indicates that, compared with lightweight Python software projects, deep learning projects tend to have substantially higher dependency complexity. Such complexity arises not only from the large-scale nature of ML frameworks themselves but also from the diverse library requirements associated with multiple stages of model development, such as data preprocessing, visualization, model training, and evaluation. This complex dependency structure further highlights the necessity and challenge of performing compatibility analysis during TPL upgrades.

\textbf{Upgrade Procedure.} 
As shown in Figure~\ref{all} (motivating study), for each of the 103 project-library pairs, we performed TPL upgrade experiments. For each target version in the upgrade sequence, we first created a new conda virtual environment using Anaconda (23.5.2)~\cite{Anaconda} to ensure complete isolation. The environment was initialized with the project's original dependencies from \mintinline{python}{requirements.txt} (installed via \mintinline{bash}{pip install -r requirements.txt}), after which we executed the upgrade to the specific target version using \mintinline{bash}{pip install --upgrade <library>==new version}.

The upgraded versions of each target TPL range from the next version of the starting version (listed in \mintinline{python}{requirements.txt}) to the latest stable version as of July 23, 2024. For example, the target TPL torch in the \texttt{crnn.pytorch} project originally uses version 1.2.0, with upgrades available for versions ranging from 1.3.0 to 1.13.1. 
Each intermediate version upgrade was performed in a separate, newly created conda environment to avoid cross-version contamination. 

To accurately characterize project execution under real-world usage scenarios, we followed the execution information provided by the original dataset~\cite{lei2023deep}. The dataset systematically collects all execution commands mentioned in project README files and automatically executes them in a unified environment to obtain a set of successfully runnable scripts. It further measures the project-level code coverage of each runnable script using the Python coverage tool~\cite{coverage}. For each project, we selected the script with the highest coverage as the final execution script to maximize the execution of the project's core logic and its key interactions with the target TPL.

After each upgrade, the selected script is executed to trigger the project's main functionality and its actual calls to the target library. 
Table~\ref{tab-information} shows the total lines of project code, the executed lines, and the corresponding line coverage. 
Beyond this project-level coverage, execution also triggers not only explicit API calls to the target TPL, but also implicitly invoked internal library code and dependency code involved in library-library interactions, thereby forming practical library-level execution coverage.

\begin{table}[!t]
    \centering
    \caption{Selected TPLs and their domain classification}
    \vspace{-4mm}
    \begin{tabular}{|l|c|c|l|c|c|}
        \hline
        \textbf{Domain} & \textbf{Library} & \textbf{APIs} &
        \textbf{Domain} & \textbf{Library} & \textbf{APIs} \\
        \hline

        \multirow{5}{*}{ML} 
            & torch              & 495 & \multirow{2}{*}{Scientific Computing} & numpy  & 109 \\
            & tensorflow         & 97   &                                       & scipy  & 7  \\ \cline{4-6}
            & scikit\_learn      & 20    & \multirow{3}{*}{Visualization}       & matplotlib    & 20 \\
            & transformers       & 5    &                                       & tensorboardX  & 6 \\
            & pytorch\_lightning & 3     &                                       & tqdm          & 5 \\
        \hline

        \multirow{3}{*}{Speech}
            & torchaudio & 3 & \multirow{2}{*}{Computer Vision} & opencv\_python & 42 \\
            & librosa    & 2 &                                   & torchvision   & 33 \\ \cline{4-6}
            & pesq       & 1 & Data Analysis                     & pandas        & 2   \\
        \hline

        \multirow{2}{*}{NLP}
            & seqeval & 4 & \multirow{2}{*}{Image Processing} & pillow        & 9 \\
            & nltk    & 1 &                                   & scikit\_image & 2   \\
        \hline
    \end{tabular}
    \label{tab-domain}
    \vspace{-6mm}
\end{table}

\begin{table}[!t]
    \centering
    \caption{Line coverage information of the selected projects}
     \vspace{-4mm}
    \begin{tabular}{|l|c|c|c|c|c|c|c|}
        \hline
         \textbf{Project} & \textbf{Loc.} & \textbf{Covered} & \textbf{Coverage} 
    & \textbf{Project} & \textbf{Loc.} & \textbf{Covered} & \textbf{Coverage}  \\
    \hline
   
    Sim   & 496 & 232 & 46.77\%  & Sty   & 928 & 644 & 69.40\% \\ \hline
    cnn   & 250 & 207 & 82.80\%  & dee   & 709 & 343 & 48.38\% \\ \hline
    crn   & 261 & 146 & 55.94\%  & Deep   & 1,525 & 863 & 56.66\% \\ \hline
    spe   & 1,665 & 436 & 26.19\%  & Kiu   & 908 & 372 & 40.97\% \\ \hline
    MAS   & 1,634 & 888 & 54.33\%  & nlp   & 332 & 224 & 67.47\% \\ \hline
    3d   & 54 & 54 & 100.00\%  & sep   & 267 & 188 & 70.41\% \\ \hline
    hif   & 526 & 435 & 82.74\%  & pyto   & 248 & 222 & 89.52\% \\ \hline
    Joi   & 558 & 357 & 63.98\%  & Dee   & 2,648 & 810 & 30.58\% \\ \hline
    BER   & 384 & 345 & 89.84\%  & Fed   & 405 & 169 & 41.73\% \\ \hline
    Ber   & 1,397 & 693 & 49.61\%  & sia   & 214 & 174 & 81.31\% \\ \hline
    pt   & 514 & 423 & 82.33\%  & PyT   & 722 & 374 & 51.80\% \\ \hline    
    Con   & 928 & 284 & 30.60\%  & sir   & 83 & 82 & 98.80\% \\ \hline
    Sen   & 394 & 250 & 63.45\%  & pyt   & 63 & 63 & 100.00\% \\ \hline
    svo   & 1,125 & 449 & 39.91\%  & GLC   & 317 & 197 & 62.14\% \\ \hline
    Ped   & 142 & 134 & 94.37\%  & Ret   & 1,227 & 663 & 54.03\% \\ \hline
    gra   & 511 & 341 & 66.73\%  & Dex   & 860 & 385 & 44.77\% \\ \hline
    R-B   & 425 & 300 & 70.59\%  & pytor   & 91 & 91 & 100.00\% \\ \hline
    \end{tabular}
    \label{tab-information}
     \vspace{-4mm}
\end{table}

During the upgrade experiment, we recorded detailed logs of the upgrade process and subsequent project execution, capturing any errors or unusual behavior encountered. Note that we fixed all CUDA/cuDNN-related issues during execution, as they are irrelevant to CCIs or VCIs.  
For example, the installation package of torch 1.8.0 in the PyPI repository lacks the CUDA/cuDNN runtime libraries for Nvidia GPU architecture \mintinline{python}{sm-75} (i.e., Nvidia RTX 2080Ti cards in our experiment environment)~\cite{cudnnerror}. As a result, all the PyTorch projects cannot be executed normally in version 1.8.0. Therefore, we changed the CUDA version to solve this issue.

\textbf{Identifying Compatibility Issues.} We monitored each upgrade attempt for two types of failures: (a) version compatibility issues, and (b) code compatibility issues. Version compatibility issues are identified if the ``\mintinline{bash}{pip install --upgrade}'' command fails to resolve a dependency and an error occurs. As shown in Figure~\ref{piperror}, the upgraded version 1.5.0 of torch violates torchvision's version constraint. Such failures manifest themselves as installation errors. In our study, we counted and analyzed all occurrences of these pip installation errors. 
After the pip upgrade, we run the project regardless of installation errors, because pip forces the installation of the target version of the TPL, even if an error occurs. Any runtime crashes that occur after the upgrade are categorized as code compatibility issues. We investigated error messages and stack traces to diagnose the cause of each runtime failure.

\subsection{Results and Analysis}

\subsubsection{Installation and Runtime Status Analysis}
As depicted in Figure~\ref{all} (motivating study), out of the 2,095 upgrades, there are 140 (6.68\%) instances where the pip installation throws errors, and 1,955 (93.32\%) instances where the pip installation succeeds. 
By analyzing the pip installation logs of all 2,095 upgrade attempts, we further classified the outcomes into three categories: (1) no dependency conflict, (2) dependency conflict that is successfully resolved by pip, and (3) dependency conflict that remains unresolved.

The classification criteria are as follows. If no conflict message appears during dependency resolution and the installation finishes normally without errors, the case is classified as having no dependency conflict. If conflict messages occur but pip successfully resolves them through backtracking and version adjustment and completes the installation, the case is classified as a successfully resolved conflict. If pip fails to find a valid version combination and terminates with installation errors, the case is classified as an unresolved conflict.

The results show that 1,415 cases complete successfully without any dependency conflict. The remaining 680 cases involve dependency conflicts, among which pip resolves 540 cases (79.41\%), while 140 cases (20.59\%) fail due to unresolved conflicts.

\begin{mdframed}[hidealllines=false,backgroundcolor=gray!10,roundcorner=3pt,skipabove=2pt]


\textbf{Finding~1:} In the 2,095 Python TPL upgrade attempts, pip performs well in installing the upgraded library versions, achieving a success rate of 93.32\%. Among the 680 attempts that involve dependency conflicts, pip's dependency resolution capability is able to resolve the vast majority of them (79.41\%).

\end{mdframed}

\begin{figure}[!t]
\centering
\includegraphics[width=4in]{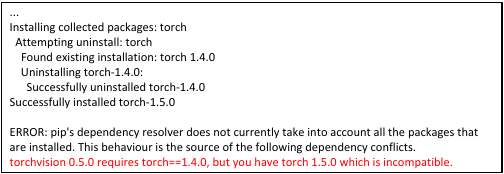}%
\Description{A screenshot of a pip installation log reporting an unresolved dependency conflict: upgrading a package (e.g., torch) violates another installed package's version constraint (e.g., torchvision), and pip prints an error message indicating the incompatibility.}
\vspace{-4mm}
\caption{An example of pip installation error.}
\vspace{-5mm}
\label{piperror}
\end{figure}

If the dependency conflict cannot be resolved during pip installation, pip will report an error. For example, given a working environment where torch 1.4.0 and torchvision 0.5.0 are installed, when performing \mintinline{bash}{pip install --upgrade torch==1.5.0} in the environment, an error will appear (Figure~\ref{piperror}). 
From the error message, we can see that torchvision 0.5.0 requires torch 1.4.0. However, torch's dependency does not require torchvision to be installed.  

Note that even if a dependency conflict during installation cannot be resolved by pip, pip will still force a user-specified upgrade of the TPL. For the example presented in Figure~\ref{piperror}, torch 1.5.0 will be installed into the environment. In the 140 instances where pip has an error during installation, the project is found to crash in 38 (27.14\%) instances, while in 102 (72.86\%) instances the project can run normally. This implies that there is no necessary connection between errors during installation and runtime crashes in program operation.

\begin{mdframed}[hidealllines=false,backgroundcolor=gray!10,roundcorner=3pt,skipabove=2pt]
\textbf{Finding 2:} Version dependency conflicts resulting from unmet constraints do 
not necessarily lead to project execution failures. 

\end{mdframed}

As shown in Figure~\ref{all} (motivating study), of the 1,955 times that pip is installed without errors, 1,587 (81.18\%) times the project runs normally, and 368 (18.82\%) times the project crashes while running. This implies that satisfying the version constraints does not necessarily mean the project will always run properly.




In the runtime environments with version dependency conflicts (140), 102 (72.86\%) times the project can run normally. The success rate of running without version dependency conflicts is higher than that of running with version dependency conflicts. This indicates that the runtime environment without version dependency conflicts is more reliable than the environment with version dependency conflicts.

\begin{mdframed}[hidealllines=false,backgroundcolor=gray!10,roundcorner=3pt,skipabove=2pt]
\textbf{Finding 3:} Although resolving version dependency conflicts alone does not guarantee proper project execution, runtime environments without version dependency conflicts are more reliable than environments with version dependency conflicts.

\end{mdframed}

\subsubsection{Analysis of Runtime Failure Patterns}\label{runtimefailurepattern}
In the following, we analyze the reasons why the project runtime crashes after version upgrades, since the version of the target TPL is upgraded regardless of whether or not an error occurs during the pip installation process.

According to the traceback information, we categorized runtime failure patterns into two-level origins (i.e., Project-TPL and TPL-TPL). Specifically, we performed a systematic manual analysis of all crash instances generated after upgrading the target libraries. For each failure, we first examined the bottom-most entry of the traceback to identify the precise source code line that triggered the exception. We then carefully reviewed the traceback segment by segment, focusing on library names, module paths, and the caller-callee relationships in the call chain. This analysis allowed us to determine whether a failure was directly triggered by project code invoking an upgraded TPL (Project-TPL) or caused by a dependent library internally invoked by that TPL (TPL-TPL). 
Note that all pip installations, project executions, and traceback collections were conducted automatically using a unified script. The subsequent independent annotation was performed on the same set of collected crash reports, rather than on reports reproduced in separate environments. 
To ensure consistency, two authors independently annotated all samples without consulting each other, followed by cross-checking. Disagreements were resolved through joint re-examination and discussion based on the traceback, stack structure, call-chain behavior, and inter-library relationships. This process took approximately one week.

\begin{figure}[!t]
    \centering

   \begin{subfigure}[b]{0.475\textwidth}
        \centering
        \includegraphics[width=\linewidth]{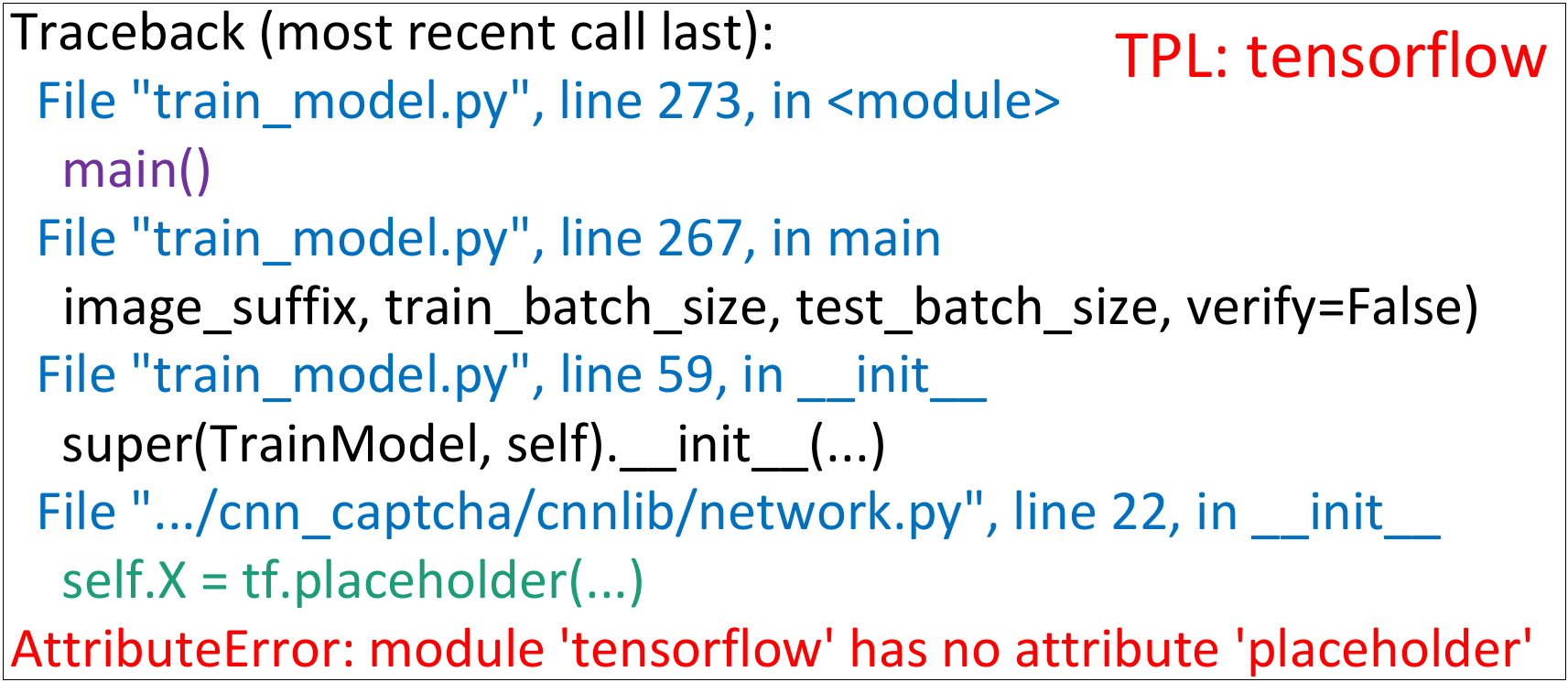}
        \caption{Project-TPL}
        \label{fig:stage:1}
    \end{subfigure}
    \hfill
    \begin{subfigure}[b]{0.475\textwidth}
        \centering
        \includegraphics[width=\linewidth]{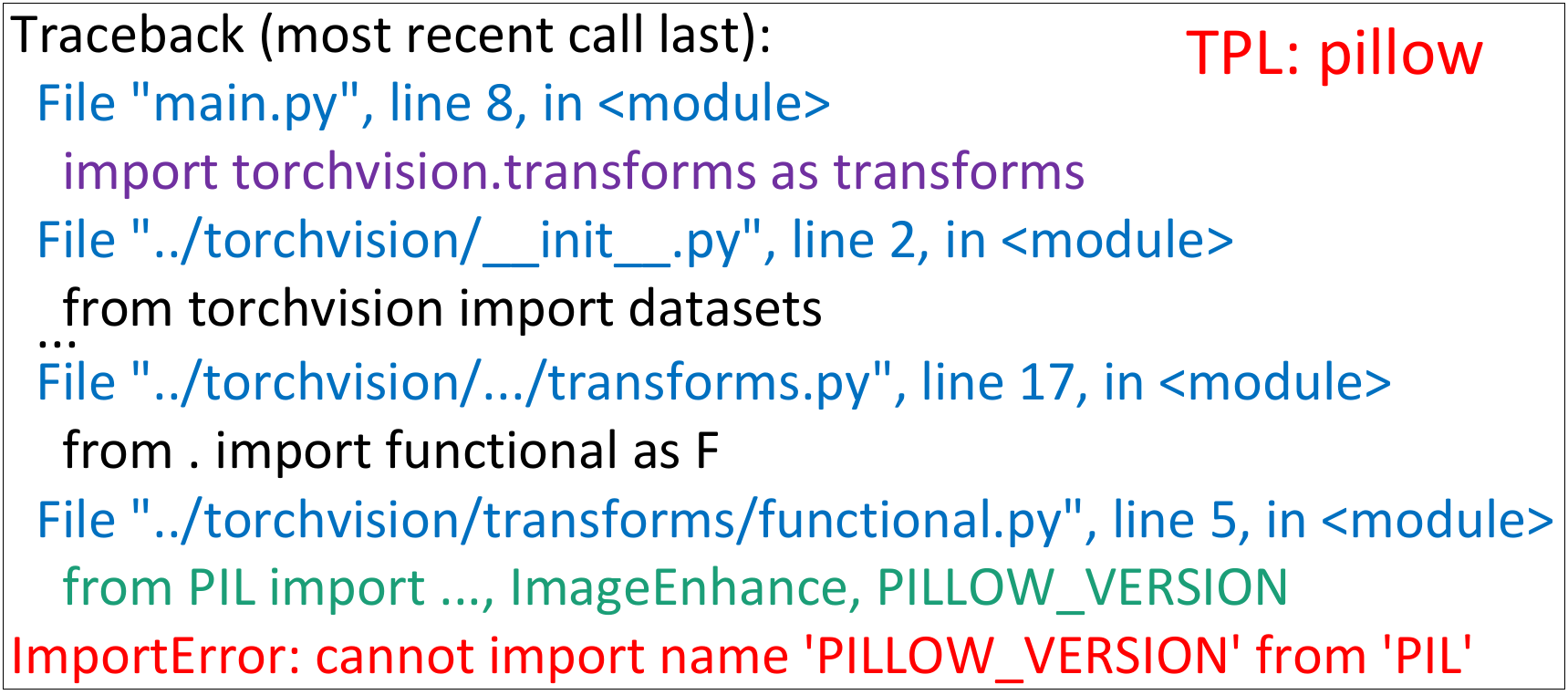}
        \caption{TPL-TPL}
        \label{fig:stage:2}
    \end{subfigure}
\Description{Two side-by-side schematic examples of code compatibility issues. Left: the project directly calls an API in an upgraded third-party library (Project-TPL) and fails due to a breaking change. Right: the project fails indirectly because one library calls into another library (TPL-TPL) where a breaking change occurs.}
\vspace{-4mm}
    \caption{Examples of Project-TPL and TPL-TPL code compatibility issues.}
    \vspace{-4mm}
    \label{fig:stage}
\end{figure}

\begin{figure}[!t]
\centering
\includegraphics[width=\linewidth]{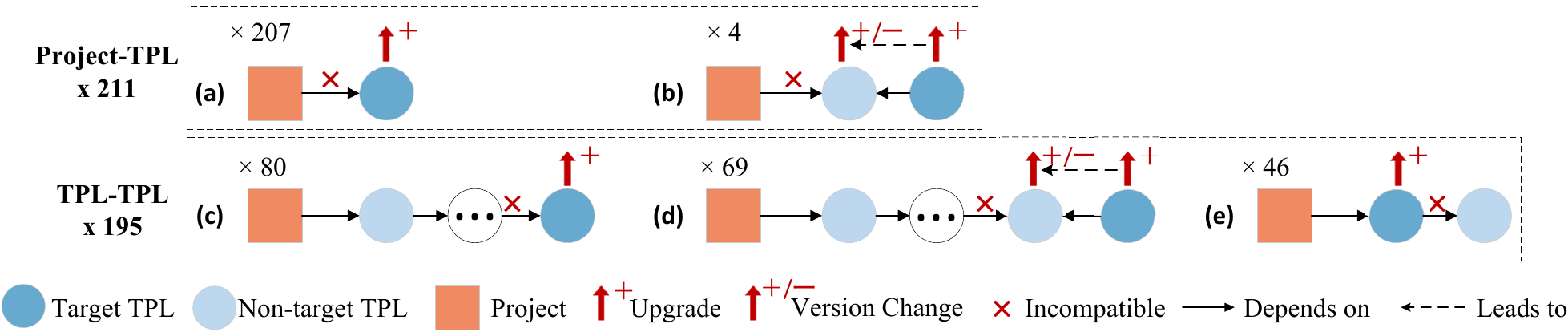}%
\Description{A diagram summarizing the categorized runtime failure patterns observed during third-party library upgrades, illustrating how incompatibilities propagate through direct and transitive dependencies and where the breaking change originates.}
\vspace{-4mm}
\caption{Runtime failure patterns in Python TPL upgrades.}
\vspace{-4mm}
\label{rc}
\end{figure}

\textbf{Level 1: Project-TPL} 
refers to 
the direct use of TPL code in the project leads, which leads to code compatibility issues. A typical example is shown in Figure~\ref{fig:stage:1}. The \texttt{cnn\_captcha} project uses a tensorflow API called \mintinline{python}{tensorflow.placeholder}. However, when tensorflow was upgraded from version 1.7 to 2.0.0, the removal of \mintinline{python}{tensorflow.placeholder} caused the program to crash. As shown in Figure~\ref{rc}, there are a total of 211 Project-TPL issues out of 406 runtime failures.

\textbf{Level 2: TPL-TPL} 
refers to the indirect use of TPL code in the project, which leads to code compatibility issues. As depicted in Figure~\ref{fig:stage:2}. The \texttt{PyTorch-ENet} project indirectly uses the pillow API \mintinline{python}{PIL.PILLOW_VERSION}. After upgrading pillow from 6.2.0 to 9.0.0, since \mintinline{python}{PIL.PILLOW_VERSION} has been removed, an \textit{ImportError: cannot import name `PILLOW\_VERSION' from `PIL'} error occurs. As shown in Figure~\ref{rc}, there are a total of 195 TPL-TPL issues out of 406 runtime failures.

\begin{mdframed}[hidealllines=false,backgroundcolor=gray!10,roundcorner=3pt,skipabove=2pt]

\textbf{Finding 4:} 195 (48.03\%) of the 406 runtime failures are not caused by TPLs that the project directly depends on (Project-TPL), but instead stem from incompatible changes to TPLs in indirect dependencies (TPL-TPL). 

\end{mdframed}

Furthermore, we classified each crash into a fine-grained failure pattern by examining its incompatibility propagation path, as shown in Figure~\ref{rc}. Specifically, we followed four steps:
(1) locating the root exception by inspecting the bottom-most traceback entry to determine which library and API triggered the incompatibility;
(2) analyzing the correspondence between the traceback call chain and the project's dependency graph to understand how execution reached the affected library;
(3) identifying the propagation path of the incompatibility (e.g., whether the project directly invoked a changed API, whether an upstream library triggered a downstream incompatible API, or whether multi-hop dependency propagation caused the failure); and
(4) grouping the failures into five distinct runtime failure patterns based on the position of the incompatible change (direct or indirect dependency), the depth of propagation, and whether single-source or multi-source incompatibilities were involved.

Following the same procedure as above, two authors independently assigned each failure to one of the five patterns based on a shared set of criteria. The annotation results were cross-checked, and disagreements were resolved through collective re-examination of the traceback, stack structure, call-chain behavior, and dependency positions. This process took approximately three weeks.

\begin{itemize}
    \item \textbf{Pattern (a): }
The direct dependency target TPL undergoes incompatible code changes after the upgrade, causing the project to fail at runtime.

\item \textbf{Pattern (b): }
After the target TPL is upgraded, it causes a version change to another TPL that the project directly depends on. The passive version change (upgrade or downgrade) of the TPL results in an incompatible change, which causes the project to fail at runtime.

\item \textbf{Pattern (c): }
The project indirectly uses the target TPL through other TPLs. After the target TPL is upgraded, it undergoes incompatible changes that affect the project's indirect calls, causing it to crash at runtime.

\item \textbf{Pattern (d): }
A TPL indirectly dependent on the project is forced to version change (upgrade or downgrade) due to the upgrade of the target TPL. Consequently, the indirectly dependent TPL undergoes incompatible changes in the new version, resulting in the project failing to run at runtime.

\item \textbf{Pattern (e): }
After upgrading the target TPL, even if it does not cause any changes to the TPL version dependencies, the paths or methods used to access other TPLs through the target TPL in the project may become invalid due to internal structural adjustments, which may in turn cause runtime errors. 

\end{itemize}

\begin{figure}[!t]
\centering
\includegraphics[width=\textwidth]{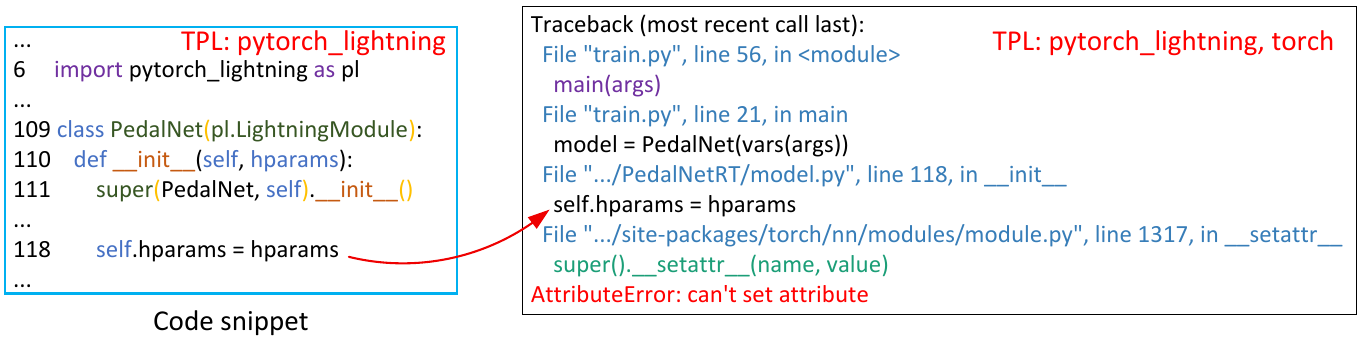}%
\Description{An illustrative example of failure pattern (e), showing how upgrading a target library changes internal structure or attribute-handling behavior, making a previously valid access path invalid and triggering an exception in downstream framework code during execution.}
\vspace{-4mm}
\caption{An example of pattern (e).}
\vspace{-4mm}
\label{patterne}
\end{figure}

Since patterns (a)--(d) are more intuitive, we present an example of pattern (e) for better understanding. Figure~\ref{patterne} illustrates a failure caused by changes in how attributes are managed after upgrading pytorch\_lightning from version 1.1.0 to 1.5.0. In earlier versions, users could directly assign hyperparameters using ``\mintinline{python}{self.hparams = hparams}''. However, in newer versions, \mintinline{python}{hparams} is defined as a read-only attribute, and direct assignment is no longer permitted. When the code attempts to execute \mintinline{python}{self.hparams = hparams}, Python invokes the \mintinline{python}{ __setattr__} method. Because \mintinline{python}{LightningModule} inherits from \mintinline{python}{torch.nn.Module}, this call is delegated to torch's \mintinline{python}{__setattr__} implementation, which detects the attribute as read-only and raises an \textit{AttributeError}. Although this error originates from pytorch\_lightning's updated attribute protection mechanism, it ultimately manifests within torch's internal code. Note that the version of torch does not change during the upgrade of pytorch\_lightning.


\begin{mdframed}[hidealllines=false,backgroundcolor=gray!10,roundcorner=3pt,skipabove=2pt]

\textbf{Finding 5:} Real-world project upgrade failures follow complex incompatibility propagation patterns, which pose significant challenges to the compatible requirements inference for Python TPL upgrades. 
\end{mdframed}

\subsubsection{Fine-grained Analysis of Code Compatibility Issues}
To further analyze the incompatible code changes, we categorized the types of code compatibility issues that introduce the 406 runtime failures. We identified four major categories of code compatibility issues: module, API name, API parameter, and API body. Table \ref{cdi} shows the distributions of code compatibility issues across different types and levels. Below, we describe each category one by one and provide corresponding concrete examples. 

\begin{figure}[!t]
    \centering

   \begin{subfigure}[b]{0.475\textwidth}
        \centering
        \includegraphics[width=\linewidth]{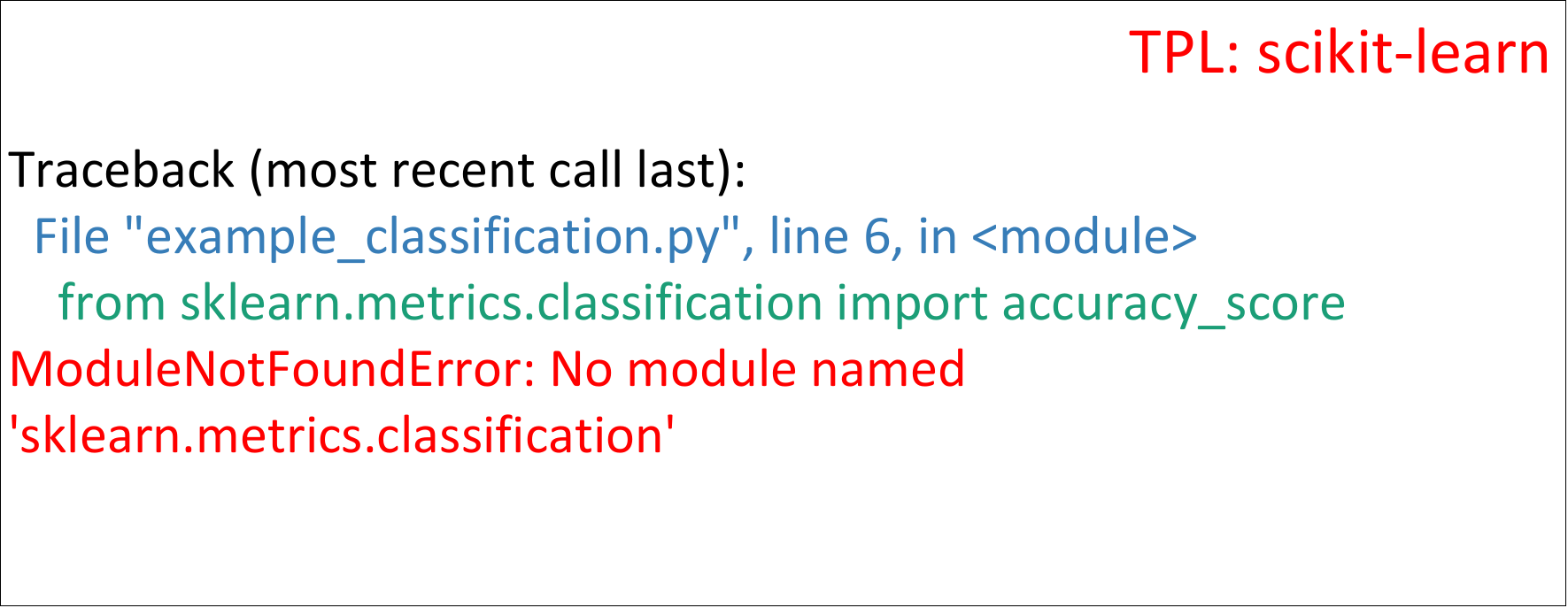}
        \caption{Module}
        \label{fig:fine-grained:module}
    \end{subfigure}
    \hfill
    \begin{subfigure}[b]{0.475\textwidth}
        \centering
        \includegraphics[width=\linewidth]{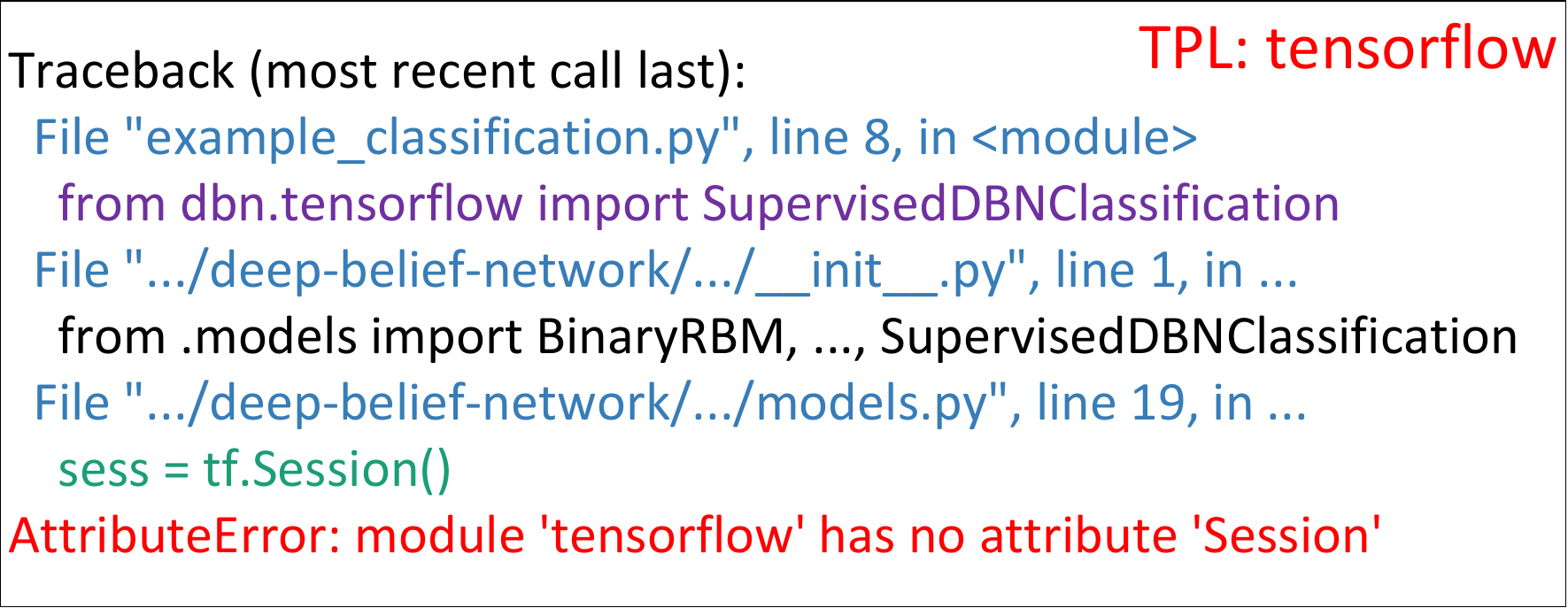}
        \caption{API Name}
        \label{fig:fine-grained:name}
    \end{subfigure}


     \begin{subfigure}[b]{0.475\textwidth}
        \centering
        \includegraphics[width=\linewidth]{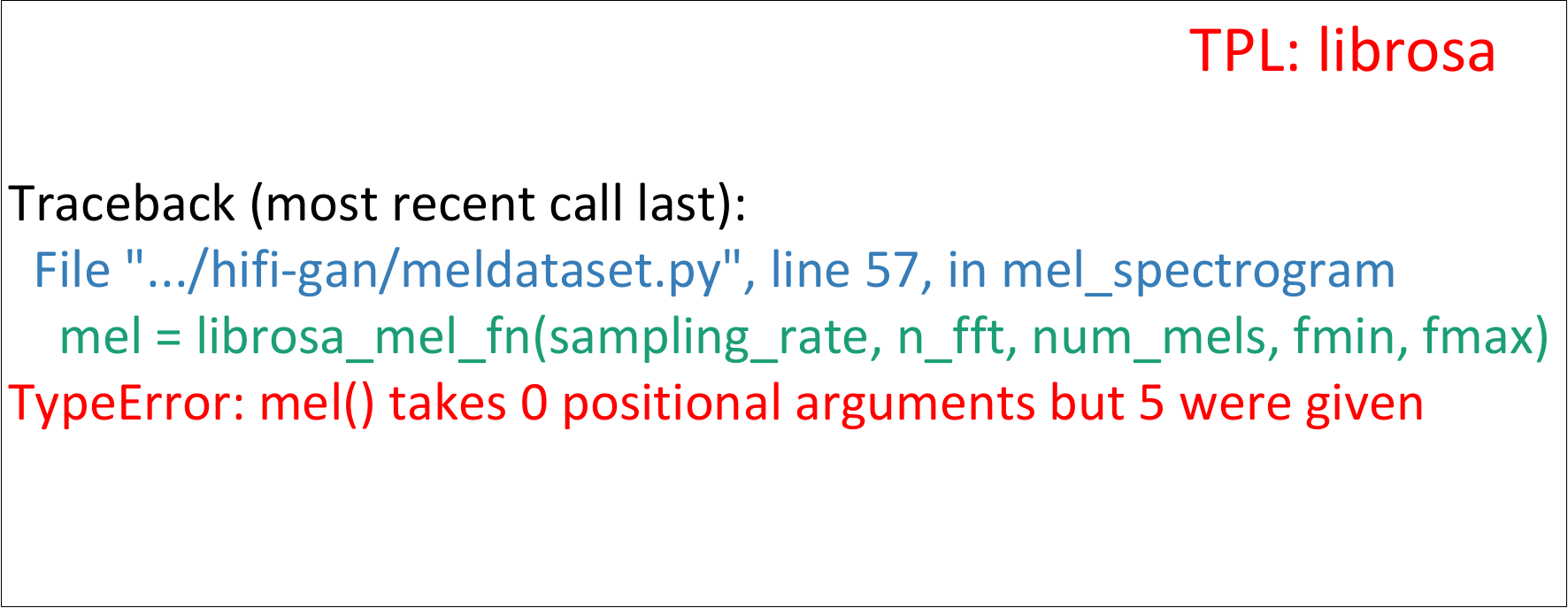}
        \caption{API Parameter}
        \label{fig:fine-grained:paramter}
    \end{subfigure}
    \hfill
    \begin{subfigure}[b]{0.475\textwidth}
        \centering
        \includegraphics[width=\linewidth]{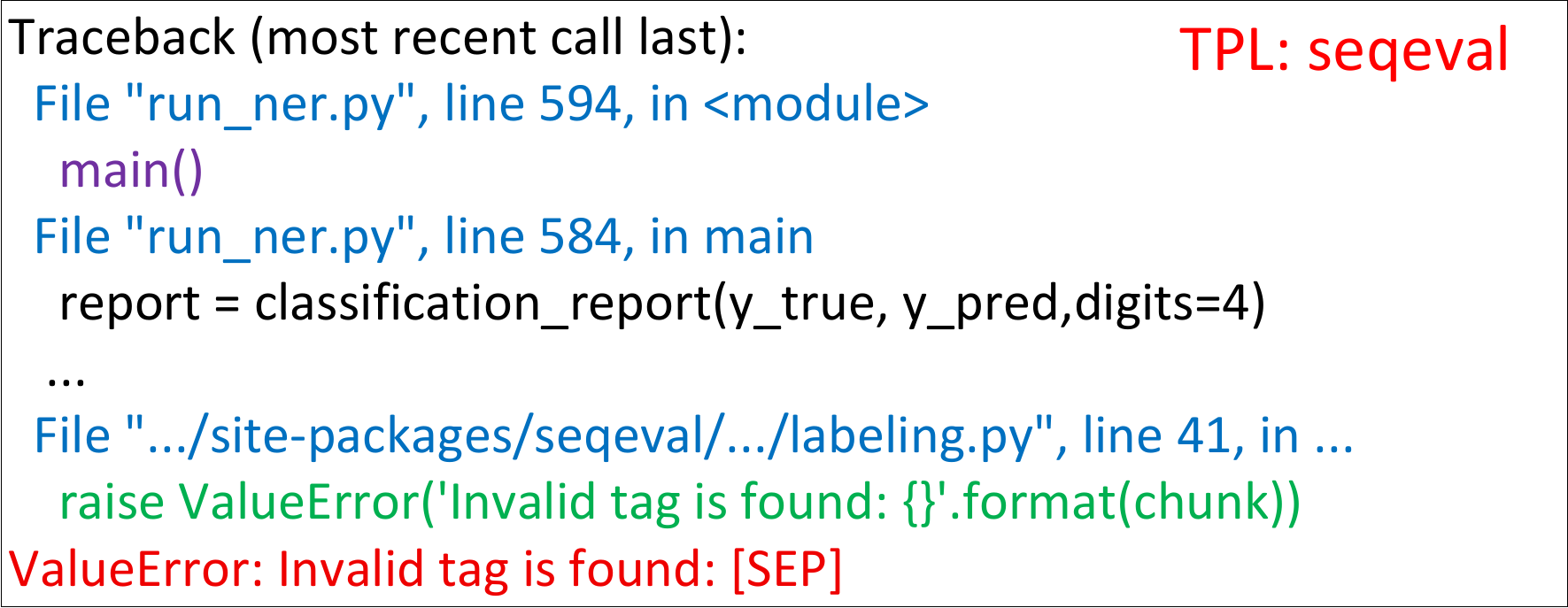}
        \caption{API Body}
        \label{fig:fine-grained:body}
    \end{subfigure}
\Description{Four subfigures illustrating representative code compatibility issue types: a removed or relocated module import; an API renamed or removed; an API signature change (e.g., positional vs. keyword-only parameters); and an internal behavior change that raises new runtime errors without changing the API surface.}
\vspace{-4mm}
    \caption{Examples of module, API name, API parameter, and API body code compatibility issues.}
    \vspace{-4mm}
    \label{fig:fine-grained}
\end{figure}

\textbf{Code Compatibility Issue 1: Module.} 
The code compatibility issue is related to module-breaking changes.  
Figure~\ref{fig:fine-grained:module} illustrates an example of code compatibility issues related to module changes. When the project \texttt{deep-belief-network} uses the statement \mintinline{python}{from sklearn.metrics.classification import accuracy_score} to access the \mintinline{python}{sklearn.metrics.classification} module in scikit-learn version 0.18.1, upgrading scikit-learn to version 0.24.0 results in the removal of the entire \mintinline{python}{sklearn.metrics.classification} module. Consequently, running the project throws a \textit{ModuleNotFoundError: No module named 'sklearn.metrics.classification'} exception. 

\textbf{Code Compatibility Issue 2: API Name.} 
The code compatibility issue is related to API name breaking changes. Figure~\ref{fig:fine-grained:name} shows an example of a code compatibility issue related to API name. In this case, the project \texttt{deep-belief-network} uses a tensorflow API called \mintinline{python}{tensorflow.Session}. However, when tensorflow is upgraded from version 1.5 to 2.0.0, the removal of the API \mintinline{python}{tensorflow.Session} causes the program to crash.

\textbf{Code Compatibility Issue 3: API Parameter.} 
The code compatibility issue is related to API parameter breaking changes. 
As shown in Figure~\ref{fig:fine-grained:paramter}, the project \texttt{hifi-gan} has a parameter compatibility issue due to its source code calls to the librosa's \mintinline{python}{librosa_mel_fn} interface. When librosa is upgraded from 0.7.2 to 0.9.0, the underlying \mintinline{python}{librosa_mel_fn} function undergoes breaking parameter changes, i.e., all positional arguments are removed and keyword arguments are mandatory. This results in a \textit{TypeError: mel() takes 0 positional arguments but 5 were given} exception when calling \mintinline{python}{librosa_mel_fn}, as the new API signature no longer accepts positional arguments.

\textbf{Code Compatibility Issue 4: API Body.}
The code compatibility issue is related to API behavior breaking changes. 
Figure~\ref{fig:fine-grained:body} shows a traceback from the \texttt{BERT-NER} project. This error is introduced by a breaking change when upgrading seqeval from version 0.0.5 to 0.0.15. The new version strictly validates whether the labels are valid when calculating \mintinline{python}{classification_report}, while the old version does not raise an error. Special tags like \mintinline{python}{[SEP]} are ignored in the old version but are now treated as invalid labels and raise a \textit{ValueError} in the new version. This change is not reflected in the API signature. Since the error occurs in the internal validation logic, this makes it difficult to detect or catch in advance.

\begin{table}[!t]
    \centering
     \caption{Code compatibility issues and levels}  
     \vspace{-4mm}
\scalebox{0.8}{
    \begin{tabular}{|l|c|c|c|c|c|}
        \hline
        \textbf{Level} & \textbf{Module} & \textbf{API Name} & \textbf{API Parameter} & \textbf{API Body} & \textbf{Total}\\ \hline
        Project-TPL & 33 & 136 & 20  & 22 & 211\\ \hline
        TPL-TPL & 22 & 111 & 1  & 61 & 195\\ \hline
        Total & 55 & 247 & 21  & 83 & 406\\ \hline
    \end{tabular}
    }
    \label{cdi}
    \vspace{-4mm}
\end{table}

\begin{mdframed}[hidealllines=false,backgroundcolor=gray!10,roundcorner=3pt,skipabove=2pt]


\textbf{Finding 6:} Code compatibility issues are distributed at different code levels, including 
breaking changes in modules, API names, API parameters, and API bodies. 

\end{mdframed}

\subsection{Survey of Related Tools}\label{sec:survey}
\subsubsection{Selection of Survey Tools}

To comprehensively review existing tools for inferring compatible runtime environments for Python programs, we examined those referenced in the SOTA tool, ReadPyE, 
as it offers a current and representative overview of relevant tools in this domain. 
In addition, to reflect the latest technical advances in this area, we also included PLLM~\cite{bartlett2025raiders}, a recently proposed LLM-based compatible environment inference tool, for discussion and comparison. 
For each tool, we outlined its approach, scope, and limitations, and compare its capabilities at both the Project-TPL and TPL-TPL levels. Specifically, we assessed how well each tool addresses version compatibility issues (VCIs) and code compatibility issues (CCIs). 

\begin{itemize}
    \item \textbf{DockerizeMe}~\cite{horton2019dockerizeme} infers a Python code snippet's dependencies by constructing an inter-dependency graph from import statements and package metadata, then generates a corresponding Dockerfile for the required environment. This ensures that the project's required packages are identified (addressing Project-TPL compatibility issues) but does not account for inter-TPL version conflicts or API mismatches.
    
    \item \textbf{V2}~\cite{horton2019v2} detects and mitigates ``configuration drift'' caused by dependency upgrades. It employs feedback-directed search and version-upgrade matrices to explore viable environment configurations, ensuring a stable set of package versions. V2 thereby focuses on maintaining Project-TPL compatibility under updates, but it does not analyze TPL-TPL code interactions.

    \item \textbf{SnifferDog}~\cite{wang2021restoring} restores execution environments for Jupyter notebooks by analyzing notebook code and mapping dependencies to compatible package versions. It reconstructs the required packages for a given notebook (Project-TPL compatibility), ensuring the correct versions are installed. SnifferDog is specialized for notebooks and does not address broader version conflicts between TPLs or their code-level interactions.

    \item \textbf{PyCRE}~\cite{cheng2022conflict} uses a domain-specific knowledge graph to perform conflict-aware dependency inference. By integrating package metadata with compatibility rules, PyCRE suggests package versions that avoid known incompatibilities. This approach covers both Project-TPL compatibility and some TPL-TPL version compatibility (preventing known TPL version conflicts), though it does not explicitly analyze code-level API compatibility between TPLs.

    \item \textbf{PyEGo}~\cite{ye2022knowledge} statically analyzes Python source code (syntax and import statements) to extract required TPLs without execution. It leverages a knowledge-based method to infer dependencies from code structure and uses version constraint solving to select appropriate TPL versions. PyEGo thereby ensures the project has the necessary TPLs (Project-TPL compatibility) and attempts to resolve version compatibility among them. However, it has only limited handling of code compatibility issues, since it does not deeply inspect API changes between TPLs (missing most code-level incompatibilities)

    \item \textbf{ReadPyE}~\cite{cheng2023revisiting} is a knowledge-driven environment inference tool that refines dependency predictions using historical package data and known constraints. It builds a knowledge graph of past TPL versions and compatibility fixes, and uses it to adjust project requirements until a working environment is found. This approach covers Project-TPL compatibility and many version incompatibilities between TPLs.

    \item \textbf{PLLM}~\cite{bartlett2025raiders} is an automated compatible environment inference approach based on LLMs. It uses retrieval-augmented generation and runtime feedback to iteratively infer and validate dependency configurations for Python programs. PLLM mainly focuses on recovering an executable Project-TPL environment, but does not explicitly analyze API interactions among TPLs, making it hard to cover code compatibility issues at the TPL-TPL level.
    
\end{itemize}

Some dependency conflict detection or resolution tools, such as Watchman~\cite{wang2020watchman}, PyDFix~\cite{mukherjee2021fixing}, SmartPip~\cite{wang2022smartpip}, Decide~\cite{zhao2023knowledge}, and LooCo~\cite{wang2023automatically}, only focus on version compatibility issues, not code compatibility issues. The scope of these tools is dependency conflict detection or resolution, rather than inferring compatible environments. Thus, these tools are not discussed in our paper.

\begin{table}[!t]
\centering
\caption{Comparison of related work in code compatibility and version compatibility}
\vspace{-4mm}
\label{table1}
\scalebox{0.8}{
\begin{tabular}{|c|c|cc|c|c|}
\hline
\multirow{2}{*}{\textbf{Approach}}                & \multirow{2}{*}{\textbf{VCI}} & \multicolumn{2}{c|}{\textbf{CCI}}               & \multirow{2}{*}{\textbf{Project}} & \multirow{2}{*}{\textbf{Real-time}}  \\ \cline{3-4}
                                    &                          & \multicolumn{1}{c|}{\textbf{Project-TPL}} & \textbf{TPL-TPL} &                                  &                             \\ \hline
DockerizeMe~\cite{horton2019dockerizeme} & \ding{55}            & \multicolumn{1}{c|}{\ding{51}}          & \ding{55}            & \ding{55}                           & \ding{55}                            \\ \hline
V2~\cite{horton2019v2}                   & \ding{55}            & \multicolumn{1}{c|}{\ding{51}}          & \ding{55}            & \ding{55}                           & \ding{55}                         \\ \hline
SnifferDog~\cite{wang2021restoring}      & \ding{55}            & \multicolumn{1}{c|}{\ding{51}}          & \ding{55}            & \ding{55}                           & \ding{55}                        \\ \hline
PyCRE~\cite{cheng2022conflict}           & \ding{51}            & \multicolumn{1}{c|}{\ding{51}}          & \ding{55}            & \ding{55}                           & \ding{55}                          \\ \hline
PyEGo~\cite{ye2022knowledge}             & \ding{51}            & \multicolumn{1}{c|}{\ding{51}}          & \ding{55}            & \ding{51}                           & \ding{55}                          \\ \hline
ReadPyE~\cite{cheng2023revisiting}       & \ding{51}            & \multicolumn{1}{c|}{\ding{51}}          & \ding{55}            & \ding{51}                           & \ding{55}                          \\ \hline
PLLM~\cite{bartlett2025raiders}    & \ding{51}     & \multicolumn{1}{c|}{\ding{51}}    & \ding{55}    & \ding{51}  & \ding{51} \\ \hline
\textbf{\tool}                       & \ding{52}            & \multicolumn{1}{c|}{\ding{52}}          & \ding{52}            & \ding{52}                           & \ding{52}                          \\ \hline
\end{tabular}
}
\vspace{-4mm}
\end{table}

\subsubsection{Summary of Limitations in Existing Tools} 
Table~\ref{table1} presents the detailed comparison of surveyed tools. 
DockerizeMe, V2, and SnifferDog primarily support the CCIs (Project-TPL), whereas PyCRE, PyEGo, ReadPyE, and PLLM extend their support to include both the CCIs (Project-TPL) and VCIs. Only PyEGo, ReadPyE, and PLLM support project-level inference. In contrast, the remaining tools support only the file-level inference. DockerizeMe, V2, and SnifferDog require downloading relevant knowledge locally to build offline knowledge bases, hindering real-time perception of changes in the knowledge base. Moreover, PyCRE, PyEGo, and ReadPyE build relevant knowledge into knowledge graphs. However, to keep the knowledge graphs up-to-date and complete, they require constant updates to module information and version relationships, which incurs high computational and labor costs.  
Although PLLM uses real-time PyPI metadata for version-related reasoning and dependency resolution, its traceback analysis and code-compatibility reasoning still rely primarily on the internal knowledge of the LLM. As a result, it may still be limited by outdated code knowledge and insufficient coverage when handling code-level compatibility issues.


\subsection{Challenges of Compatible Requirements Inference in Python TPL Upgrades}\label{sec:challenge}
While version constraints ensure that dependent TPLs are installed correctly, they do not guarantee code compatibility across Project-TPL and TPL-TPL dependency levels. Below, we summarize three key challenges to the inference of compatible requirements in TPL upgrade scenarios.

\begin{figure}[!t]
\centering
\includegraphics[width=\linewidth]{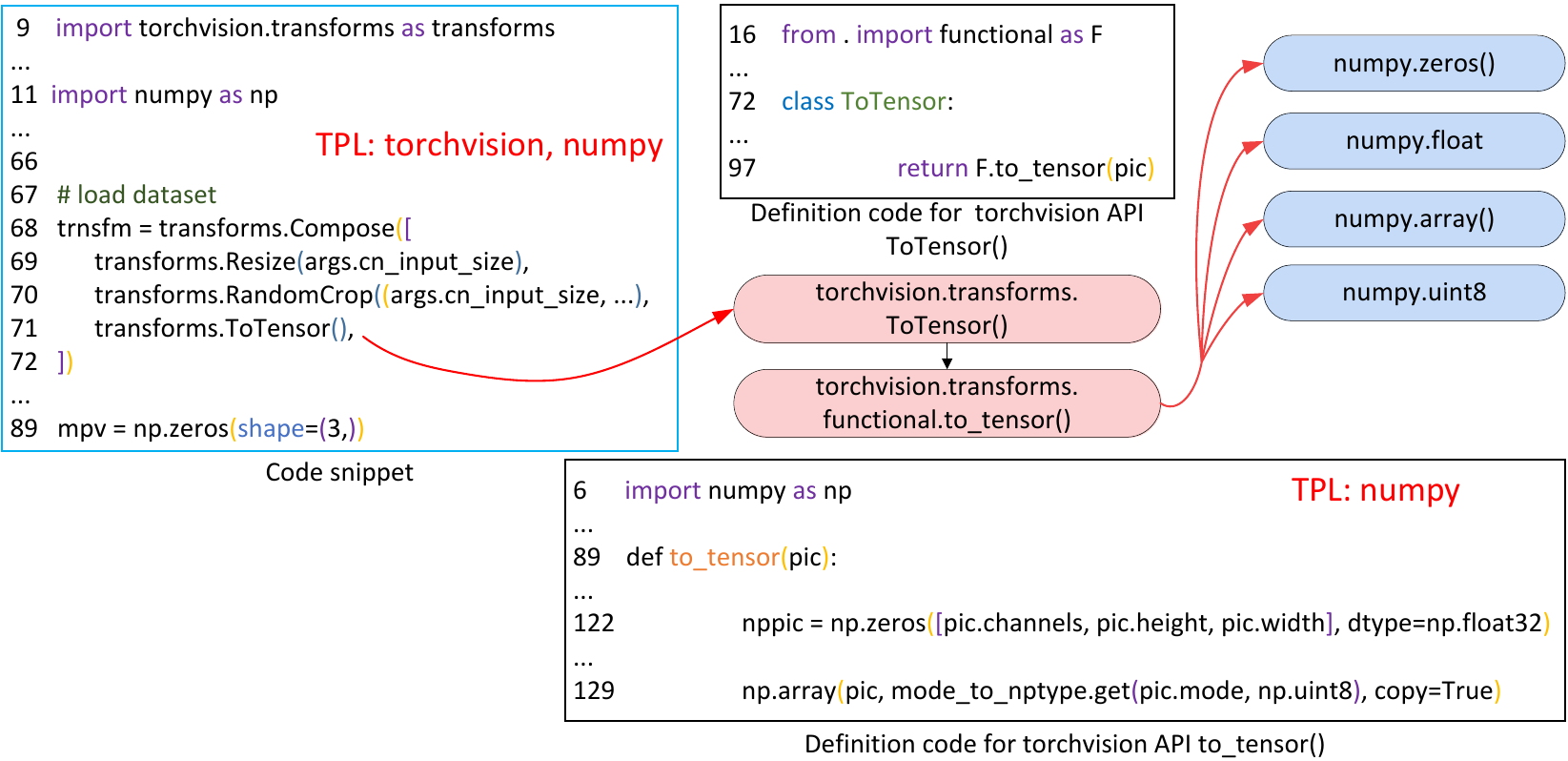}%
\Description{An example dependency chain where a project calls an API from a direct dependency, which internally invokes functions in a transitive dependency. Upgrading the transitive dependency removes an API (e.g., numpy.float), causing the project to crash even though the project never directly calls that removed API.}
\vspace{-4mm}
\caption{An example of transitive dependency incompatibility.}
\vspace{-4mm}
\label{cdg2}
\end{figure}

\textbf{Challenge 1.} \textit{Transitive Dependency Incompatibility.}
As discussed in Section~\ref{runtimefailurepattern}, implicit incompatibilities caused by TPL-TPL dependencies are the challenge. Many Python projects rely on multiple TPLs, and interdependencies between these TPLs can lead to hard-to-see compatibility issues. As the example presented in Figure~\ref{cdg2}, the source code of the project \texttt{siamese-pytorch} directly uses an API from torchvision, namely \mintinline{python}{torchvision.transforms.transforms.Compose}. Through this API, the project indirectly calls \mintinline{python}{numpy.zeros()}, \mintinline{python}{numpy.float}, \mintinline{python}{numpy.array()}, and \mintinline{python}{numpy.uint8}. After changing numpy from version 1.19.5 to 1.24.0, \mintinline{python}{numpy.float} has been removed, leading to a runtime failure, i.e., \textit{AttributeError: module `numpy' has no attribute `float'}. 

In this case, the set of code entities to be considered extends beyond the scope of the primary Python project and requires a comprehensive examination of all external TPLs and their interdependencies. This includes not only direct code entity calls made by the project to these TPLs but also indirect calls made through any other dependent TPLs. Given the complexity of these interactions, a thorough analysis of all relevant dependent code entities is required to fully understand the dependencies and behavior during the execution of the project.

\begin{figure}[!t]
\centering
\includegraphics[width=\linewidth]{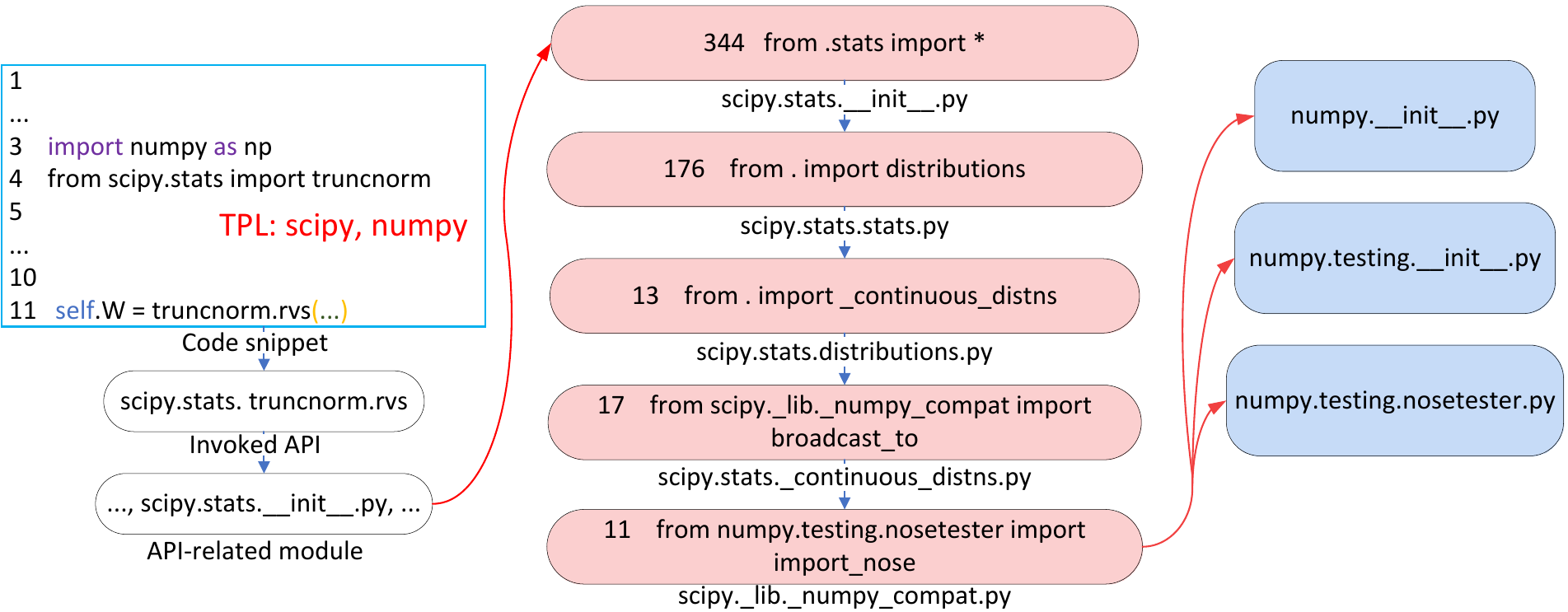}%
\Description{A schematic illustrating Python's import mechanism along an API's module path: calling a library API triggers a cascade of imports across multiple modules, and upgrading a transitively imported module (e.g., a numpy testing submodule) can break execution with a missing-module error.}
\vspace{-4mm}
\caption{An example of import mechanism 1.}
\vspace{-4mm}
\label{cdg1}
\end{figure}

\begin{figure}[!t]
\centering
\includegraphics[width=\linewidth]{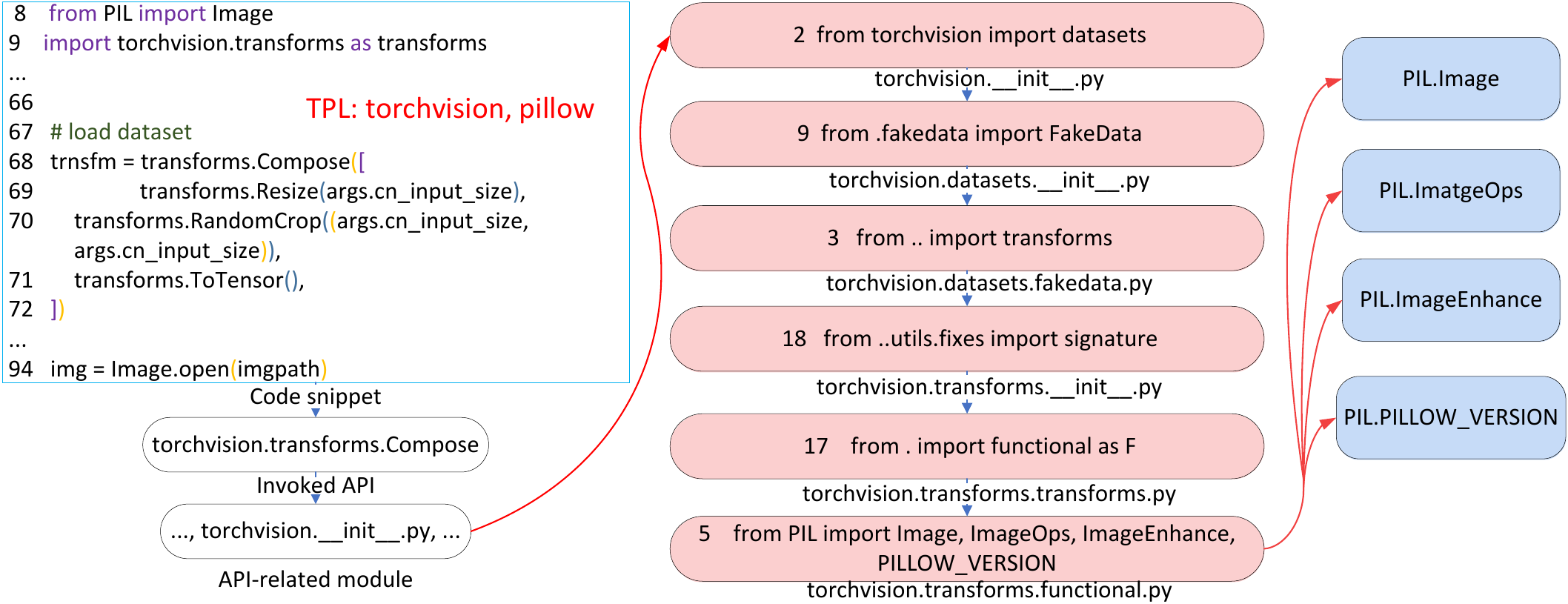}%
\Description{A schematic showing an import chain where a project calls a direct dependency API (e.g., torchvision), which imports another library's API (e.g., pillow's PIL.PILLOW_VERSION). After upgrading the indirectly imported library, the imported API is removed, leading to an import-time failure.}
\vspace{-4mm}
\caption{An example of import mechanism 2.}
\vspace{-4mm}
\label{cdg3}
\end{figure}

\textbf{Challenge 2.} \textit{The Import Mechanism in Python.}
Python's import mechanism further broadens the scope of potential breakages. When a project calls a TPL API, Python will load all modules and APIs along that API's import path. An upgrade that removes or renames anything in this chain can cause errors even if the project never directly refers to it.

As shown in Figure~\ref{cdg1},  the project \texttt{deep-belief-network} directly invokes the SciPy API \mintinline{python}{scipy.stats.truncnorm.rvs}. This call involves the \mintinline{python}{scipy.stats.__init__.py} module, which in turn imports the \mintinline{python}{scipy._lib._numpy_compat.py} module. The \mintinline{python}{scipy._lib._numpy_compat.py} module further imports \mintinline{python}{numpy.testing.nosetester.py} via an import statement. However, the transitively imported \mintinline{python}{numpy.testing.nosetester.py} module has been removed after upgrading numpy from version 1.16.4 to 1.18.0, leading to a runtime failure, i.e., \textit{ModuleNotFoundError: No module named `numpy.testing.nosetester'}.

Similarly, as depicted in Figure~\ref{cdg3}, the project \texttt{siamese-pytorch} directly calls the torchvision API \mintinline{python}{torchvision.transforms.Compose}, which involves the \mintinline{python}{torchvision.__init__.py} module. This module, in turn, imports \mintinline{python}{torchvision.transforms.functional.py}. Within \mintinline{python}{torchvision.transforms.functional.py}, the API \mintinline{python}{PIL.PILLOW_VERSION} is imported. When upgrading pillow from version 6.2.0 to 8.0.0, the project crashes because the \mintinline{python}{PIL.PILLOW_VERSION} has been removed.

\textbf{Challenge 3.} \textit{Code Evolution Complexity.}
The variety of Code changes during evolution also poses a challenge. TPLs may remove or relocate entire modules, rename APIs or parameters, or make behavior-breaking changes to existing APIs. Our fine‐grained analysis confirms that all of these change types, i.e., module, API name, API parameter, and API body, occur in practice. 
The diverse code entity changes require the inference tool to perform a thorough assessment of code compatibility across Project-TPL and TPL-TPL level dependencies.

\section{Our \tool Approach}\label{sec:approach}

\begin{figure}[!t]
\centering
\includegraphics[width=\linewidth]{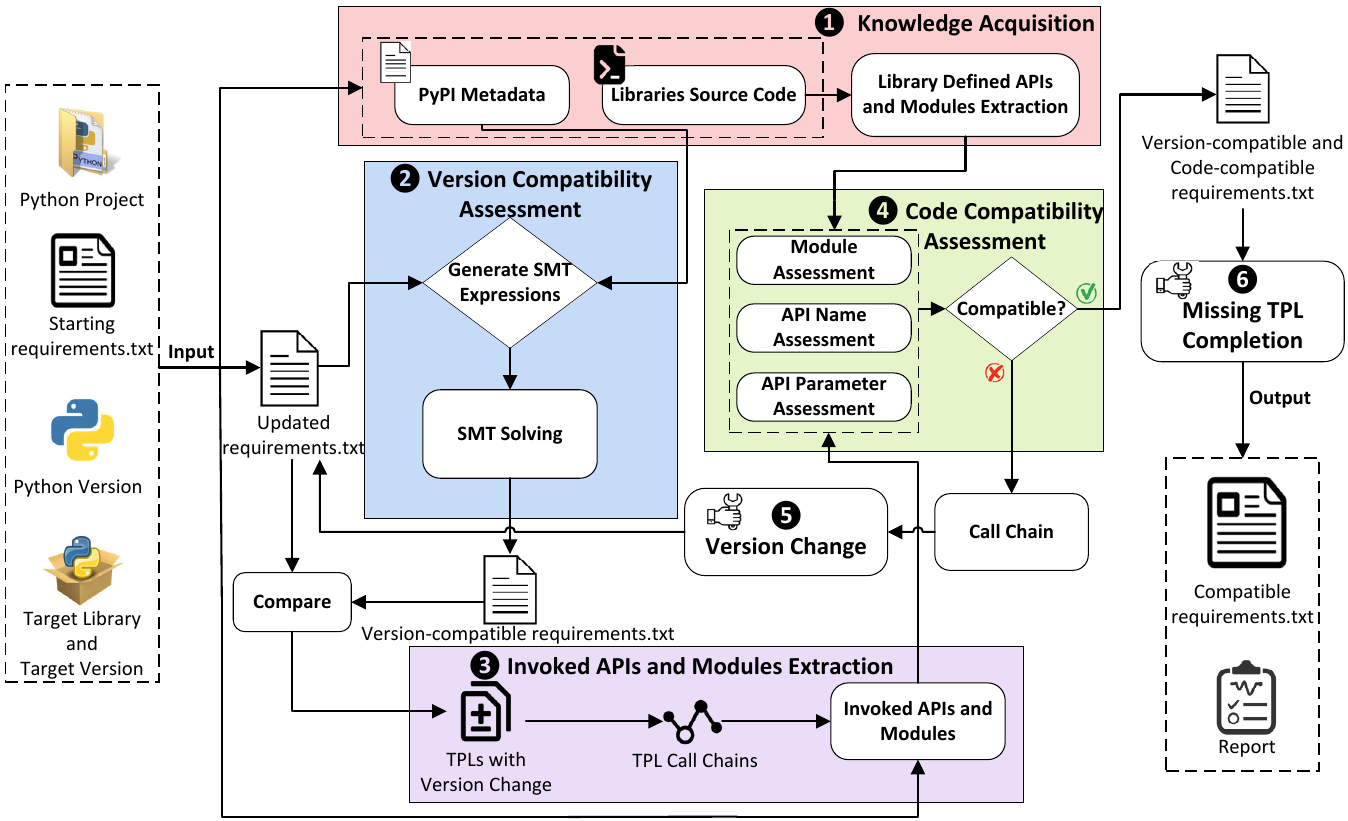}%
\Description{A high-level workflow diagram of the \tool approach, showing the main phases (knowledge acquisition, version compatibility assessment, and code compatibility assessment) and the data flow between inputs (project requirements and configuration) and outputs (inferred compatible requirements).}
\vspace{-4mm}
\caption{Overview of our \tool approach.}
\vspace{-4mm}
\label{overview}
\end{figure}

\subsection{Overview}
To address the challenges above, we introduce \tool with the following key advantages.

\begin{itemize}
    \item \textbf{Fully Automated Inference of Compatible Requirements.} \tool can automatically infer compatible requirements for Python TPL upgrades. As shown in Table \ref{table1}, \tool supports VCIs, CCIs (Project-TPL and TPL-TPL) detection, which can detect potential runtime failures due to code changes in direct or indirect TPLs, thus significantly improving the inference success rate of Python TPL upgrades. 

    \item \textbf{Real-time Knowledge.} \tool obtains the corresponding source code and version information from PyPI in real time, which ensures real-time knowledge and avoids the limitations of knowledge graphs, such as the need to constantly update module information and version relationships to maintain the timeliness and completeness of knowledge graphs, which brings higher computational and labor costs.
    
    
    \item \textbf{Support for Transitive Dependency Analysis.} \tool recursively analyzes TPLs on which the project indirectly depends, building a complete dependency path to detect code compatibility issues across TPLs, not just those on which the project directly depends.
\end{itemize}

Figure~\ref{overview} shows the overview of \tool. 
The process begins with gathering all relevant dependency knowledge, including TPLs' version constraints and candidates, and TPLs' source code. Next, \tool evaluates the version compatibility of the updated requirements after upgrading the target TPL to a new version, generating a version-compatible requirements file through satisfiability modulo theories (SMT) solving. 
Subsequently, \tool extracts all invoked APIs and modules from the project and conducts a code compatibility assessment. If the code is fully compatible, \tool outputs the finalized requirements. Otherwise, if code compatibility issues are detected, \tool adjusts the TPL versions accordingly. This iterative process continues until a requirements file that satisfies both version and code compatibility is successfully generated. In the final step, any newly required TPLs are added to complete the requirements.

Below, we elaborate on the design details of \tool, which consists of six main modules:  \ding{182} knowledge acquisition, \ding{183} version compatibility assessment, \ding{184} invoked APIs and modules extraction, \ding{185} code compatibility assessment, \ding{186} version change, and \ding{187} missing TPL completion.


\subsection{Knowledge Acquisition}
\tool needs to have all the candidate versions of all the TPLs in requirements under the specified Python version, as well as the metadata for all the candidate versions, and the source code in a subsequent step, so it downloads all the required knowledge locally in advance.

\subsubsection{Version-related Knowledge} \label{versionknowledge}

First, \tool begins by parsing the input \mintinline{python}{requirements.txt} file line by line. Each line in this file typically specifies a TPL along with its version in the format ``\textit{library==version}''. \tool splits each line using the delimiter ``=='' to separate the TPL name from its specified version. This process results in the construction of an initial list containing all the TPLs mentioned in the file, represented as $[TPL_1, TPL_2,..., TPL_n]$.

\tool then proceeds to iterate through each TPL in the list. For each TPL, \tool uses its name in conjunction with the target Python version, as specified by the user input (Figure~\ref{config}), to query PyPI. This query allows us to retrieve a comprehensive list of all the available versions of the corresponding TPL that are compatible with the specified Python interpreter version.

Once we have obtained the list of compatible versions for each TPL, \tool stores this data locally in JSON format. This step not only facilitates efficient access for subsequent processing but also provides a persistent record of version information that can be reused or analyzed later.

Following the version retrieval and storage step, \tool initiates a metadata crawling process. For every candidate version of each TPL identified in the previous step, \tool extracts detailed metadata directly from the PyPI website. Note that the relevant dependency information is in the \mintinline{python}{requires_dist} attribute of the metadata. 

\begin{figure}[!t]
    \centering

   \begin{subfigure}[b]{0.475\textwidth}
        \centering
        \includegraphics[width=\linewidth]{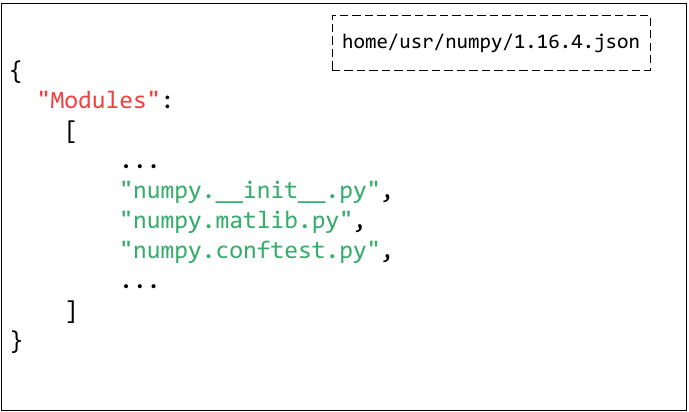}
        \caption{Module}
        \label{fig:knowledge:1}
    \end{subfigure}
    \hspace{1em}
    \begin{subfigure}[b]{0.475\textwidth}
        \centering
        \includegraphics[width=\linewidth]{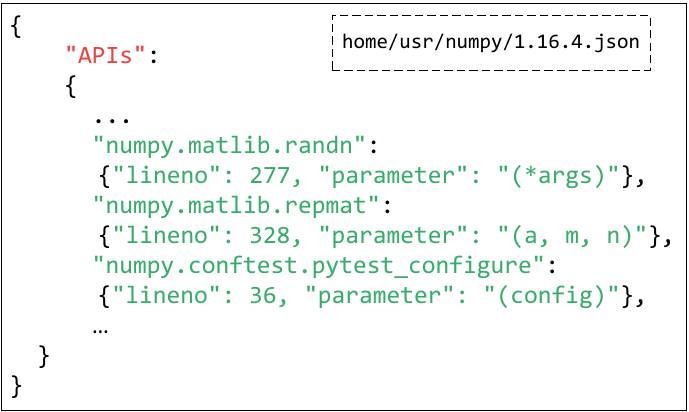}
        \caption{API}
        \label{fig:knowledge:2}
    \end{subfigure}
\Description{Two side-by-side examples of the JSON-based knowledge storage produced by \tool. The left subfigure shows how extracted module paths are recorded, and the right subfigure shows how extracted APIs are recorded with their fully qualified names and associated metadata such as definition location and parameter information.}
\vspace{-4mm}
    \caption{Storage format for extracted code elements.}
    \label{fig:knowledge}
    \vspace{-4mm}
\end{figure}

\subsubsection{Code-related Knowledge} \label{codeknowledge}
\tool obtains all the relevant source code from PyPI based on the TPLs obtained in the previous step, as well as the candidate versions. Then, \tool extracts both modules and APIs from the source code. The storage format is shown in Figure~\ref{fig:knowledge}.

\textit{(1) Extracting Modules Defined in TPLs.} 
The extraction begins once the source code of the TPL has been downloaded and decompressed. To traverse the directory hierarchy of the TPL, \tool employs Python's built-in os.walk() function. This utility enables a recursive exploration of the file system, yielding each directory and its contents in a top-down manner. Importantly, \tool retains full path information throughout the traversal process, enabling the reconstruction of module namespaces based on the original directory structure.

During the scanning phase, \tool identifies two main types of Python components. First, it detects standalone Python module files, defined as any file ending with a \mintinline{python}{.py} extension. These files typically contain classes, functions, or executable statements and represent the fundamental building blocks of a TPL's functionality. Second, it identifies Python package directories that include an \mintinline{python}{__init__.py} file, which signals to the interpreter that the directory should be treated as a package and enables nested imports within that namespace.

To ensure that the extraction focuses solely on relevant source code, \tool implements a set of filtering heuristics. These filters exclude directories and files that are unrelated to the TPL's core logic. For example, it ignores caching folders such as \mintinline{python}{__pycache__}, test directories, documentation, build artifacts, and any non-Python resources (e.g., images, configuration files, or compiled binaries). This filtering step reduces noise and enhances the accuracy of the extracted module list.

The output of this process is a structured inventory of all identifiable Python modules within the TPL, including their fully qualified import paths derived from their location in the directory tree, as shown in Figure~\ref{fig:knowledge:1}. 

\textit{(2) Extracting APIs Defined in TPLs.} 
First, \tool parses each TPL source file into an abstract syntax tree (AST). Through traversing the AST, all \mintinline{python}{FunctionDef}, \mintinline{python}{ClassDef}, and \mintinline{python}{Assign} nodes are identified, corresponding to the definition (signature) statements of functions, classes, and global variables in the code, respectively. One complicated form of API definitions is the nested definitions, i.e., classes defined within classes and functions defined within functions. 
It is essential to accurately discern the hierarchical relationships between classes and the affiliations among APIs to construct the correct path for each API within the source code. Thus, the depth-first search (DFS) algorithm is employed for navigation.

In addition, regarding the built-in APIs, i.e., C extension APIs, developers usually declare their definitions in stub files (\mintinline{python}{.pyi}). 
Hence, \tool attempts to parse \mintinline{python}{.pyi} files to acquire the definitions of built-in APIs. 
Finally, by considering the class to which an API belongs, the module that contains the class, and the package that encompasses the module, \tool constructs the fully qualified path of each API within the source code.

\begin{figure}[!t]
\centering
\includegraphics[width=\linewidth]{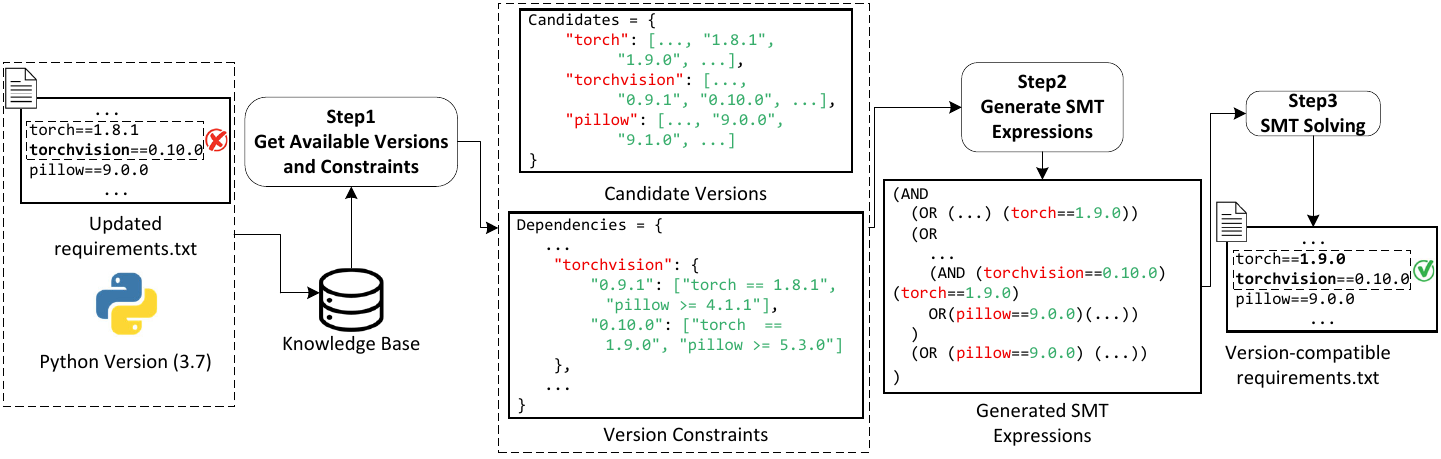}%
\Description{An illustrative example of encoding dependency version constraints into an SMT problem: library versions are represented as ordered variables, constraints capture specifiers (e.g., >= and <), and the solver searches for a version assignment that satisfies all constraints to produce a version-compatible requirements set.}
\vspace{-4mm}
\caption{An example of version compatibility assessment.}
\vspace{-4mm}
\label{smt}
\end{figure}

Typically, to reuse the functionality of a TPL, Python projects follow the official documentation to call the TPL's API. However, most of the APIs in the official documentation are shortened API paths. For example, the fully qualified name of an API is \mintinline{python}{Lib.A.B.C.API}, but the call name in the official documentation is \mintinline{python}{Lib.API}. This is because the TPL developers often use \mintinline{python}{__init__.py} modules to adjust the API call name for the convenience of the user.

To address this issue, \tool recursively traverses the TPL's source code directory structure, focusing on \mintinline{python}{__init__.py} files at each level. By analyzing the import statements in these files, it simplifies the fully qualified names of the APIs in the source code to match the call names in the TPL's official API documentation. For details of the API fully qualified name simplification process, please refer to our previous work~\cite{zhang2024pcart}. 
The knowledge file uses a JSON format to store API information. The overall structure begins with the key ``APIs'', which maps to a dictionary. Each entry in this dictionary represents a specific API, using its API name as the key. The value is another dictionary containing metadata, including the line number where the API is defined (``lineno'') and its parameter list (``parameter'').

\subsection{Version Compatibility Assessment}
After updating the TPL to a new version, dependency conflicts may arise between the various dependent TPLs within the project. 
Similar to existing studies~\cite{ye2022knowledge, cao2024diagnosis, cheng2023revisiting}, \tool models the dependency conflict resolution problem as an SMT problem by first generating SMT expressions and then using SMT solving to output a version-compatible requirements file. 

As the example presented in Figure~\ref{smt}, after the target TPL torchvision is changed from version 0.9.1 to 0.10.0, \tool first obtains all the candidate versions of TPLs and the corresponding dependent TPLs and their version constraints in the updated requirements through the PyPI metadata as well as the specified Python version, and then generates the SMT expressions, and finally outputs the version-compatible requirements file by SMT solving.

Note that when using \mintinline{python}{pip install --upgrade TPL==version} in an environment, pip will report an error and force the installation even if the specified version of the TPL to be upgraded conflicts with some TPLs in the existing environment. However, suppose there is a dependency conflict in a \mintinline{python}{requirements.txt} file, in this case, pip will report an error and not install any of the TPLs in the requirements when users install using \mintinline{python}{pip install -r requirements.txt}. Since the output of \tool is requirements, it must be free of dependency conflicts.

\SetArgSty{}
\begin{algorithm}[!t]
\caption{Generate SMT Expression from Candidate Versions and Dependency Constraints}
\label{alg:smt}
\footnotesize
\SetAlgoLined
\SetKwIF{If}{ElseIf}{Else}{if}{then}{}{}{}
\SetKwFor{ForEach}{foreach}{}{}
\SetKw{in}{in}
\SetKw{or}{or}
\SetKw{and}{and}
\SetKwProg{Fn}{Function}{:}{end}

\KwIn{Candidates: Dict[TPL $\rightarrow$ List[Versions]], \\
\hspace{2.7em} Dependencies: Dict[TPL $\rightarrow$ Dict[Version $\rightarrow$ List[(DepTPL, VersionConstraint)]]]}
\KwOut{The generated expression: EXP}

\Fn{\textnormal{generateFinalEXP(Candidates, Dependencies)}}{
    EXP $\gets$ TRUE\;
    \ForEach{TPL \in Candidates}{
        EXPC $\gets$ FALSE\;
        \ForEach{version \in Candidates[TPL]}{
            EXPCv $\gets$ \textnormal{generateTPLEXP}(TPL, version, Candidates, Dependencies)\;
            EXPC $\gets$ OR(EXPC, EXPCv)\;
        }
        EXP $\gets$ AND(EXP, EXPC)\;
    }
    \Return{EXP}\;
}

\vspace{0.5em}
\Fn{\textnormal{generateTPLEXP(TPL, version, Candidates, Dependencies)}}{
    EXP $\gets$ (TPL == version)\;
    depList $\gets$ Dependencies[TPL][version]\;
    \ForEach{(dep, constraint) \in depList}{
        EXPd $\gets$ FALSE\;
        \ForEach{v \in Candidates[dep]}{
            \If{\textnormal{satisfies}(v, constraint)}{
                subEXP $\gets$ \textnormal{generateTPLEXP}(dep, v, Candidates, Dependencies)\;
                EXPd $\gets$ OR(EXPd, subEXP)\;
            }
        }
        EXP $\gets$ AND(EXP, EXPd)\;
    }
    \Return{EXP}\;
}
\end{algorithm}

\subsubsection{Get Available Versions and Constraints} 
As shown in Figure~\ref{smt}, \tool parses the \mintinline{python}{requirements.txt} to generate a list of elements that represent all the TPLs in the requirements, i.e., [$...$, $torch$, $torchvision$, $pillow$, $...$]. All candidate versions of each TPL and the version constraints for all candidate versions of the TPL are then queried from the knowledge base. 

\subsubsection{Generate SMT Expressions}

\tool takes all the obtained data to generate SMT expressions. 
As shown in Algorithm \ref{alg:smt}, \tool recursively generates a complete SMT expression describing all possible combinations of versions that satisfy all dependency constraints based on the list of candidate versions (\mintinline{python}{Candidates}) for each TPL and the dependencies of each version (\mintinline{python}{Dependencies}). First, the algorithm initializes a global expression ``\mintinline{python}{EXP}'' (line 2) with an initial value set to the logical constant \mintinline{python}{TRUE}, and then it traverses each TPL (line 3) and constructs an expression ``\mintinline{python}{EXPC}'' (line 4) for that TPL that captures all combinations of its feasible versions. In lines 5-8, the algorithm further traverses all candidate versions of the current TPL. For each version, the subfunction ``\mintinline{python}{generateTPLEXP}'' (line 6) is called to recursively generate a logical expression containing the version and its dependencies, and the logical ``\mintinline{python}{OR}'' is used to combine these version expressions into ``\mintinline{python}{EXPC}'' (line 7). This step means that a TPL only needs to select one of the versions that satisfy the dependencies. After all versions have been processed, the algorithm performs a logical ``\mintinline{python}{AND}'' (line 8) between the TPL-level expression ``\mintinline{python}{EXPC}'' and the global expression ``\mintinline{python}{EXP}'', indicating that the constraints of all TPLs in the system must be satisfied simultaneously. Finally, the algorithm returns the constructed global SMT expression ``\mintinline{python}{EXP}'' (line 9).

The subfunction ``\mintinline{python}{generateTPLEXP}'' handles the expression generation for a specific TPL and one of its versions (line 11). It first constructs a basic expression ``\mintinline{python}{EXP}'' indicating that the current version of that TPL is selected (line 12), and subsequently extracts a list of dependencies for that version (line 13). For each dependency, the algorithm initializes a temporary expression ``\mintinline{python}{EXPd}'' (line 15) and then iterates through the candidate versions of that dependent TPL (line 16). For the candidate versions that satisfy the version constraints (line 17), ``\mintinline{python}{generateTPLEXP}'' is called recursively again to construct its expression (line 18) and merge it into ``\mintinline{python}{EXPd}'' (line 19) with a logical ``\mintinline{python}{OR}'', indicating that the dependency can be satisfied as long as one of the versions is satisfied. The dependency expression ``\mintinline{python}{EXPd}'' (line 19) is merged into the main expression ``\mintinline{python}{EXP}'' (line 20) with a logical ``\mintinline{python}{AND}'' operation to ensure that all dependencies are satisfied. Finally, an expression containing the current TPL version and its complete dependency chain logic is returned (line 21).

\subsubsection{SMT Solving}
After obtaining the SMT expression, \tool leverages the widely used SMT solver Z3 ~\cite{z3} to generate feasible solutions. Based on the principle that pip installs the latest version, \tool also follows this principle and chooses the latest version. In addition, to minimize the number of modified versions of the TPL, \tool sets the priority of the version of the TPL in requirements to the highest. This implies that if the TPL's version satisfies the version constraints, its version will be fixed. The optimization goal of SMT solving is that the TPL's version should preferably be the version in the \mintinline{python}{requirements.txt}; otherwise, the newer the version, the higher the priority. Specifically, for each feasible version of a TPL, \tool arranges V = < $V_1$, $V_2$, $...$, $V_m$, $V_x$ >, where $V_x$ is the version in the \mintinline{python}{requirements.txt}, $V_1$ is the oldest version, and $V_m$ is the latest version. 
Therefore, the optimization objective $O$ is as follows: 

\begin{equation}
O = \max \sum_{n \in N} \left[ 1(n = V_x) \cdot (|V| + 1) + \text{index}(n) \right],
\label{eq}
\end{equation}

\noindent where $1(n=V_x)$ is an indicator function that indicates that when $n = V_x$ the value is 1 and 0 otherwise.  $|V| + 1$ is an additional priority weight that is added to ensure that $V_x$ is always prioritized over all other versions. $index(n)$ denotes the index of this version. After SMT solving, \tool obtains the version-compatible requirements file. 

To formalize and efficiently solve version constraints, \tool applies several technical designs to version representation and solver configuration. First, each library version string (e.g., ``1.8.1'') is encoded into a numeric representation that the SMT solver can handle. Specifically, all candidate versions of a library (e.g., 1.8.0, 1.8.1, 1.9.0) are sorted by release time and mapped to integer indices such as 0, 1, and 2. In this way, comparison constraints (e.g., $>=$, $<$) can be directly imposed on integer variables, precisely representing version range requirements such as ``$>=1.8.1$''.

In addition, \tool fully supports special version specifiers in dependency declarations, including the compatible release operator ($\sim=$), wildcards, and environment markers, following the standard semantics defined in PEP 440 ~\cite{pep440} and PEP 508~\cite{pep508}. In implementation, \tool uses the \texttt{SpecifierSet} class in the Python \texttt{packaging} library to parse and match version constraints. Specifically, \tool first normalizes dependency specifiers when necessary, and then constructs a \texttt{SpecifierSet} object to determine which candidate versions satisfy the given constraints. The matched versions are further mapped to integer indices and encoded as logical conditions in the SMT model. For example, $\sim=1.8.0$'' is parsed as the version range $>=1.8.0$ and $<1.9.0$'', and $==1.8.*$'' is expanded into $>=1.8.0$ and $<1.9.0$''. 
For dependencies associated with environment markers, \tool evaluates their conditions according to the current execution environment (e.g., Python version and operating system), and only incorporates the corresponding dependency constraints into the SMT model when the conditions are satisfied. 

We configure the Z3 solver with a timeout of 8~s to avoid long blocking. \tool applies a backtracking-based search mechanism to traverse the version combination space and adopts a parallel solving strategy to explore different solution paths using multi-core computing resources. Once conflicts are detected in an initial solution, \tool backtracks and adjusts the related library versions according to the encountered constraint conflicts and re-solves the SMT problem, until a dependency version configuration that satisfies all compatibility constraints is obtained.

\subsection{Invoked APIs and Modules Extraction}
\label{sec:Approach:Invoked}
In the invoked APIs and modules extraction phase, \tool compares the starting requirements with the version-compatible requirements file (obtained in the phase \ding{183}) to obtain a set of TPLs that have undergone version changes, which is formulated as a triple $<l, V_1, V_2>$. $l$ denotes the TPL, $V_1$ denotes the version of the TPL in the starting requirements, and $V_2$ denotes the version of the TPL in the version-compatible requirements. For each triple $<l, V_1, V_2>$, \tool extracts all TPL $l$-related APIs and modules.

\begin{figure}[!t]
\centering
\includegraphics[width=\linewidth]{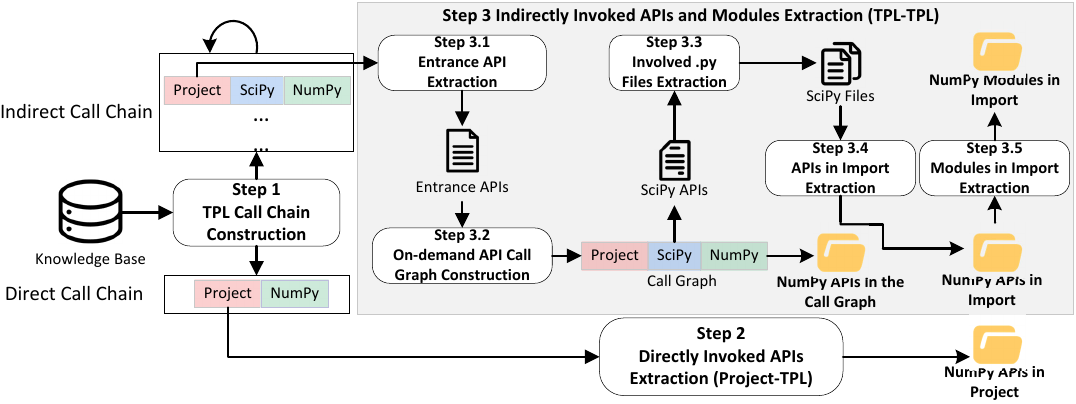}%
\Description{A process overview for extracting invoked APIs and modules related to a target third-party library, covering both direct calls from the project and indirect calls through other libraries, and summarizing the main steps from building call chains to constructing on-demand call graphs and collecting relevant import modules.}
\vspace{-4mm}
\caption{Overview of invoked APIs and modules extraction.}
\label{codeextraction}
\vspace{-4mm}
\end{figure}

As shown in Figure~\ref{codeextraction}, the extraction process considers both directly and indirectly invoked APIs and modules. Given a Python project and a TPL, it is important to recognize that the APIs and modules associated with the TPL not only originate from the Python project itself but also stem from other TPLs that interact with the project during execution. 
In step 1, \tool constructs TPL call chains by querying the knowledge base (obtained in Section~\ref{versionknowledge}), identifying both direct and indirect paths. In step 2, \tool extracts TPL APIs directly invoked in the project. For indirect invocation, \tool follows the steps outlined 
in Figure~\ref{codeextraction}. Step 3.1 extracts the entrance APIs involved in the call chains. Step 3.2 constructs an on-demand API call graph based on these entrances. Step 3.3 identifies the relevant TPL module files involved in the chain. In Step 3.4, APIs used in import statements are extracted, and in Step 3.5, the related modules in import statements are identified.


As such, we have now extracted all API set $S_a = \{api_0, api_1, ..., api_n\}$ and the module set $S_m = \{module_0, module_1, ..., module_n\}$ involved in the TPL $l$ 
of the project $p$. Each element in $S_a$ corresponds to a specific API, identified by its fully qualified name, directly or indirectly invoked by $p$. In contrast, each element in $S_m$ corresponds to a specific module, directly or indirectly invoked by the project $p$. 
In the following, we present details of each step.

\SetArgSty{}
\begin{algorithm}[!t]
\caption{Find All TPL Call Chains}
\label{alg:findchains}
\footnotesize
\SetAlgoLined
\SetKwIF{If}{ElseIf}{Else}{if}{then}{else if}{else}{}  
\SetKwFor{ForEach}{foreach}{}{}
\SetKw{in}{in}
\SetKw{or}{or}
\SetKw{and}{and}
\SetKwProg{Fn}{Function}{:}{end}

\KwIn{project: Root node name, tpl: TPL name, allDependencies: Global dependency dictionary}
\KwOut{allChains: All call chains from project to tpl}
\Fn{\textnormal{findAllCallChains(project, tpl, allDependencies)}}{
    allChains $\gets$ $\emptyset$\;
    queue $\gets$ [ [project] ]\;
    
    \While{queue $\neq$ $\emptyset$}{
        currentPath $\gets$ queue.get()\;
        lastNode $\gets$ currentPath[-1]\;
        
        \ForEach{dep \in allDependencies.get(lastNode)}{
            newPath $\gets$ currentPath\;
            newPath.append(dep)\;
            
            \eIf{dep $=$ tpl}{
                allChains.add(newPath)\;
            }{
                queue.add(newPath)\;
            }
        }
    }
    \Return{allChains}\;
}
\end{algorithm}

\subsubsection{TPL Call Chain Construction}
Given a target TPL, 
\tool obtains the call chain that starts with the project and ends with the TPL, by recursively checking all dependencies of the project, i.e., the TPLs in requirements. 
For example, the project \texttt{deep-belief-network} directly relies on three TPLs: scikit-learn, scipy, and numpy. According to \mintinline{python}{requirements.txt} and PyPI metadata (version knowledge base), four TPL call chains would be constructed: 
1) $project \rightarrow numpy$, 2) $project \rightarrow scipy \rightarrow numpy$, 3) $project \rightarrow scikit-learn \rightarrow numpy$, and 4) $project \rightarrow scikit-learn \rightarrow scipy \rightarrow numpy$.

Algorithm \ref{alg:findchains} shows the process of obtaining all TPL call chains. 
The algorithm takes the project name, a TPL, and a global dependency dictionary (\mintinline{python}{requirements.txt} and PyPI metadata) as input, then outputs all possible TPL call chains from the project to the TPL. 
First, the algorithm initializes an empty result set and a queue containing the project root (lines 2-3). Subsequently, it employs a breadth-first search (BFS) approach, where it dequeues the current path and examines the dependencies of the last node (lines 4-6). For each dependency, the algorithm creates a new path by appending the dependency to the current path (lines 7-9). If the dependency matches the TPL, the new path is added to the result set; otherwise, the path is enqueued for further exploration (lines 10-13). This process continues until the queue is exhausted, ensuring all possible call chains are discovered, and finally returns the complete collection of valid paths (line 15).

\subsubsection{Directly Invoked APIs Extraction (Project-TPL)}\label{sec:Approach:Invoked:Direct}
For direct call chains such as $project \rightarrow numpy$, \tool directly extracts the numpy-related APIs called in the project. 

\begin{figure}[!t]
\centering
\includegraphics[width=4in]{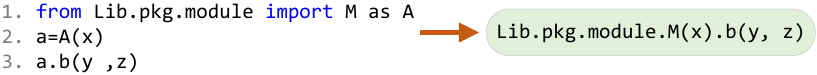}%
\Description{An example illustrating how \tool converts a shorthand or aliased API call into its fully qualified call path by resolving assignments and import aliases, mapping the original call expression to the corresponding library/package/module/class-level API.}
\vspace{-4mm}
\caption{Conversion of an API call.}
\vspace{-4mm}
\label{regularAPI}
\end{figure}




\textit{(1) Extracting API Calls in Project.} 
To extract TPL-related API calls from a Python project, \tool first parses each source file into an AST using Python's built-in AST module. This allows it to systematically traverse the code structure and identify relevant nodes such as \mintinline{python}{Assign}, \mintinline{python}{Import}, and \mintinline{python}{ImportFrom}. By applying DFS on the AST branches, \tool extracts complex API call patterns, such as 
direct invocation, class object invocation, return value invocation, argument invocation, and inheritance invocation. 

To accurately identify the TPL each API belongs to, \tool reconstructs the full API call path by combining assignment and import information, standardizing paths into the format \mintinline{python}{Lib.Package.Module.Class.API}, ``\mintinline{python}{Lib}'' is the target TPL to be upgraded in \tool's configuration file (Figure~\ref{config}). As the example shown in Figure~\ref{regularAPI}, a call like \mintinline{python}{a.b(y, z)} is traced back to its origin through assignment \mintinline{python}{a = A(x)} and import statements \mintinline{python}{from pkg.module import M as A}, resulting in the fully qualified form \mintinline{python}{Lib.pkg.module.M(x).b(y, z)}. 

\textit{(2) Restoring API Fully Qualified Names.}
After extracting all TPL APIs used in the project, \tool converts each API into its fully qualified name. For example, an API initially referenced as \mintinline{python}{Lib.API} will be transformed into a more complete path such as \mintinline{python}{Lib.A.B.C.API}. It's important to note that the fully qualified name used here is based on the version of the TPL specified in the starting \mintinline{python}{requirements.txt}.

\tool first tries to match the simplified API names found in the project with simplified forms of fully qualified names stored in the knowledge base (obtained in Section~\ref{codeknowledge}). If the knowledge base contains an entry whose simplified form matches the project's usage, \tool directly retrieves the corresponding original fully qualified name. 
If the knowledge base does not have an exact match for the simplified API form, \tool retrieves all candidate fully qualified names from the knowledge base that have the same API name (i.e., the last part, such as the function or class name). It then performs fuzzy matching between the actual API used in the project and the list of candidate fully qualified names. 
This fuzzy matching is done using the Levenshtein distance to measure the similarity between strings. \tool selects the candidate fully qualified name with the highest similarity score as the most likely fully qualified name for the API used in the project.

Note that \tool does not extract the modules directly called by the project, because \tool has converted the APIs directly called by the project to their fully qualified names. In \ding{185} code compatibility assessment, if the corresponding module is deleted, the corresponding API must be deleted as well.

\subsubsection{Indirectly Invoked APIs and Modules Extraction (TPL-TPL)} 
\tool follows a top-down hierarchical approach to extract indirectly invoked APIs and modules. As shown in Figure~\ref{codeextraction}, the process consists of three steps. In the following, to better illustrate the extraction process, we use the TPL call chain, i.e., $project \rightarrow scipy \rightarrow numpy$, as an example. 
\textit{(1) Entrance API Extraction.} 
\tool begins by parsing the project's source code to extract the set of scipy APIs that are directly invoked. These extracted APIs serve as entry points for constructing an internal on-demand API call graph within scipy. Details of entrance API extraction can be referred to Section~\ref{sec:Approach:Invoked:Direct}. 

\textit{(2) On-demand API Call Graph Construction.} 
To extract call traces containing target TPL APIs, \tool analyzes each acquired entry API and constructs an on-demand call graph by merging the call trajectories originating from these entry points. Since the objective is to identify all affected external APIs, it is necessary to capture the direct dependencies of the TPL (denoted as $TPL_1$) that may be invoked within the extracted call traces.

For each node in the call trajectory, \tool maintains a structured tuple consisting of essential attributes: the node name, caller, callee, and dependency location. For example, an internal node implemented and invoked by $TPL_1$ (e.g., scipy) itself is represented as <$API_1$, $API_2$, $A.B.C.API_1$, $TPL_1$>, indicating that $API_1$ is called by $API_2$ and defined within $TPL_1$ at the location $A.B.C.API_1$. Conversely, an externally called node, such as <$API_3$, $API_4$, -, $TPL_2$>, signifies that $API_3$ is invoked by $API_4$ but implemented by $TPL_2$ (e.g., numpy, a dependency of $TPL_1$), with the precise location remaining unspecified. 

To achieve this, \tool leverages an established call graph generation tool, i.e.,  code2flow~\cite{wang2023automatically, code2flow}, which facilitates the analysis of potential call traces by examining all feasible method invocations within the AST structure, starting from the identified entry APIs. Taking the API call graph (\mintinline{python}{scipy.Package.Module.API} $\rightarrow$ $...$ $\rightarrow$ \mintinline{python}{numpy.API}) as an example, \tool extracts \mintinline{python}{numpy.API}, then converts it to a fully qualified name.

Note that the fully qualified name of numpy APIs is defined in the version specified in the starting \mintinline{python}{requirements.txt}, 
while the scipy version is determined by the version-compatible \mintinline{python}{requirements.txt}.

\textit{(3) Involved \mintinline{python}{.py} Files Extraction.}
As mentioned in challenge 2 (Section~\ref{sec:challenge}), Python's import mechanism can cause a project to crash, which is impacted by the APIs imported in the \mintinline{python}{.py} files through import statements, although they are not called through the API call chain. 
Therefore, after obtaining all the scipy APIs called by the project, \tool also processes them. 

Taking the API call graph (\mintinline{python}{scipy.Package.Module.API} $\rightarrow$ $...$ $\rightarrow$ \mintinline{python}{numpy.API}) as an example, \tool first processes the entry API, i.e., \mintinline{python}{scipy.Package.Module.API}, to obtain all directly related \mintinline{python}{.py} files, such as \mintinline{python}{scipy.__init__.py}, \mintinline{python}{scipy.Package.__init__.py}, and \mintinline{python}{scipy.Package.Module.py}. These initial files are obtained based on the structure of the API name and are treated as entry points. 

Then, using these \mintinline{python}{.py} files as entry points, \tool examines their import statements to identify all \mintinline{python}{.py} files involved in the API \mintinline{python}{scipy.Package.Module.API}. 
Next, \tool recursively analyzes the import statements in these entry files to find all other involved \mintinline{python}{.py} files. At this stage, \tool uses a predefined directory path, called \mintinline{python}{libraryRoot} (e.g., \mintinline{python}{home/usr/scipy}), to help resolve imported module names into actual file paths. For each import statement encountered, \tool interprets the module name, constructs its corresponding path under \mintinline{python}{libraryRoot}, and checks whether the corresponding \mintinline{python}{.py} file exists. If the file exists and has not been visited yet, it is added to the processing queue.

The process of obtaining all relevant \mintinline{python}{.py} files is shown in Algorithm \ref{alg:findpys}. 
The algorithm takes a list of initial Python files (\mintinline{python}{EntryFiles}) and a base directory path (\mintinline{python}{libraryRoot}) as input, and outputs a set of all related \mintinline{python}{.py} files by resolving import statements. It initializes an empty set visited and an empty queue (lines 2-3). For each entry file, it resolves the absolute path and adds it to the visited and queue if not already present (lines 4-8). The main loop (lines 9-17) processes each file in the queue, extracts its imports, resolves their paths, and adds them to the queue if they are new. Finally, the visited set containing all related \mintinline{python}{.py} files is returned (line 18).

\SetArgSty{}
\begin{algorithm}[!t]
\caption{Find All Related \mintinline{python}{.py} Files}
\label{alg:findpys}
\footnotesize
\SetAlgoLined
\SetKwIF{If}{ElseIf}{Else}{if}{then}{}{}{}
\SetKwFor{ForEach}{foreach}{}{}
\SetKw{in}{in}
\SetKw{or}{or}
\SetKw{and}{and}
\SetKwProg{Fn}{Function}{:}{end}

\KwIn{EntryFiles: List of initial .py files, Root: Base directory path}
\KwOut{Set of all related .py files}
\Fn{\textnormal{findAllRelatedFiles(EntryFiles, libraryRoot)}}{
    visited $\gets$ $\emptyset$\;
    queue $\gets$ $\emptyset$\;  
    \ForEach{file \in EntryFiles}{
        absPath $\gets$ \textnormal{resolveAbsolutePath}(file, libraryRoot)\;
        \If{absPath $\notin$ Visited}{
            visited.add(absPath)\;
            queue.add(absPath)\;
        }
    }   
    \While{queue $\neq$ $\emptyset$}{
        currentFile $\gets$ queue.get()\;
        imports $\gets$ \textnormal{extractImports}(currentFile)\;       
        \ForEach{module \in imports}{
            modulePath $\gets$ \textnormal{resolveModulePath}(module, currentFile, libraryRoot)\;
            \If{modulePath $\neq$ null \and modulePath $\notin$ Visited}{
                visited.add(modulePath)\;
                queue.add(modulePath)\;
            }
        }
    }
    \Return{visited}\;
}
\end{algorithm}


\textit{(4) APIs in Import Extraction.}
After obtaining all relevant \mintinline{python}{.py} files, \tool parses the source code into an AST for each \mintinline{python}{.py} file. Then, \tool uses BFS to search the AST in hierarchical order, identifying type nodes, i.e.,  \mintinline{python}{Import} and \mintinline{python}{ImportFrom}, which correspond to import and from-import statements, respectively. 
For all elements extracted from the import statement, \tool performs string matching on these elements based on the TPL name, such as ``numpy'', to obtain all numpy-related APIs in the import statement.

Note that while import statements may reference modules rather than APIs, we conservatively treat all extracted elements as APIs. This ensures comprehensive coverage, as subsequent compatibility assessment stages will naturally filter invalid entries.
At the same time, \tool does not convert the extracted APIs into their fully qualified forms. Instead, it outputs the raw extracted elements for further processing. This is because 
certain extracted ``APIs'' might represent modules. Enforcing strict fully qualified name matching could lead to false positives, thereby misclassifying unused APIs as utilized, which could ultimately degrade \tool's overall performance.

\textit{(5) Modules in Import Extraction.} 
Finally, \tool splits the APIs obtained from the previous step based on the character ``.'' to identify the modules involved. For example, \mintinline{python}{numpy.Package.Module.API} is split into \mintinline{python}{numpy.__init__.py}, \mintinline{python}{numpy.Package.__init__.py}, \mintinline{python}{numpy.Package.py}, \mintinline{python}{numpy.Package.Module.__init__.py}, \mintinline{python}{numpy.Package.Module.py}, \mintinline{python}{numpy.Package.Module.API.__init__.py} and \mintinline{python}{numpy.Package.Module.API.py}. This is done by progressively combining each prefix of the API path and appending a "\mintinline{python}{.py}" suffix to treat each as a potential module. Through this process, \tool reconstructs the full module structure implied by the import statement and captures all potentially relevant numpy-related modules for further compatibility checking.

Note that this decomposition serves two main purposes. First, some intermediate modules such as \mintinline{python}{numpy.Package.py} may not exist in the source code of the starting version of numpy. These will be filtered out later during the code compatibility assessment phase. Second, in cases where the final API component might itself be a module (e.g.,  \mintinline{python}{numpy.Package.Module.API}), \tool includes the full path \mintinline{python}{numpy.Package.Module.API.py} to ensure completeness. Just like the intermediate paths, if this full path does not correspond to a real module, it will also be excluded during later filtering. This conservative inclusion ensures that no potentially valid module reference is missed during analysis.

\subsection{Code Compatibility Assessment}
To systematically detect code compatibility issues, \tool first analyzes the code changes of the TPL $l$ from $V_1$ to $V_2$, and then conducts a comprehensive compatibility assessment based on the actual modules and APIs invoked by the project.


\subsubsection{Module Compatibility Assessment} 
\begin{figure}[!t]
\centering
\includegraphics[width=4.25in]{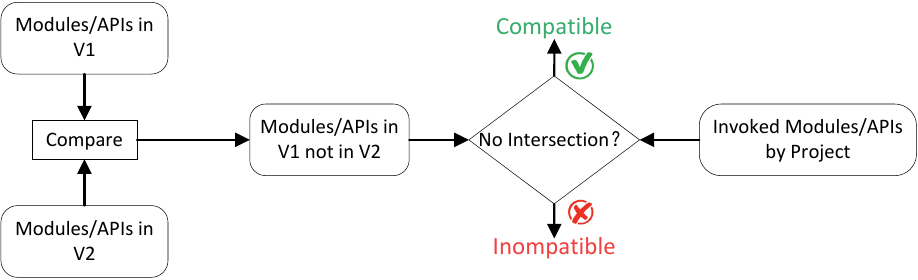}%
\Description{A step-by-step illustration of \tool's code compatibility checks for module and API-name breaking changes between two library versions: it computes removed modules/APIs by diffing versions, intersects them with the project's extracted used-module set and used-API set, and flags potential incompatibilities when an intersection is non-empty.}
\vspace{-4mm}
\caption{Module and API name compatibility assessment.}
\label{codeassessment}
\vspace{-4mm}
\end{figure}

As shown in Figure~\ref{codeassessment}, \tool first compares all modules of the TPL $l$ between versions $V_1$ and $V_2$ to obtain a set $S_1$, which contains modules that exist in $V_1$ but were removed in $V_2$. Here, $S_m$ represents the set of modules used by the project (obtained in Section ~\ref{sec:Approach:Invoked}). \tool then checks for any intersection between $S_m$ and $S_1$. If an intersection exists, it indicates that some modules the project depends on have been removed or modified in the new version $V_2$, which will lead to code compatibility issues.

\subsubsection{API Name Compatibility Assessment} 
Similarly, as shown in Figure\ref{codeassessment}, \tool first compares all APIs of the TPL $l$ between versions $V_1$ and $V_2$ to obtain a set $S_2$, which contains APIs that exist in $V_1$ but were removed in $V_2$. Here, $S_a$ represents the set of APIs called by the project (obtained in Section ~\ref{sec:Approach:Invoked}). \tool then checks for any intersection between $S_a$ and $S_2$. If an intersection exists, it indicates that some APIs the project calls have been removed or modified in the new version $V_2$, which will introduce code compatibility issues.

\subsubsection{API Parameter Compatibility Assessment} 
API parameter compatibility is influenced not only by changes in parameter definitions but also by how parameters are passed during actual usage. Therefore, we follow our previous work, i.e., PCART~\cite{zhang2024pcart}, to evaluate API parameter compatibility. 

\textit{(1) Parameter Change Analysis.} \tool begins by distinguishing between positional and keyword parameters based on API definitions. It then maps parameters between two versions (i.e., $V_1$ and $V_2$) using attributes such as name, position, and type. The process consists of three steps: 
First, \tool establishes the mapping relationship between parameters based on the consistency of
parameter names. For positional parameters, \tool analyzes type and position changes. For keyword parameters, only type changes are considered. 
Then, \tool detects conversions between positional and keyword parameters by using parameter name consistency. 
Finally, remaining unmapped positional parameters are mapped by considering the consistency of both position and type. 
The remaining keyword parameters with undetermined mappings are mapped by type consistency. 
Parameters in $V_1$ without a match in $V_2$ are considered removed; unmatched parameters in $V_2$ are considered newly added.


\textit{(2) Parameter Passing Method Analysis.} In Python, parameters can be passed in three ways: positionally, by keyword, or not at all (if default values are used). Positional parameters can be passed by position or name, keyword parameters must be passed by name, and default-valued parameters can be omitted.
\tool uses AST analysis to examine \mintinline{python}{ast.args} (positional) and \mintinline{python}{ast.keywords} (keyword) nodes in API calls. Parameters not appearing in either node are treated as not passed.

\textit{(3) API Parameter Compatibility Evaluation.} After identifying both the parameter changes and the actual parameter passing methods, \tool evaluates compatibility using an API parameter compatibility model, proposed in our previous work PCART~\cite{zhang2024pcart}. The compatibility assessment model considers three dimensions: the parameter type (positional or keyword), the type of change (such as remove or renaming), and the passing method (positional, keyword, or not passed). 
The model defines a compatibility function $f(P, E, M)$, where $P$ represents the parameter type, $E$ represents the change type, and $M$ represents the passing method. This function determines whether a parameter is compatible given its change and usage pattern. The overall compatibility of an API invocation is determined by evaluating the function $f$ for each parameter in the call. The invocation is considered compatible only if all parameters are compatible.

\subsection{Version Change}
The primary goal of the version change phase is to resolve code compatibility issues that arise when changing TPL versions. During the code compatibility assessment phase, two levels of code compatibility issues would be identified, i.e., Project-TPL and TPL-TPL code compatibility issues. 
According to different levels of code compatibility issues, \tool adopts different strategies to resolve code compatibility issues by changing the corresponding TPL's version, selected from the version candidates (obtained in \ding{182} knowledge acquisition phase). After the version change, \tool re-enters stages \ding{183}, \ding{184}, and \ding{185} to assess version and code compatibility.

\begin{figure}[!t]
\centering
\includegraphics[width=\linewidth]{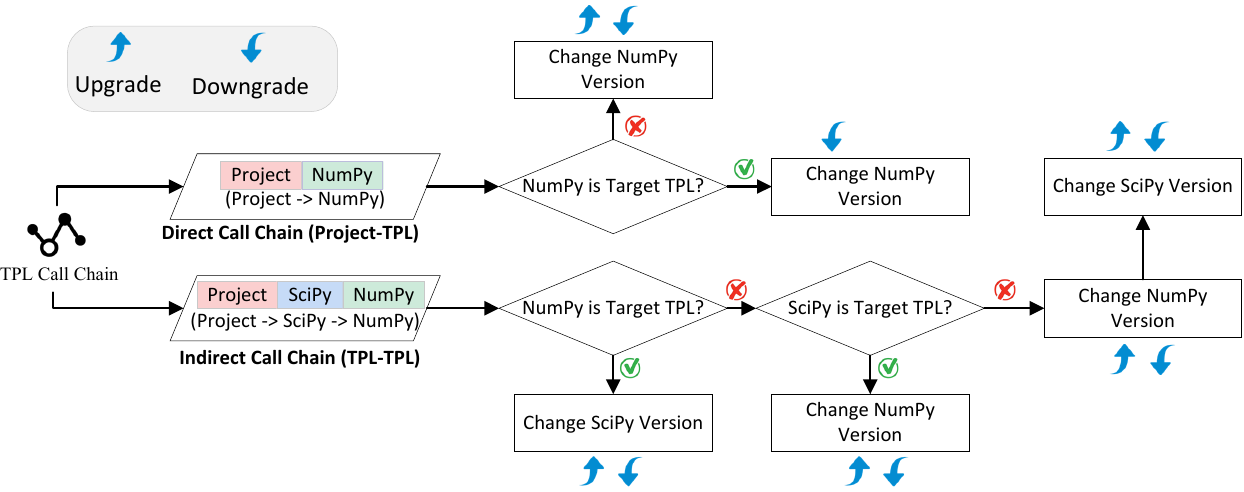}%
\Description{A workflow diagram for the version-change phase in \tool: after detecting Project-TPL or TPL-TPL incompatibilities, the method selects candidate versions for the relevant library (upgrade/downgrade search), updates the requirements accordingly, and loops back to re-check version and code compatibility until a compatible configuration is found.}
\vspace{-4mm}
\caption{Workflow of version change.}
\label{fig:change}
\vspace{-4mm}
\end{figure}

\subsubsection{Project-TPL Incompatibility Resolving} 
As shown in Figure~\ref{fig:change},  taking the TPL call chain ($Project \rightarrow numpy$) as an example, 
regardless of whether numpy is used as a target TPL (the upgraded one) or not, \tool resolves code compatibility issues by changing its version. 

When numpy is specified as the target TPL, \tool performs a version downgrade policy. The set of candidate versions is defined as an ordered list $[V_1,...,V_{x-2}, V_{x-1}]$, where $V_x$ denotes the version number in the version-compatible \mintinline{python}{requirements.txt} and $V_1$ denotes the oldest version. The strategy starts from the previous stable version ($V_{x-1}$) of the current version and tests each candidate version in decreasing order of version number until the first compatible version that satisfies all conditions is found. This degradation strategy is based on the assumption that newer versions may have introduced incompatible API changes, while earlier versions usually maintain better backward compatibility.  
Note that \tool will only downgrade the target TPL version if the target TPL is incompatible with the project during the upgrade. 
Until \tool finds a version that meets the requirements, it will keep the target TPL version unchanged. 

In addition, when numpy is not the target TPL, \tool employs a bidirectional version search strategy with two candidate version lists: the upgrade list $[V_{x+1}, V_{x+2}, ..., V_n]$ containing newer versions up to the latest stable version $V_n$, and the downgrade list $[V_1, V_2, ..., V_{x-1}]$ consisting of older versions. The search first attempts versions in the upgrade list from $V_{x+1}$ to $V_n$ in ascending order. If no suitable version is found, it then tries versions in the downgrade list from $V_{x-1}$ to $V_1$ in descending order until a compatible version is identified. If no compatible version is found in either direction, \tool stops further attempts and retains the original target TPL version and all other TPL versions unchanged. 


\subsubsection{TPL-TPL Incompatibility Resolving} 
As shown in Figure~\ref{fig:change},  taking the TPL call chain ($Project \rightarrow scipy \rightarrow numpy$) as an example, 
when the target TPL is numpy, \tool employs a version-holding strategy that maintains the current numpy version $V_x$ while dynamically adjusting scipy's version to resolve code compatibility issues. 

\tool uses two distinct version candidate lists for scipy: an upgrade list $[V_{y+1}, V_{y+2}, ..., V_n]$ containing newer versions up to the latest stable version $V_n$, and a downgrade list $[V_1, V_2, ..., V_{y-1}]$ consisting of older versions, where $V_y$ represents the originally specified scipy version in the version-compatible requirements file, and $V_1$ is the oldest available version. 
The search process follows a two-phase approach. In the first phase, \tool tests versions from the upgrade list, examining each candidate from $V_{y+1}$ to $V_n$ in ascending order to identify a compatible version. If this upgrade path fails to yield a satisfactory solution, \tool transitions to the second phase, where it explores the downgrade list, evaluating versions from $V_{y-1}$ to $V_1$ in descending order until it discovers the first compatible scipy version that works with the fixed numpy version $V_x$. 

In addition, when the target TPL is scipy, \tool also adopts a version-holding strategy that keeps the current scipy version $V_y$ fixed and resolves code compatibility issues by adjusting numpy's version. Similarly, \tool uses a bidirectional search algorithm with two separate version lists: an upgrade list containing newer versions $[V_{x+1}, V_{x+2}, ..., V_n]$ and a downgrade list containing older versions $[V_1, V_2, ..., V_{x-1}]$, where $V_x$ is the originally specified numpy version, $V_n$ is the latest stable version, and $V_1$ is the oldest available version.
The search proceeds in two stages. First, \tool attempts version upgrades from $V_{x+1}$ to $V_n$ in ascending order. If no compatible version is found through upgrading, it then tries version downgrades from $V_{x-1}$ to $V_1$ in descending order until it finds the first compatible numpy version. 

Moreover, there are scenarios where the target TPL is neither numpy nor scipy. Consider the case where the target TPL is tensorflow with a starting version of 1.9.0 being upgraded to 1.10.0 while numpy version 1.19.5 is installed. During this tensorflow version upgrade, 
since tensorflow 1.10.0 specifies a dependency constraint on numpy (i.e., <=1.14.5, >=1.13.3), the numpy version will consequently be modified. 
In this situation, when examining the call chain $project \rightarrow scipy \rightarrow numpy$, neither scipy nor numpy represents the target TPL being upgraded. 

Therefore, \tool adopts a cascading version adjustment approach. It maintains two version candidate lists for each TPL: an upgrade path containing newer versions $[V_{x+1}, V_{x+2}, ..., V_n]$ and a downgrade path with older versions $[V_1, V_2, ..., V_{x-1}]$, where $V_n$ represents the latest stable version and $V_1$ the oldest available version.
The adjustment process occurs in two sequential stages. First, \tool attempts to resolve code compatibility issues by exploring numpy versions, systematically testing newer versions from $V_{x+1}$ to $V_n$ before falling back to older versions from $V_{x-1}$ to $V_1$ if necessary. Should this initial adjustment fail, \tool automatically progresses to the second stage, where it applies the same bidirectional search strategy to scipy versions while keeping the previously determined numpy version. 
Note that if no compatible version is identified through either upgrade or downgrade, \tool ends further resolution attempts and preserves the original versions of all TPLs in the \mintinline{python}{requirements.txt}.

\subsection{Missing TPL Completion}
After obtaining the new \mintinline{python}{requirements.txt} that meets both version and code compatibility, \tool examines all TPLs and their corresponding versions specified within it. It then verifies the associated version constraints. For TPLs that are referenced in the version constraints but absent from the new  \mintinline{python}{requirements.txt}, \tool adds them along with the oldest available versions that satisfy the specified constraints. 
Such TPLs are not included in the starting \mintinline{python}{requirements.txt}, making it impossible for \tool to perform compatibility checks on them. Therefore, to minimize the risk of compatibility issues that may arise with newer versions, \tool deliberately selects the oldest versions that still conform to the given constraints.

\subsection{Usage Scenario of \tool}

\begin{figure}[!t]
\centering
\includegraphics[width=3in]{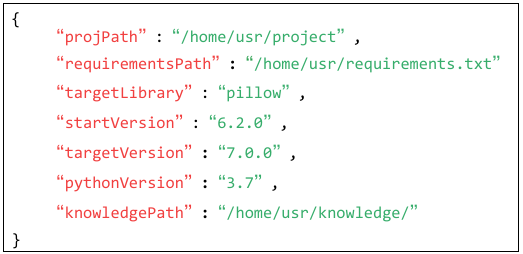}%
\Description{A screenshot/example of \tool's JSON configuration file, showing key fields such as project path, requirements file path, target library name, current and target versions, Python version, and the knowledge-base path used to run the tool.}
\vspace{-4mm}
\caption{The input configuration of \tool.}
\vspace{-4mm}
\label{config}
\end{figure}

\begin{figure}[!t]
\centering
\includegraphics[width=\textwidth]{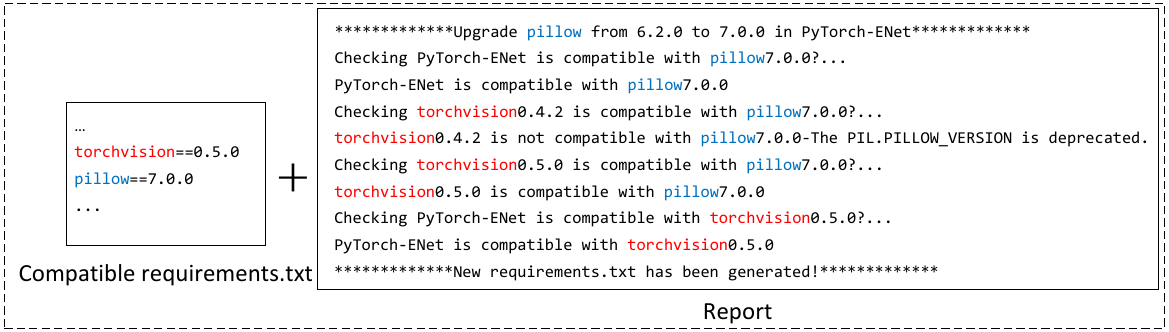}%
\Description{A screenshot/example of \tool's output, including an updated requirements file with inferred compatible dependency versions and a report summarizing detected incompatibilities and the version adjustments made to resolve them.}
\vspace{-4mm}
\caption{The output of \tool.}
\vspace{-4mm}
\label{output}
\end{figure}

\tool takes a configuration file as input (Figure~\ref{config}), which specifies the project path, the path to \mintinline{python}{requirements.txt}, the target TPL name, the current and target versions of the target TPL, and the path to the knowledge base. Based on this configuration, \tool outputs a compatible \mintinline{python}{requirements.txt} file and an inference report (Figure~\ref{output}). The updated \mintinline{python}{requirements.txt} records all TPLs and their inferred compatible versions, while the inference report documents the performed version and code compatibility checks and corresponding resolutions.

To illustrate how and when \tool is used, we provide a concrete example based on the \texttt{PyTorch-ENet} project. \tool is designed to be used in scenarios where a developer intends to upgrade a specific TPL in a Python project, and needs to automatically infer a new \textit{requirements.txt} that remains both version-compatible and code-compatible after the upgrade. The project originally relied on pillow 6.2.0. When upgrading pillow to 7.0.0, the project failed with the error \textit{ImportError: cannot import name `PILLOW\_VERSION' from `PIL'}, because the \mintinline{python}{PILLOW_VERSION} API was removed in newer versions of pillow and was indirectly invoked through torchvision.

To perform compatibility inference using \tool, the developer first prepares a JSON configuration file (as shown in Figure~\ref{config}), specifying the project path, the path to \mintinline{python}{requirements.txt}, the target library (``pillow''), the current and desired versions (i.e., 6.2.0 and 7.0.0), the Python version, and the path to the dependency knowledge base. After running \tool with this configuration, \tool automatically performs compatibility checking and version inference. In this example, \tool detects that the project itself is compatible with pillow 7.0.0, but torchvision 0.4.2 is not, due to its use of the removed \mintinline{python}{PILLOW_VERSION} API. \tool then applies its version-adjustment strategy and identifies torchvision 0.5.0 as a compatible replacement. All dependencies are re-validated, and the inference succeeds.

Finally, \tool outputs a new \mintinline{python}{requirements.txt}, where pillow is upgraded to 7.0.0 and torchvision is updated to 0.5.0, together with a detailed inference report (i.e., Figure~\ref{output}) documenting the detected conflict, the adjustment actions taken, and the final compatibility verification results.

\section{Evaluation}\label{sec:evaluation}
\subsection{Research Questions}
Our study mainly focuses on answering the following four research questions (RQs):

\begin{itemize}
    \item \textbf{RQ1:} How does \tool perform in compatible requirements inference?

    \item \textbf{RQ2:} How does \tool compare to SOTA tools in compatible requirements inference?
    
    \item \textbf{RQ3:} How does \tool compare to LLMs in compatible requirements inference?
    
    \item \textbf{RQ4:} What is the time cost of \tool in compatible requirements inference?

\end{itemize}

\begin{table}[!t]
    \centering
     \caption{Detailed information for \benchmark}
      \vspace{-4mm}
     \scalebox{0.85}{
    \begin{tabular}{|l|c|c|c|c|c|}
        \hline
        \textbf{Projects}     & \textbf{TPLs}   & \textbf{Pip Solved}   & \textbf{Pip Unsolved} & \textbf{Total Scenarios}   \\ \hline
        34           & 20       & 1,689        & 406 & 2,095 \\ \hline
    \end{tabular}}
    \label{reqbench}
     \vspace{-4mm}
\end{table}

\subsection{\benchmark: Benchmark for Compatible Requirements Inference in Python TPL Upgrade Scenario}
We construct a dataset named \benchmark for conducting evaluation experiments. \benchmark is derived from our motivating study (Section~\ref{sec:motivatingstudy}) and consists of real-world Python projects selected from open-source repositories (Table~\ref{reqbench}).

Each upgrade scenario consists of a Python project, including its requirements, the Python version, the target TPL, the starting and target versions of the TPL, and the associated compatibility labels. Based on the motivating study results, we assigned labels to each upgrade scenario, which fall into two categories: 

\begin{itemize}
    \item \textbf{Pip Solved}: 
    This indicates that after upgrading the target TPL to the specified version using pip, the project can be successfully executed without any crash or uncaught exception.
    
    \item \textbf{Pip Unsolved}: This indicates that upgrading the target TPL to the specified version using pip leads to a project environment that no longer functions correctly.
\end{itemize}

As depicted in Table~\ref{reqbench}, the numbers of pip solved and pip unsolved scenarios are 1,689 and 406, respectively. For the 406 pip unsolved scenarios, we further classified them 
into two levels of code compatibility issues: Project-TPL and TPL-TPL. Moreover, we categorized the code compatibility issues into four finer-grained types: module, API name, API parameter, and API body. The distribution of these 406 scenarios across the different levels and issue types is shown in Table~\ref{cdi}. 
Note that we include the ``pip solved'' scenarios in the evaluation to assess the robustness of \tool and the baseline tools (i.e., PyEGo, ReadPyE, and the LLM-based methods). For scenarios where pip can already produce a working dependency configuration, we examine whether these tools can also infer valid configurations without breaking the existing functional setups, thus verifying that they do not introduce regressions in already solvable upgrade scenarios. 

Compared with existing environment-reconstruction datasets such as Gistable~\cite{horton2018gistable} and HG2.9K~\cite{horton2019dockerizeme}, \benchmark differs fundamentally in both design goals and data composition. First, the primary objective of Gistable and HG2.9K is to reproduce the original executable environment of real Python projects. Their data mainly consists of project source code and does not include \mintinline{python}{requirements.txt} files or equivalent dependency specification information, which are essential for studies involving dependency upgrades. The absence of explicit version constraints means that these datasets can only support reproducing the execution of projects in their original environments, but cannot support research on library upgrades from an initial version to a target version, nor can they describe how dependency relationships evolve over time as libraries change.

Second, Gistable and HG2.9K do not provide fine-grained code compatibility annotations associated with library version evolution, such as module compatibility, API name compatibility, API parameter compatibility, or API behavioral compatibility. These semantic compatibility details are necessary for evaluating whether a compatibility inference method (such as \tool) can accurately identify code-level compatibility issues. However, such information dimensions are not covered in the design of these environment-reconstruction datasets. As a result, Gistable and HG2.9K cannot address the core task requirements of TPL upgrade scenarios, nor are they sufficient for systematically evaluating methods that jointly analyze version compatibility and code compatibility.

Motivated by these differences, \benchmark is constructed based on the empirical findings in Section~2 to provide a benchmark dataset specifically tailored for upgrade scenarios and equipped with fine-grained compatibility annotations, thereby filling the gap left by existing datasets in research on library-upgrade compatibility.

\subsection{Experiment Setup}
\subsubsection{Settings of Comparison Tools} 
In this study, we aim to address RQ2 and RQ3 by comparing \tool with the SOTA tools (PyEGo~\cite{ye2022knowledge} and ReadPyE~\cite{cheng2023revisiting}) and LLM (DeepSeek V3~\cite{DeepSeek}, DeepSeek R1~\cite{DeepSeek}, and ChatGPT (GPT-4o-2024-11-20)~\cite{ChatGPT}). 

\textbf{Settings of PyEGo.} 
PyEGo is a compatible environment inference tool for Python projects. 
Given a Python project, PyEGo infers a compatible runtime environment, with the latest TPL version taking priority. By contrast, given a target TPL and the target version to be upgraded, \tool infers the compatible requirements after the upgrade. To evaluate PyEGo's capabilities in our use case, we migrate its functionality with the following settings. PyEGo infers a set of candidate TPLs and an incomplete set of candidate versions. 
Accordingly, if the upgraded version of the target TPL is compatible with the project, we fix the version of the target TPL to be the target version. If the upgraded version of the target TPL is incompatible with the project, we loosen the condition by using the candidate versions inferred by PyEGo as the target version. 

Note that PyEGo will 
always generate a \mintinline{python}{requirements.txt}. Thus, in our evaluation, a successful inference is defined as one where after \mintinline{python}{pip install -r requirements.txt} the project runs correctly. If the project crashes during execution, the inference is considered a failure. The running environments are all independent virtual environments (conda) to prevent mutual interference.

\textbf{Settings of ReadPyE.} 
ReadPyE is also a Python-compatible environment inference tool with a default use case similar to PyEGo. It accepts a Python project and infers a compatible and up-to-date runtime environment. 
To evaluate the effectiveness of ReadPyE in our scenario, we adjust its behavior in the same way as PyEGo. Since ReadPyE outputs a set of candidate TPLs and their version ranges, we first determine whether the target version is within the version range inferred by ReadPyE. If it is, we fix the target TPL's version to be the target version. If the target version is not within the version range inferred by ReadPyE and the target TPL's version is incompatible with the project, we 
use the version range inferred by ReadPyE. By contrast, if the target version is not within the version range inferred by ReadPyE and the target TPL's version is compatible with the project, this indicates that ReadPyE has made an incorrect inference. 
In our evaluation, a successful inference is defined as one where after \mintinline{python}{pip install -r requirements.txt} the project runs correctly. If the project crashes during execution, the inference is considered incorrect. The running environments are all independent virtual environments (conda) to prevent mutual interference.

\begin{figure}[!t]
\centerline{\includegraphics[width=5in]{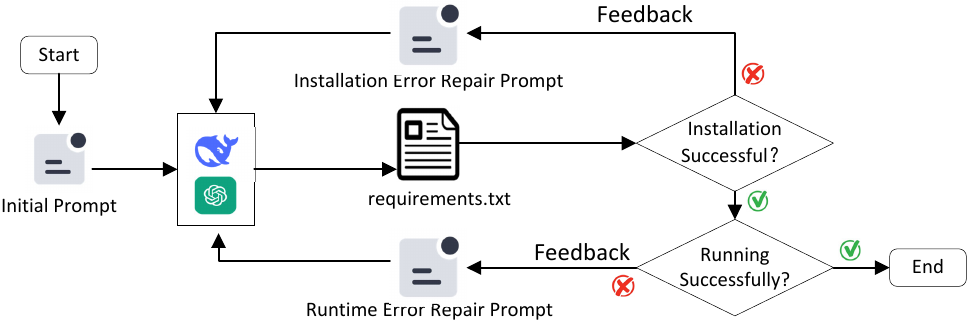}}%
\Description{A workflow diagram of the LLM-based baseline: given the original requirements and a target library upgrade, the method prompts an LLM to propose updated requirements, iteratively fixes dependency installation conflicts using error feedback, then iteratively fixes runtime failures using traceback feedback until success or a maximum number of iterations.}
\vspace{-4mm}
\caption{Workflow of our LLM-based approach.}
\label{llm}
\vspace{-4mm}
\end{figure}

\begin{figure}[!t]
\centerline{\includegraphics[width=5in]{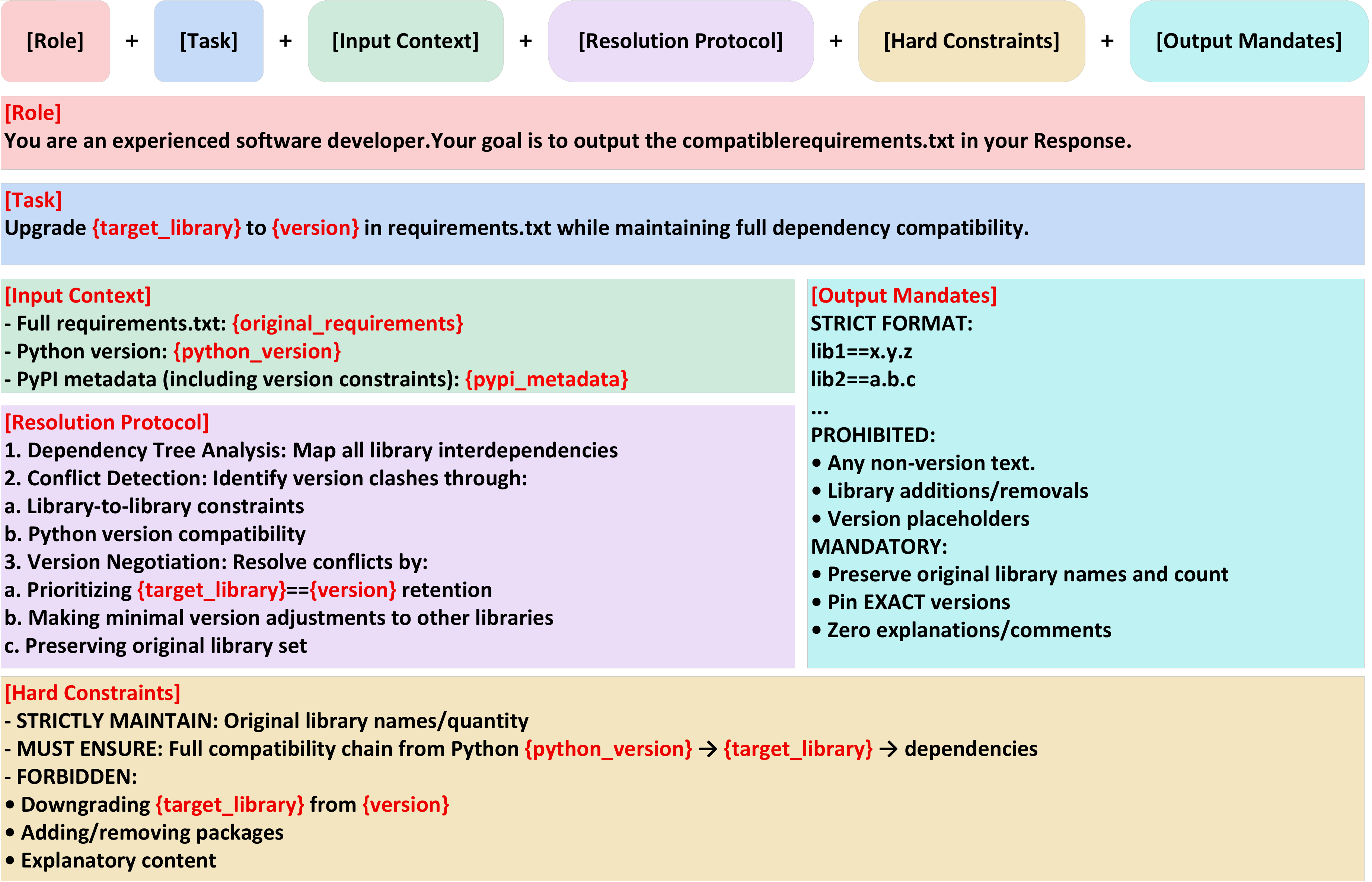}}%
\Description{The template of the initial LLM prompt used in the baseline, including structured sections (e.g., role, task, input context such as requirements and target upgrade, constraints, and output format) that instruct the model to generate an upgraded, compatible requirements file.}
\vspace{-4mm}
\caption{Initial prompt.}
\label{prompt1}
\vspace{-4mm}
\end{figure}


\textbf{Settings of LLMs.} To evaluate the capability of LLMs under Python TPL upgrade scenarios, we design an LLM-based workflow aligned with upgrade tasks. Existing LLM-based compatible environment inference approaches, such as PLLM, primarily target environment reconstruction in the absence of an existing dependency configuration, aiming to obtain a runnable environment by analyzing source code and iteratively resolving build and runtime errors using LLMs with retrieval-augmented generation. In contrast, our work focuses on proactive library upgrades, where an initial dependency configuration exists, and a target library version is explicitly fixed. The objective is to infer an upgraded configuration that simultaneously satisfies version constraints and code-level compatibility, ensuring semantic and behavioral correctness rather than arbitrary executability. Due to these fundamental differences in input assumptions, task objectives, and constraint forms, directly reproducing PLLM as a comparison baseline would be unfair. Instead, we redesign multi-stage LLM prompts following a realistic upgrade-maintenance workflow to evaluate the capability of LLM-based methods under our task setting. 

\begin{figure}[!t]
\centerline{\includegraphics[width=5in]{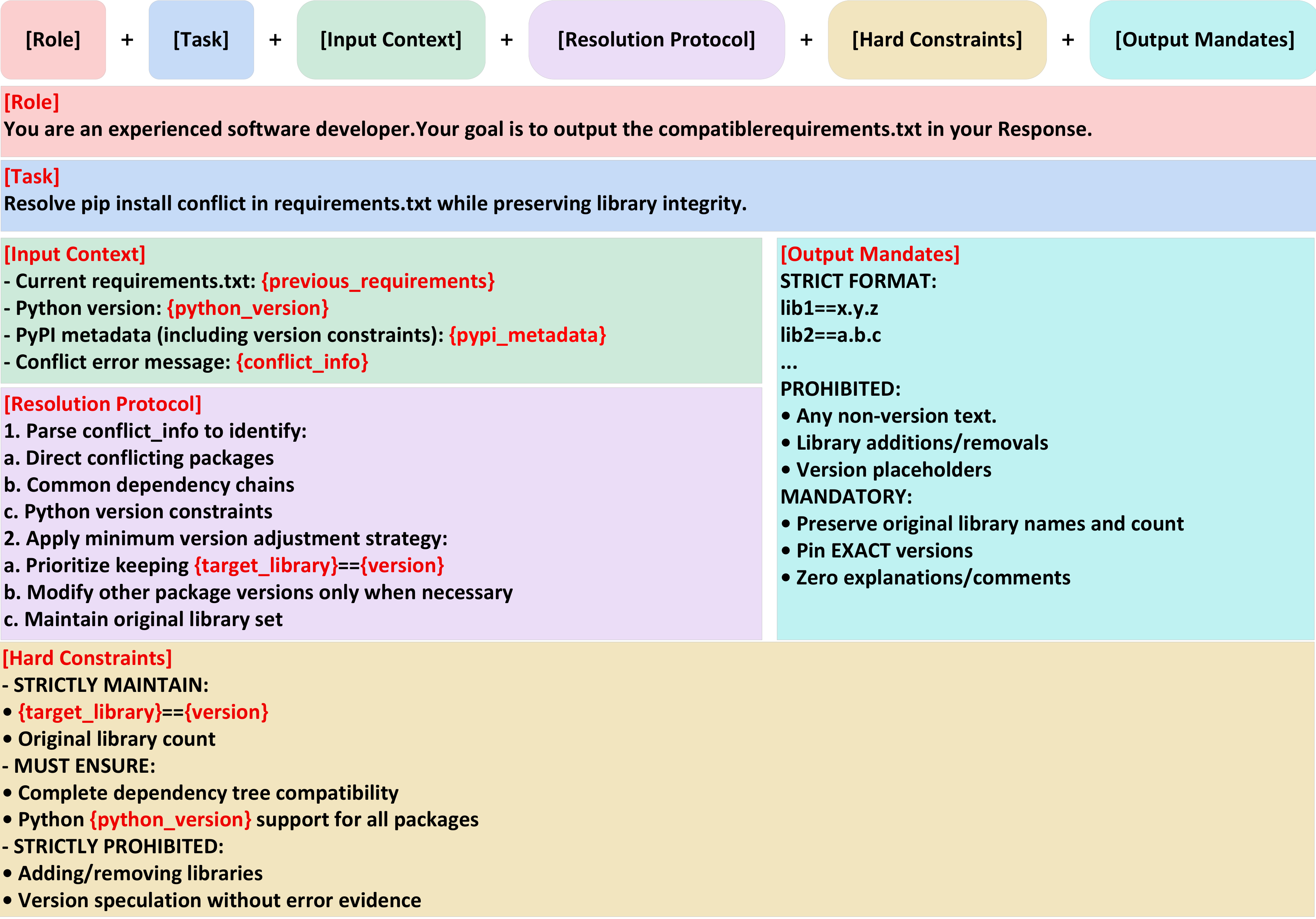}}%
\Description{The template of the installation-error repair prompt, where pip dependency resolver error messages are provided as feedback and the LLM is instructed to modify dependency versions/constraints to produce an installable requirements file while respecting hard constraints.}
\vspace{-4mm}
\caption{Installation error repair prompt.}
\label{prompt2}
\vspace{-4mm}
\end{figure}

\begin{figure}[!t]
\centerline{\includegraphics[width=5in]{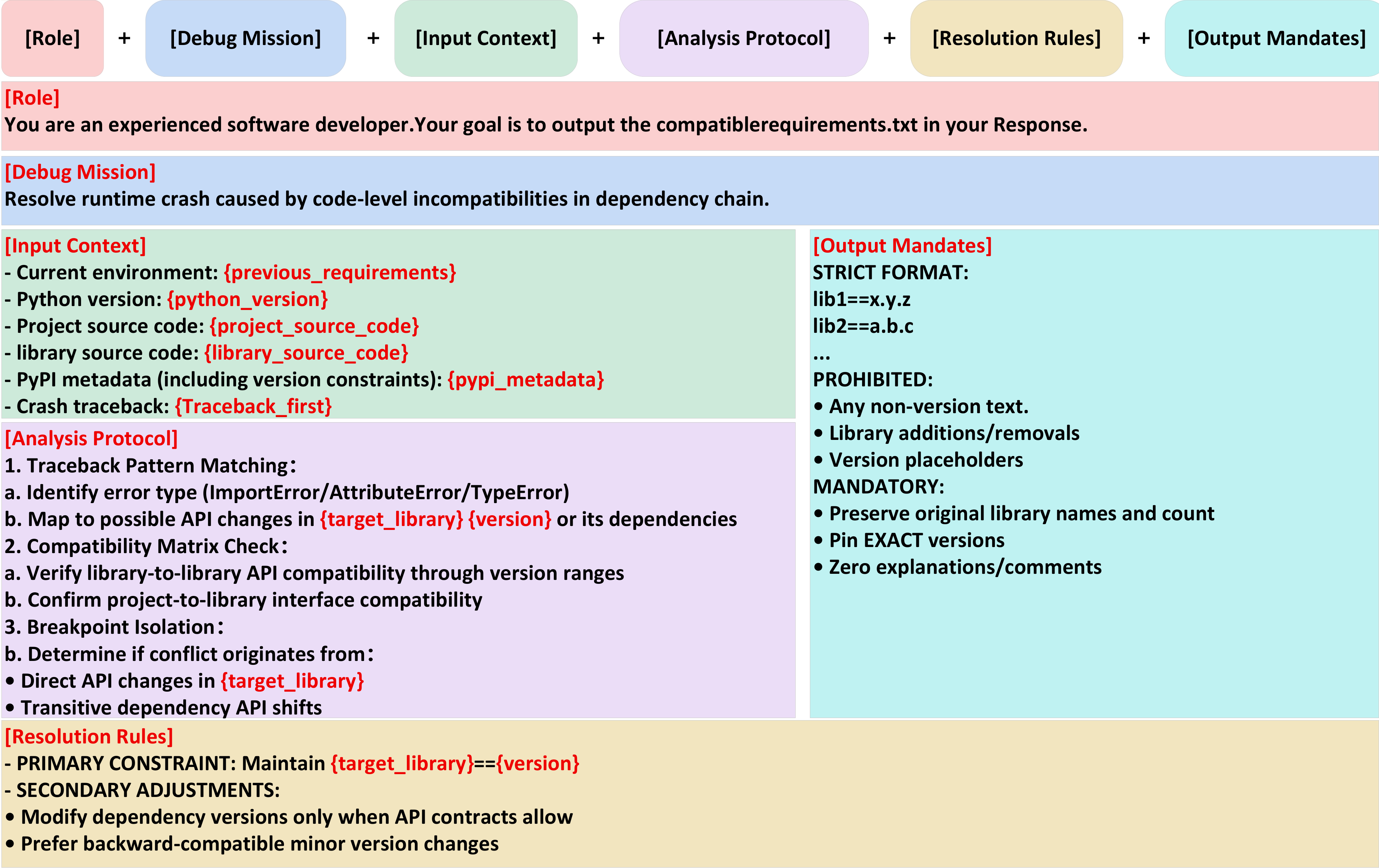}}%
\Description{The template of the runtime-error repair prompt, where a project crash traceback and relevant project/TPL source context are provided and the LLM is instructed to adjust dependency versions to eliminate the runtime error without violating specified upgrade goals and constraints.}
\vspace{-4mm}
\caption{Runtime error repair prompt.}
\label{prompt3}
\vspace{-4mm}
\end{figure}

As illustrated in Figure~\ref{llm}, our process begins by providing the original requirements, the target TPL, the Python version, and the desired upgrade version. As the initial prompt depicted in Figure~\ref{prompt1}, we then prompt the LLM to generate compatible requirements that reflect the intended upgrade. Subsequently, we attempt to install the dependencies specified in the LLM-generated \mintinline{python}{requirements.txt} file. If a dependency conflict arises during installation, we feed the error messages back to the LLM and prompt it to regenerate a conflict-free version of requirements (Figure~\ref{prompt2}). Once the installation succeeds, we proceed to run the project. If runtime errors occur, we similarly return the error tracebacks to the LLM for further correction (Figure~\ref{prompt3}). 
If the generated requirements finally allow the project to execute successfully, the process is considered complete, and the result is deemed correct. Otherwise, the inference result is incorrect. The running environments are all independent virtual environments (conda) to prevent mutual interference.

Figures~\ref{prompt1}-\ref{prompt3} show the initial prompt, the installation error repair prompt, and the runtime error repair prompt, respectively. Each prompt is composed of six structured components that define the behavior and output constraints for LLMs. These components include: \textbf{Role}, which specifies the model's assumed identity and responsibility; \textbf{Task} or \textbf{Debug Mission}, which outlines the specific objective (e.g., upgrade, conflict resolution, or runtime debugging); \textbf{Input Context}, which provides relevant environmental inputs such as \mintinline{python}{requirements.txt}, Python version, PyPI metadata, project source code, library source code, or error traces; \textbf{Resolution Protocol} or \textbf{Analysis Protocol}, which defines the procedural steps for resolving compatibility issues; \textbf{Hard Constraints}, which impose non-negotiable rules on allowed actions; and \textbf{Output Mandates}, which strictly govern the format and content of the response. 
Note that our LLM-based method adopts zero-shot prompting~\cite{brown2020language}.

\begin{figure}[!t]
    \centering
    \includegraphics[width=0.7\linewidth]{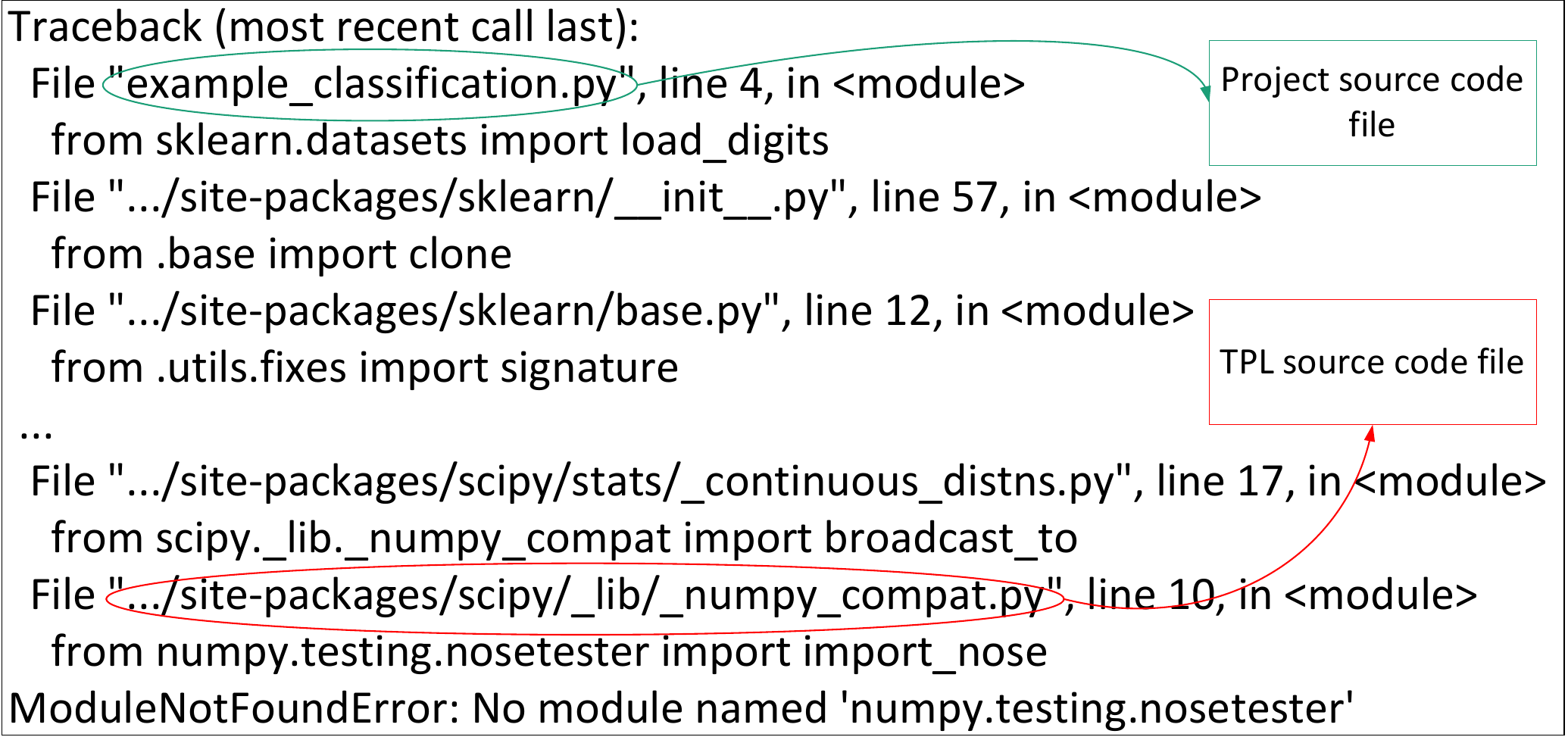}%
    \Description{An example traceback annotated to show how we select context for the LLM: the closest project source file outside site-packages at the project-library boundary, and the deepest third-party library source file inside site-packages where the exception is raised.}
    \caption{Example of project and TPL source code file acquisition.}
    \label{fig:llm-example}
\end{figure}

To ensure a fair comparison between \tool and LLM-based methods, the inputs provided to the LLMs are aligned with those used by \tool and include all available information: the project source code, the project's \mintinline{python}{requirements.txt}, the Python version, as well as PyPI metadata and source code of the involved TPLs. Note that the project code provided to the LLM refers to the project file that is closest to the TPL code in the traceback. As illustrated in Figure~\ref{fig:llm-example}, this file typically appears at the boundary in the traceback where execution transitions from project code to library code, and is usually the last file not located under the \mintinline{python}{site-packages} directory. This file directly invokes or triggers the execution of the library functions. Correspondingly, the library code provided to the LLM is the deepest TPL file in the traceback, i.e., the code location where the exception is actually raised, which is typically located under the \mintinline{python}{site-packages} directory. 

The selected LLMs for the experiments include the SOTA open-source models \textbf{DeepSeek V3} and \textbf{DeepSeek R1}, as well as the closed-source model \textbf{ChatGPT} (GPT-4o-2024-11-20), all accessed via their respective APIs on June 2, 2025. The initial prompt is queried only once, while the installation error repair prompt and runtime error repair prompt are iterated seven times each. 
In our preliminary experiments, we randomly sampled 200 upgrade scenarios and compared repair limits of five, seven, and ten iterations. Seven iterations achieve the highest inference success rate and are therefore used in the main experiments. This number is a maximum limit rather than a fixed requirement: once a repair succeeds, the process stops and proceeds to the next stage.

\subsubsection{Evaluation Datasets} 
Since both PyEGo and ReadPyE are knowledge graph-based methods and neither of their published knowledge graphs can fully cover \benchmark, we selected the subsets of \benchmark that can be covered by the knowledge graphs (i.e., \benchmark-PyEGo and \benchmark-ReadPyE) as the evaluation datasets for PyEGo and ReadPyE, respectively. 
For the evaluation of LLMs, we use \benchmark as the dataset. 

\textbf{\benchmark-PyEGo.} As shown in Tables~\ref{tab:graphcoverage1}, \ref{tab:graphcoverage2}, and \ref{tab:graphcoverage3}, 
\benchmark-PyEGo is a subset of \benchmark, as the knowledge graph constructed by PyEGo does not encompass all versions included in \benchmark. To ensure an accurate evaluation, we examine the official knowledge graph released by PyEGo and extract \benchmark-PyEGo, which includes only those versions of projects that are covered by PyEGo's knowledge graph.

\begin{table}[!t]
    \centering
     \caption{Coverage of knowledge graphs for PyEGo and ReadPyE}
      \vspace{-4mm}
     \scalebox{0.8}{
    \begin{tabular}{|l|c|c|c|}
        \hline
         \textbf{Approach} & \textbf{Pip Solved (1,689)} & \textbf{Pip Unsolved (406)} & \textbf{Total (2,095)} \\ \hline
        PyEGo & 1,064 & 146 & 1,210\\ \hline
        ReadPyE & 1,523 & 323 & 1,846 \\ \hline     
    \end{tabular}}
    \label{tab:graphcoverage1}
    \vspace{-4mm}
\end{table}

\begin{table}[!t]
    \centering
     \caption{Coverage of knowledge graphs for PyEGo and ReadPyE in Project-TPL code compatibility issues}
      \vspace{-4mm}
     \scalebox{0.8}{
    \begin{tabular}{|l|c|c|c|c|c|}
        \hline
        \textbf{Approach} & \textbf{Module (33)} & \textbf{API Name (136)} & \textbf{API Parameter (20)} & \textbf{API Body (22)} & \textbf{Total (211)} \\ \hline
        PyEGo & 23 & 50 & 3 & 3 & 79 \\ \hline
        ReadPyE & 33 & 105 & 14 & 15 & 167 \\ \hline
    \end{tabular}}
    \label{tab:graphcoverage2}
     \vspace{-4mm}
\end{table}

\begin{table}[!t]
    \centering
     \caption{Coverage of knowledge graphs for PyEGo and ReadPyE in TPL-TPL code compatibility issues}
      \vspace{-4mm}
     \scalebox{0.8}{
    \begin{tabular}{|l|c|c|c|c|c|}
        \hline
        \textbf{Approach} & \textbf{Module (22)} & \textbf{API Name (111)} & \textbf{API Parameter (1)} & \textbf{API Body (61)} & \textbf{Total (195)} \\ \hline
        PyEGo & 13 & 41 & 0 & 13 & 67 \\ \hline
        ReadPyE & 22 & 92 & 1 & 41 & 156 \\ \hline
    \end{tabular}}
    \label{tab:graphcoverage3}
     \vspace{-4mm}
\end{table}
To determine the coverage range of the knowledge graph, we take a representative TPL, such as TensorFlow, as an example. We create a test Python file that contains the statement \mintinline{python}{import tensorflow} and use PyEGo to infer the TPL version based on this file. Since PyEGo infers the latest version available in its knowledge graph, the returned version indicates the most recent version that PyEGo can support. This helps us determine the versions supported by PyEGo's knowledge graph.

\textbf{\benchmark-ReadPyE.} Similarly, as presented in Tables~\ref{tab:graphcoverage1}, \ref{tab:graphcoverage2}, and \ref{tab:graphcoverage3}, \benchmark-ReadPyE is a subset of \benchmark. Since the knowledge graph provided by ReadPyE does not include all versions present in \benchmark, we analyze the knowledge graph released by ReadPyE and extract \benchmark-ReadPyE to ensure that the selected examples fall within ReadPyE's coverage.

To evaluate the coverage of ReadPyE's knowledge graph, we follow the same procedure as above. Using the TensorFlow TPL as an example, we generate a test Python file containing \mintinline{python}{import tensorflow} and apply ReadPyE to infer the TPL version. Since ReadPyE, like PyEGo, defaults to inferring the latest version supported in its knowledge graph, the result represents the upper bound of ReadPyE's version coverage. This helps us determine the versions supported by ReadPyE's knowledge graph.

\subsubsection{Metrics for Evaluation}
In our evaluation, we use two key metrics: inference success rate and inference time, as shown in formulas \eqref{successrate} and \eqref{executiontime}. The inference success rate measures the proportion of successfully processed scenarios among all upgrade scenarios, while the inference time reflects the amount of time the system takes to complete each inference task.

\begin{equation}
\label{successrate}
Inference\ Success\ Rate=\frac{\#success\ scenarios}{\#all\ scenarios},
\end{equation}

\begin{equation}
\label{executiontime}
Inference\ Time={end\_time-start\_time}.
\end{equation}

\subsubsection{Experiment Environment} 
Our experiments were conducted on a server running a 64-bit Ubuntu 18.04.1 OS, equipped with two Intel Xeon Gold 6230R CPUs at 2.10GHz (26 cores with 52 threads), three Nvidia RTX 2080Ti GPUs, 160GB of RAM, 256 GB SSD, and 8 TB HDD storage. \tool is implemented using Python 3.9.

\section{Results and Analysis}\label{sec:resultanalysis}

\subsection{RQ1: How Does \tool Perform in Compatible Requirements Inference?}\label{sec:rq1}

Table~\ref{tab:performance} shows the evaluation results of \tool on \benchmark. \tool achieves a significantly higher success rate in inferring compatible requirements for upgrades compared to the baseline pip approach. 
Although pip is not a classical SAT-based dependency resolver that formulates dependency resolution as a Boolean satisfiability problem, it is the officially recommended and most widely used package manager in the Python ecosystem. Since version 20.3~\cite{pip20.3}, pip has adopted a significantly enhanced dependency resolution mechanism that integrates candidate version enumeration, backtracking, and priority-based heuristics, substantially improving its ability to handle dependency conflicts. Therefore, we select pip as the baseline package manager to reflect practical dependency upgrade scenarios in real-world development. 
We note that conda is also a relevant package manager. However, conda mainly targets environment-level package management and resolves dependencies over a broader space than our setting. In contrast, our work focuses on compatible \mintinline{python}{requirements.txt} inference for Python third-party library upgrades. Since this task setting is more directly aligned with pip and other \mintinline{python}{requirements.txt}-based tools, we do not include conda in our evaluation.

Out of 2,095 upgrade operations in \benchmark, \tool successfully produces compatible requirements for 1,970 scenarios, yielding an overall inference success rate of 94.03\%, with an improvement of 13.41\%  over the baseline pip's inference success rate (80.62\%). \tool resolves 72.17\% of the originally failing upgrade scenarios (293 out of 406 failures) by analyzing fine-grained code compatibility issues. Meanwhile, \tool maintains a significantly high inference success rate (99.29\%) on those upgrades that are already passing with pip. As a result, the overall 
upgrade-induced crashes drops substantially when using \tool. In summary, \tool greatly improves upgrade reliability, significantly increasing the inference success rate and reducing the failure rate relative to the standard pip workflow.

\begin{table}[!t]
    \centering
     \caption{Performance of \tool in inferring compatible requirements on \benchmark}
      \vspace{-4mm}
     \scalebox{0.85}{
    \begin{tabular}{|c|c|c|}
        \hline
          \textbf{Pip Solved (1,689)} & \textbf{Pip Unsolved (406)} & \textbf{Total (2,095)} \\ \hline
         1,677 (99.29\%) &293 (72.17\%) & 1,970 (94.03\%)\\ \hline
    \end{tabular}}
    \label{tab:performance}
     \vspace{-4mm}
\end{table}

\begin{table}[!t]
    \centering
    \caption{Performance of \tool in different levels and types of code compatibility issues on \benchmark}
     \vspace{-4mm}
    \scalebox{0.8}{
    \begin{tabular}{|l|r|r|r|}
        \hline
        \textbf{Type} & \textbf{Project-TPL} & \textbf{TPL-TPL} & \textbf{Total} \\ \hline
        Module        & (33 / 33) 100.00\%  & (21 / 22) 95.45\%  & (54 / 55) 98.18\% \\ \hline
        API Name      & (132 / 136) 97.06\% & (92 / 111) 82.88\%  & (224 / 247) 90.69\% \\ \hline
        API Parameter & (4 / 20) 20.00\%    & (0 / 1) 0.00\%      & (4 / 21) 19.05\% \\ \hline
        API Body      & (0 / 22) 0.00\%     & (11 / 61) 18.03\%   & (11 / 83) 13.25\% \\ \hline
        Total         & (169 / 211) 80.09\% & (124 / 195) 63.59\% & (293 / 406) 72.17\% \\ \hline
    \end{tabular}}
    \label{tab:performance1}
     \vspace{-6mm}
\end{table}

Table \ref{tab:performance1} categorizes \tool's performance across the levels (i.e., Project-TPL and TPL-TPL) and types (i.e., module, API name, API parameter, and API body) of code compatibility issues. 
For Project-TPL,  
\tool resolves 80.09\% of the issues. Specifically, \tool achieves a 100.00\% inference success rate for module-related issues and a 97.06\% success rate for API name-related issues. However, \tool handles only 20.00\% of API parameter issues and none of the API body behavior changes. 
In addition, 
\tool resolves 63.59\% of the TPL-TPL issues. Overall, \tool performs exceptionally well (more than 90\%) on module and API name issues but shows limited effectiveness for API parameter changes (19.05\%) and API body changes (13.25\%).

For module-related compatibility issues, \tool resolves 54 out of 55 scenarios, and one scenario remains unresolved. The reason is that pytorch\_lightning with version 1.2.9 requires the TPL packaging. However, in pytorch\_lightning's version constraints (1.2.9), there is no mandatory requirement of  
packaging and its version. 
Therefore, \tool infers that the requirements do not include packaging, leading to the 
project \texttt{PedalNetRT} encountering a \textit{ModuleNotFoundError: No module named ``packaging''}. Note that pytorch\_lightning has fixed the issue by adding packaging to version constraints in versions 1.3.0 and later.

For compatibility issues related to API name, \tool resolved 224 out of 247 scenarios. There are still 23 scenarios that remain unresolved. Although these 23 scenarios expose compatibility issues related to API names, there are also issues related to API body. Therefore, even after \tool resolves the API name issues, problems related to API body arise, leading to program crashes. 
For example, for the project \texttt{PedalNetRT}, its target TPL is pytorch\_lightning, with the target version set to 1.3.0. In \benchmark, since pytorch\_lightning 1.3.0 requires torchmetrics to be >= 0.4.0, and the version constraint of pytorch\_lightning 1.1.0 (specified in the starting \mintinline{python}{requirements.txt}) does not include torchmetrics. During the upgrade, pip will install the latest version of torchmetrics (i.e., 0.11.4). However, the API \mintinline{python}{get_num_classes} was removed in 0.11.4, which caused the project to crash. Since torchmetrics is a newly added TPL required by pytorch\_lightning 1.3.0, \tool specifies the oldest version of torchmetrics (0.4.0), resolving the API name-related issue. However, pytorch\_lightning 1.3.0 introduces an API body-related issue, causing the project to encounter an \textit{AttributeError: can't set attribute error}, as shown 
in Figure~\ref{patterne}.

For compatibility issues related to API parameters, \tool resolves 6 out of 21 scenarios, while 15 scenarios remain unresolved, primarily due to two reasons: (1) parameter type constraint change, and (2) API signature mismatch. 
For example, for the project \texttt{graphSAGE-pytorch}, its target TPL is torch, with a starting version of 1.0.1, and the target version is 1.5.0. 
The parameter \mintinline{python}{tensors} requirement of the API \mintinline{python}{torch.stack} in 1.5.0 is a non-empty \mintinline{python}{TensorList}, but the project used position parameters to pass an empty \mintinline{python}{TensorList} to tensors, resulting in \textit{RuntimeError: stack expects a non-empty TensorList}. For issues related to parameter type constraint changes, \tool cannot resolve them. 

Another inference failure case made by \tool is due to the API signature mismatch. For example, for the project \texttt{svoice}, its target TPL is torchaudio, with a starting version of 0.6.0 and a target version of 0.8.0. The API \mintinline{python}{torchaudio.load}'s fully qualified name in 0.6.0 is \mintinline{python}{torchaudio.backend.sox_backend.load}, while in 0.8.0 its fully qualified name has been changed to \mintinline{python}{torchaudio.backend.sox_io_backend.load}. Since \mintinline{python}{torchaudio.backend.sox_io_backend.load} does not have the `offset' keyword argument, this leads to a \textit{TypeError: load() got an unexpected keyword argument `offset'}. Importantly, the original backend (\mintinline{python}{sox_backend}) is not removed in 0.8.0, while it remains accessible via its fully qualified name (\mintinline{python}{torchaudio.backend.sox_backend.load}) instead of 
\mintinline{python}{torchaudio.load}. However, since the old API signature has not changed, \tool statically matches the old signature and considers the upgrade to be compatible, resulting in an incorrect inference. 


Furthermore, we analyze compatibility issues related to API body changes and perform empirical validation. Our findings reveal that static analysis methods for detecting such compatibility issues tend to produce a high false-positive rate, i.e., incorrectly labeling compatible changes as incompatible. This high false-positive rate makes \tool unreliable in practice, as misjudgments significantly degrade its overall effectiveness by incorrectly blocking viable upgrades. To mitigate this issue, we opt to exclude certain API body compatibility checks from our detection pipeline, thereby improving the accuracy of inferred compatible requirements. 
Note that the results reported in Table~\ref{tab:performance1} were obtained under the setting where the API behavioral compatibility checks were excluded from the detection pipeline. All RQs in this study consistently adopt the same setting; therefore, the results reported in the later RQs are not affected by this exclusion. 

Finally, we observe a small number of scenarios (12 in total) that pip could handle but \tool could not. These scenarios arose because some TPLs, for backward compatibility, allow access to deprecated or removed APIs via internal import mechanisms. 
For example, in scikit-learn 0.18.1, the function \mintinline{python}{accuracy_score} is defined in \mintinline{python}{sklearn.metrics.classification.py}. Intuitively, it can be accessed via \mintinline{python}{sklearn.metrics.classification.accuracy_score}. However, in scikit-learn 0.22.1, \mintinline{python}{accuracy_score} is defined in \mintinline{python}{sklearn.metrics._classification.py}, and \mintinline{python}{sklearn.metrics.classification.py} does not contain the \mintinline{python}{accuracy_score} function. Nevertheless, for backward compatibility, the TPL developers used \mintinline{python}{import _classification} in \mintinline{python}{sklearn.metrics.classification.py}. Therefore, users can still access this API via \mintinline{python}{sklearn.metrics.classification.accuracy_score}. However, \tool determines the API 
has been removed, leading to a misjudgment.

\begin{mdframed}[hidealllines=false,backgroundcolor=gray!10,roundcorner=3pt,skipabove=2pt]

\textbf{Answer to RQ1:} \tool achieves an inference success rate of 94.03\% in the compatible requirements inference task on \benchmark, significantly outperforming pip (80.62\%). 
\end{mdframed}

\subsection{RQ2: How Does \tool Compare to SOTA Tools in Compatible Requirements Inference?}

To address RQ2, we evaluate \tool against two SOTA tools, PyEGo and ReadPyE, focusing on their ability to infer compatible requirements for the TPL upgrade scenario. We use the respective benchmark subsets, i.e., \benchmark-PyEGo and \benchmark-ReadPyE, to ensure a fair comparison, as both PyEGo and ReadPyE rely on knowledge graphs that cover only part of \benchmark. 

\begin{table}[!t]
    \centering
     \caption{Comparison of \tool and PyEGo on \benchmark-PyEGo}
      \vspace{-4mm}
     \scalebox{0.8}{
    \begin{tabular}{|l|r|r|r|r|}
        \hline
        \textbf{Approach} & \textbf{Pip Solved (1,064)}  & \textbf{Pip Unsolved (146)} & \textbf{Total (1,210)} \\ \hline
        PyEGo  & 434 (40.79\%)   & 14 (9.59\%) & 448 (37.02\%)\\ \hline
        \textbf{\tool} & \textbf{1,057 (99.34\%)}  & \textbf{132 (90.41\%)} & \textbf{1,189 (98.26\%)}\\ \hline
        Impro. &  58.55\% &  80.82\% & 61.24\%\\ \hline
    \end{tabular}}
    \label{tab:performance2}
     \vspace{-4mm}
\end{table}

\textbf{Comparison Between \tool and PyEGo.} 
Tables~\ref{tab:performance2},~\ref{tab:performance3}, and~\ref{tab:performance4} report the comparison results between \tool and PyEGo on \benchmark-PyEGo. Specifically, Table~\ref{tab:performance2} presents the overall performance comparison, while Tables~\ref{tab:performance3} and~\ref{tab:performance4} further compare the two tools on Project-TPL and TPL-TPL code compatibility issues, respectively. 
As shown in Table~\ref{tab:performance2}, \tool significantly outperforms PyEGo in terms of pip solved, pip unsolved, and overall, with improvements of 58.55\%, 80.82\%, and 61.24\%, respectively. 
The primary reason for PyEGo's poor performance is that it primarily relies on static analysis (parsing import statements) to construct a knowledge graph and infer dependency versions, which results in deficiencies in its version compatibility analysis. Under \tool's experimental setup (forcing an upgrade to the target version), PyEGo often outputs an empty dependency list, indicating that it cannot adapt to the newly specified version. Second, PyEGo lacks sufficient code compatibility detection. It does not thoroughly check whether the APIs used by the project have changed, but merely ensures that the project includes the required TPL. As a result, many project-level API changes (such as method signature or module path changes) cannot be detected, leading to compatibility issues at the Project-TPL level. Finally, PyEGo does not address code compatibility issues between libraries (TPL-TPL), and cannot detect implicit code compatibility issues caused by cross-library calls.

\begin{table}[!t]
    \centering
    \caption{Comparison of \tool and PyEGo in Project-TPL code compatibility issues on \benchmark-PyEGo}
     \vspace{-4mm}
    \scalebox{0.8}{
    \begin{tabular}{|l|c|c|c|c|c|}
        \hline
        \textbf{Approach} & \textbf{Module (23)} & \textbf{API Name (50)} & \textbf{API Parameter (3)} & \textbf{API Body (3)} & \textbf{Total (79)}\\ \hline
        PyEGo & 1 & 0 & 0 & *\textbf{3} & 4\\ \hline
        \textbf{\tool} & \textbf{23} & \textbf{46} & 0 & 0 & \textbf{69}\\ \hline
    \end{tabular}}
    \label{tab:performance3}
     \vspace{-4mm}
\end{table}

\begin{table}[!t]
    \centering
    \caption{Comparison of \tool and PyEGo in TPL-TPL code compatibility issues on \benchmark-PyEGo}
     \vspace{-4mm}
    \scalebox{0.8}{
    \begin{tabular}{|l|c|c|c|c|c|}
        \hline
        \textbf{Approach} & \textbf{Module (13)} & \textbf{API Name (41)} & \textbf{API Parameter (0)} & \textbf{API Body (13)} & \textbf{Total (67)} \\ \hline
        PyEGo & 0 & 10 & 0 & 0 & 10 \\ \hline
        \textbf{\tool} & \textbf{12} & \textbf{40} & 0 & \textbf{11} & \textbf{63} \\ \hline
    \end{tabular}}
    \label{tab:performance4}
     \vspace{-4mm}
\end{table}

Tables \ref{tab:performance3} and \ref{tab:performance4} show the ability of \tool and PyEGo to resolve fine-grained code compatibility issues at different levels. It can be seen that \tool is superior to PyEGo in all aspects except for its ability to handle API body-related compatibility issues at the Project-TPL level. 
Taking 
the project \texttt{BERT-NER} as an example, 
the target TPL seqeval was upgraded from version 0.0.12 to 0.0.15. Due to PyEgo's experimental setup, 
we didn't fix the seqeval version to be 0.0.15, as the upgraded version is incompatible with the project. Thus, we directly used the PyEGo inferred environment, i.e., seqeval 1.2.2. Under this inferred environment, the project works normally. 

\begin{figure}[!t]
    \centering

     \begin{subfigure}[b]{0.4\textwidth}
        \centering
        \includegraphics[width=\linewidth]{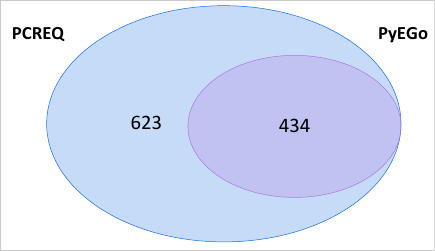}%
        \caption{Pip Solved}
        \label{fig:venn_pyego:1}
    \end{subfigure}
    \hfill
   \begin{subfigure}[b]{0.4\textwidth}
        \centering
        \includegraphics[width=\linewidth]{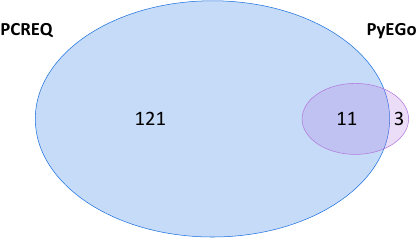}%
        \caption{Pip Unsolved}
        \label{fig:venn_pyego:2}
    \end{subfigure}
\Description{Two Venn diagrams comparing which upgrade scenarios are successfully solved by \tool versus PyEGo on \benchmark-PyEGo, separated into cases where pip can resolve dependency conflicts (Pip Solved) and cases where pip cannot (Pip Unsolved).}
\vspace{-4mm}
    \caption{Venn diagram between \tool and PyEGo on \benchmark-PyEGo.}
    \label{fig:venn_pyego}
    \vspace{-4mm}
\end{figure}

Moreover, as shown in Figure~\ref{fig:venn_pyego}, we present the intersection of compatible requirements inference between \tool and PyEGo.  
Figure~\ref{fig:venn_pyego:1} shows that in the pip solved category, all of the scenarios solved by PyEGo are fully covered by \tool. 
In addition, Figure~\ref{fig:venn_pyego:2} shows that in the pip unsolved category, 
\tool covers most of the scenarios solved by PyEGo. 

\textbf{Comparison Between \tool and ReadPyE.} 
Tables~\ref{tab:performance5},~\ref{tab:performance6}, and~\ref{tab:performance7} report the comparison results between \tool and ReadPyE on \benchmark-ReadPyE. Specifically, Table~\ref{tab:performance5} presents the overall performance comparison, while Tables~\ref{tab:performance6} and~\ref{tab:performance7} further compare the two tools on Project-TPL and TPL-TPL code compatibility issues, respectively. 
As shown in Table~\ref{tab:performance5}, \tool significantly outperforms ReadPyE in terms of pip solved, pip unsolved, and overall, with improvements of 62.37\%, 37.46\%, and 58.02\%, respectively. 
The reason for ReadPyE's poor performance is that it uses an iterative optimization method based on historical data to derive the compatible version range of the TPL. However, this mechanism is too strict in its requirements for the target version. Specifically, the version constraints obtained by ReadPyE often do not include the actual target version that needs to be upgraded to, making it unable to find a compatible working environment. In addition, ReadPyE also lacks in-depth analysis of code-level changes and has no detection mechanism for code changes at the TPL-TPL level.

\begin{table}[!t]
    \centering
     \caption{Comparison of \tool and ReadPyE on \benchmark-ReadPyE}
      \vspace{-4mm}
     \scalebox{0.85}{
    \begin{tabular}{|l|r|r|r|r|}
        \hline
        \textbf{Approach} & \textbf{Pip Solved (1,523)} & \textbf{Pip Unsolved (323)} & \textbf{Total (1,846)} \\ \hline
        ReadPyE & 563 (36.97\%) & 123 (38.08\%) & 686 (37.16\%)\\ \hline
        \textbf{\tool} & \textbf{1,513 (99.34\%)} & \textbf{244 (75.54\%)} & \textbf{1,757 (95.18\%)}\\ \hline
        Impro.  &  62.37\% &  37.46\% & 58.02\%\\ \hline
    \end{tabular}}
    \label{tab:performance5}
     \vspace{-4mm}
\end{table}

\begin{table}[!t]
    \centering
     \caption{Comparison of \tool and ReadPyE in Project-TPL code compatibility issues on \benchmark-ReadPyE}
      \vspace{-4mm}
     \scalebox{0.8}{
    \begin{tabular}{|l|c|c|c|c|c|}
        \hline
        \textbf{Approach} & \textbf{Module (33)} & \textbf{API Name (105)} & \textbf{API Parameter (14)} & \textbf{API Body (15)} & \textbf{Total (167)} \\ \hline
        ReadPyE & 0 & 100 & 0 & \textbf{12} & 112 \\ \hline
        \textbf{\tool} & \textbf{33} & \textbf{105} & \textbf{1} & 0 & \textbf{139} \\ \hline
    \end{tabular}}
    \label{tab:performance6}
     \vspace{-4mm}
\end{table}

\begin{table}[!t]
    \centering
     \caption{Comparison of \tool and ReadPyE in TPL-TPL code compatibility issues on \benchmark-ReadPyE}
      \vspace{-4mm}
     \scalebox{0.8}{
    \begin{tabular}{|l|c|c|c|c|c|}
        \hline
        \textbf{Approach} & \textbf{Module (22)} & \textbf{API Name (92)} & \textbf{API Parameter (1)} & \textbf{API Body (41)} & \textbf{Total (156)} \\ \hline
        ReadPyE & 1 & 0 & \textbf{1} & 9 & 11 \\ \hline
        \textbf{\tool} & \textbf{21} & \textbf{73} & 0 & \textbf{11} & \textbf{105} \\ \hline
    \end{tabular}}
    \label{tab:performance7}
     \vspace{-4mm}
\end{table}

\begin{figure}[!t]
    \centering
    \begin{subfigure}[b]{0.4\textwidth}
        \centering
        \includegraphics[width=\linewidth]{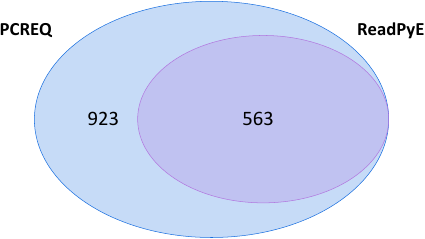}%
        \caption{Pip Solved}
        \label{fig:venn_readpye:1}
    \end{subfigure}
    \hfill
   \begin{subfigure}[b]{0.4\textwidth}
        \centering
        \includegraphics[width=\linewidth]{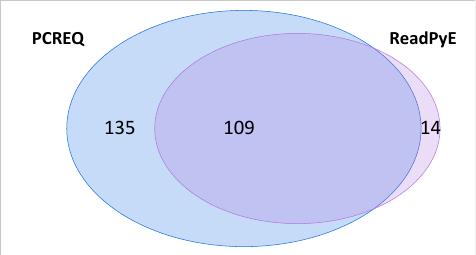}%
        \caption{Pip Unsolved}
        \label{fig:venn_readpye:2}
    \end{subfigure}
\Description{Two Venn diagrams comparing which upgrade scenarios are successfully solved by \tool versus ReadPyE on \benchmark-ReadPyE, separated into cases where pip can resolve dependency conflicts (Pip Solved) and cases where pip cannot (Pip Unsolved).}
\vspace{-4mm}
    \caption{Venn diagram between \tool and ReadPyE on \benchmark-ReadPyE.}
    \label{fig:venn_readpye}
    \vspace{-4mm}
\end{figure}

In addition, Tables \ref{tab:performance6} and \ref{tab:performance7} show the ability of \tool and ReadPyE to resolve fine-grained code compatibility issues at different levels. 
Except for the inferior ability to handle API body-related compatibility issues at the Project-TPL level and API parameter-related compatibility issues at the TPL-TPL level compared to ReadPyE, \tool outperforms ReadPyE in all other aspects. 

As shown in Table~\ref{tab:performance6}, 12 API body-related scenarios are successfully resolved by ReadPyE. 
These 12 scenarios are all from the project \texttt{pt.darts}, with a target TPL of torch. The starting version is 1.0.0, and the target versions range from 1.7.0 to 1.12.1 (12 in total). All scenarios encounter the same code compatibility issues. In the following, we use the target version 1.7.0 as an example to explain why ReadPyE can solve these scenarios. For these scenarios, ReadPyE infers that the torch version range is (>=1, <=1.3.1). Since the target version 1.7.0 is not compatible with the project, i.e., a Project-TPL issue exists. 
Therefore, according to our experimental setup, in this case, we use the version inferred by ReadPyE as the target version (1.3.1), which 
is compatible with the project. 

For the TPL-TPL code compatibility issues (Table \ref{tab:performance7}), ReadPyE solves one module-related scenario and one API parameter-related scenario, while \tool fails to solve them.  
Below, we present details of these two scenarios. 
For the module-related scenario, the project is \texttt{PedalNetRT} with a target TPL of pytorch\_lightning. The target version is 1.2.9, where the version does not specify packaging in its metadata. However, the source code of pytorch\_lightning imports packaging, resulting in a \textit{ModuleNotFoundError: No module named `packaging'} error occurring in \benchmark. Since the project's source code uses matplotlib, ReadPyE infers that the requirements include the matplotlib TPL, with the inferred version being 3.5.3. The dependency relationships of matplotlib 3.5.3 include packaging. 
Therefore, when using pip to install the requirements inferred by ReadPyE, packaging is also installed, successfully resolving this issue. By contrast, \tool retains the initial version of matplotlib as 3.3.3 in the \mintinline{python}{requirements.txt} file, which does not include packaging, leading to an incorrect inference. 

For the API parameter-related scenario (Table~\ref{tab:performance7}), the project is \texttt{GLCIC-PyTorch}, which aims to upgrade the target TPL pillow from version 8.2.0 to 8.3.0. However, the project used torchvision 0.10.0 (declared in the \mintinline{python}{requirements.txt}), which has known code compatibility issues with pillow 8.3.0~\cite{torchvisionproblem}. Therefore, after upgrading pillow using pip, the following error is raised during execution: \textit{TypeError: \_\_array\_\_() takes 1 positional argument but 2 were given}.
The root cause is that the implementation of the \mintinline{python}{__array__()} method for image objects in pillow 8.3.0 is incompatible with numpy's calling method. When numpy executes \mintinline{python}{np.array(pic, dtype=..., copy=True)}, it automatically calls \mintinline{python}{pic.__array__(dtype)}. However, pillow's image objects do not properly handle this interface, accepting only one positional argument, resulting in a mismatch in the number of arguments and finally triggering an exception. Since this is an implicit behavior in the underlying interaction between numpy and pillow, such errors cannot be explicitly detected and fixed by \tool.
By contrast, ReadPyE successfully avoids this issue, as it infers that the project depends on torch versions >=1.0.0 and <=1.8.1, and prioritizes 1.8.1. Under this inferred environment, the torchvision version is set to 0.9.1, which is compatible with torch 1.8.1, rather than the original 0.10.0. Since torchvision 0.9.1 does not have this compatibility issue with pillow 8.3.0, the project can run successfully.

Moreover, as shown in Figure~\ref{fig:venn_readpye}, we present the intersection of compatible requirements inference between \tool and ReadPyE. Figure~\ref{fig:venn_readpye:1} shows that in the pip solved category, all of the scenarios solved by ReadPyE are fully covered by \tool. In addition, Figure~\ref{fig:venn_readpye:2} shows that in the pip unsolved category, \tool covers most of the scenarios solved by ReadPyE, with 14 scenarios as exceptions. All the scenarios that \tool fails to handle have been previously discussed.

\begin{mdframed}[hidealllines=false,backgroundcolor=gray!10,roundcorner=3pt,skipabove=2pt]

\textbf{Answer to RQ2:} \tool significantly outperforms PyEGo (37.02\%) and ReadPyE (37.16\%) in terms of compatible requirements inference on \benchmark-PyEGo and \benchmark-ReadPyE, with improvements of inference success rate 
by 61.24\% and 58.02\%, respectively.

\end{mdframed}

\subsection{RQ3: How Does \tool Compare to LLMs in Compatible Requirements Inference?}

To further evaluate the effectiveness of \tool in inferring compatible requirements, 
we compare \tool with LLMs, i.e., 
ChatGPT (GPT-4o), DeepSeek V3, and DeepSeek R1 based on 2,095 upgrade scenarios from the \benchmark dataset. 

\textbf{Comparison Between \tool and LLMs.} 
As shown in Table~\ref{tab:performance8}, \tool achieves an inference success rate of 94.03\% (out of 1,970 scenarios), significantly outperforming ChatGPT (75.56\%), DeepSeek V3 (71.98\%), and DeepSeek R1 (76.32\%). Specifically, among the 406 pip unsolved scenarios, 
\tool successfully resolves 293 (72.17\%), while ChatGPT resolves 45 (11.08\%),  DeepSeek V3 resolves 45 (11.08\%), and DeepSeek R1 resolves 99 (24.38\%). In addition, \tool rarely breaks projects that were previously functional (pip solved), whereas ChatGPT and DeepSeek introduce new errors in a certain proportion of scenarios.

As shown in Tables \ref{tab:performance9} and \ref{tab:performance10}, we further analyze the performance of different approaches in handling different types of code compatibility issues. 
For Project-TPL issues, 
\tool successfully identifies 169 issues, significantly outperforming LLM-based methods. By contrast, DeepSeek R1 detects only 45, ChatGPT (GPT-4o) detects even fewer (8), and DeepSeek V3 is nearly completely ineffective, identifying only 2 issues. 
For the TPL-TPL issues, 
\tool achieves a detection rate of 124/195, far exceeding DeepSeek R1 (54/195), DeepSeek V3 (43/195), and ChatGPT (37/195).

\begin{table}[!t]
    \centering
     \caption{Performance comparison of \tool, ChatGPT, DeepSeek V3, and DeepSeek R1 on \benchmark}
      \vspace{-4mm}
     \scalebox{0.85}{
    \begin{tabular}{|l|r|r|r|r|}
        \hline
        \textbf{Approach} & \textbf{Pip Solved (1,689)}  & \textbf{Pip Unsolved (406)} & \textbf{Total (2,095)} \\ \hline
        ChatGPT  & 1,538 (91.06\%)  & 45 (11.08\%) & 1,583 (75.56\%)\\ \hline
        DeepSeek V3 & 1,463 (86.62\%) & 45 (11.08\%) & 1,508 (71.98\%)\\ \hline
        DeepSeek R1 & 1,500 (88.81\%) & 99 (24.38\%) & 1,599 (76.32\%)\\ \hline
        \textbf{\tool} & \textbf{1,677 (99.29\%)} &  \textbf{293 (72.17\%)} & \textbf{1,970 (94.03\%)}\\ \hline
    \end{tabular}}
    \label{tab:performance8}
     \vspace{-4mm}
\end{table}

\begin{table}[!t]
    \centering
    \caption{Performance comparison of \tool, ChatGPT, DeepSeek V3, and DeepSeek R1 in Project-TPL code compatibility issues on \benchmark}
     \vspace{-4mm}
    \scalebox{0.80}{
    \begin{tabular}{|l|c|c|c|c|c|}
        \hline
        \textbf{Approach} & \textbf{Module (33)} & \textbf{API Name (136)} & \textbf{API Parameter (20)} & \textbf{API Body (22)} & \textbf{Total (211)}\\ \hline
        ChatGPT     & 5 & 0  & 0 & 3 & 8\\ \hline
        DeepSeek V3 & 0 & 1 & 1 & 0 & 2\\ \hline
        DeepSeek R1 & 12 & 27 & 3 & 3 & 45\\ \hline
        \textbf{\tool} & \textbf{33} & \textbf{132} & \textbf{4} & 0 & \textbf{169} \\ \hline
    \end{tabular}}
    \label{tab:performance9}
     \vspace{-4mm}
\end{table}

\begin{table}[!t]
    \centering
    \caption{Performance comparison of \tool, ChatGPT, DeepSeek V3, and DeepSeek R1 in TPL-TPL code compatibility issues on \benchmark}
     \vspace{-4mm}
    \scalebox{0.80}{
    \begin{tabular}{|l|c|c|c|c|c|}
        \hline
        \textbf{Approach} & \textbf{Module (22)} & \textbf{API Name (111)} & \textbf{API Parameter (1)} & \textbf{API Body (61)}) & \textbf{Total (195)} \\ \hline
        ChatGPT     & 13 & 24 & 0 & 0 & 37 \\ \hline
        DeepSeek V3 & 20 & 21 & 0 & 2 & 43 \\ \hline   
        DeepSeek R1 & 12 & 41 & 0 & 1 & 54 \\ \hline
        \textbf{\tool} & \textbf{21} & \textbf{92} & 0 & \textbf{11} & \textbf{124} \\ \hline
    \end{tabular}}
    \label{tab:performance10}
     \vspace{-4mm}
\end{table}

\begin{figure}[!t]
    \centering
   \begin{subfigure}[b]{0.45\textwidth}
        \centering
        \includegraphics[width=\linewidth]{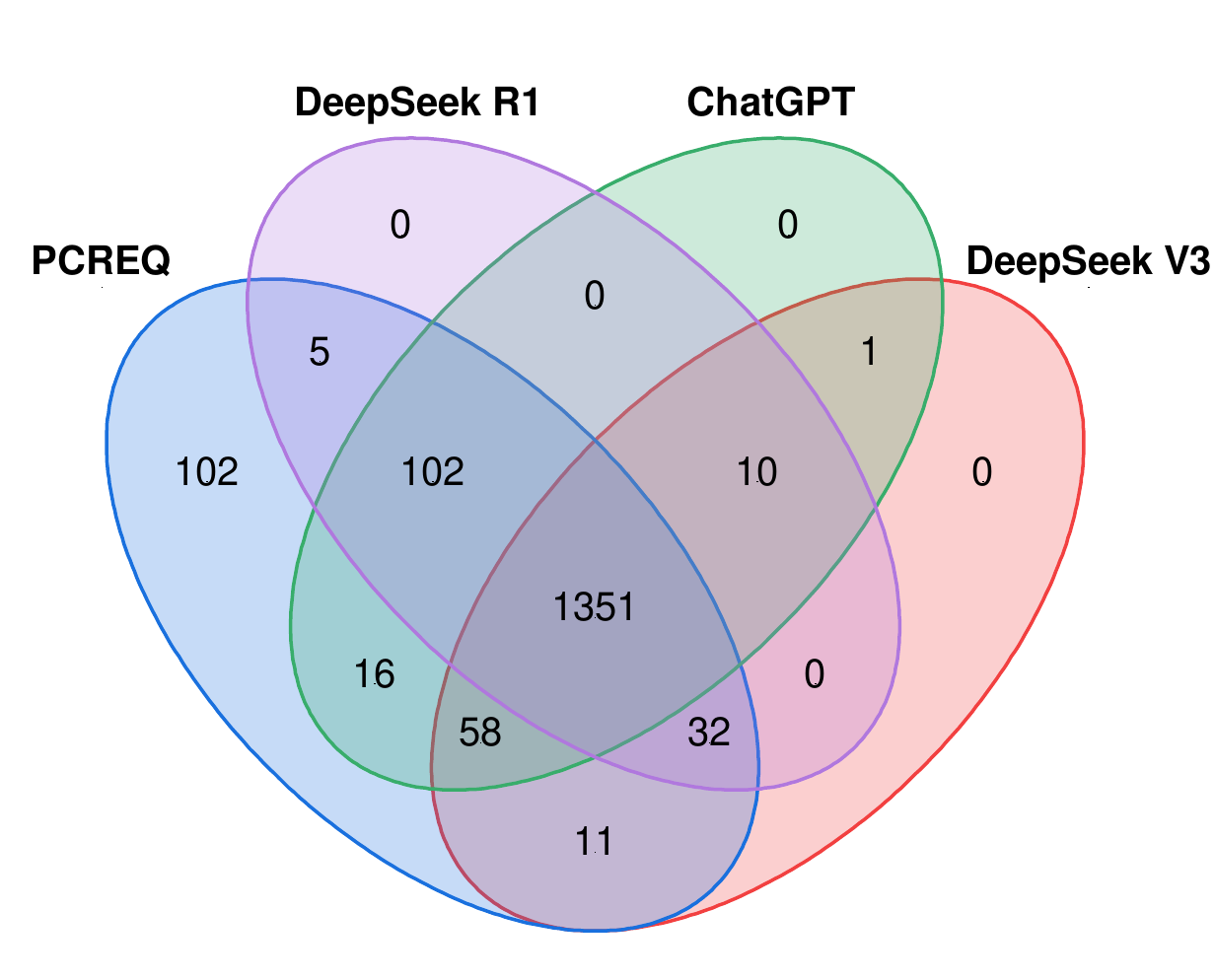}%
        \caption{Pip Solved}
        \label{fig:venn_llm:1}
    \end{subfigure}
    \hfill
   \begin{subfigure}[b]{0.45\textwidth}
        \centering
        \includegraphics[width=\linewidth]{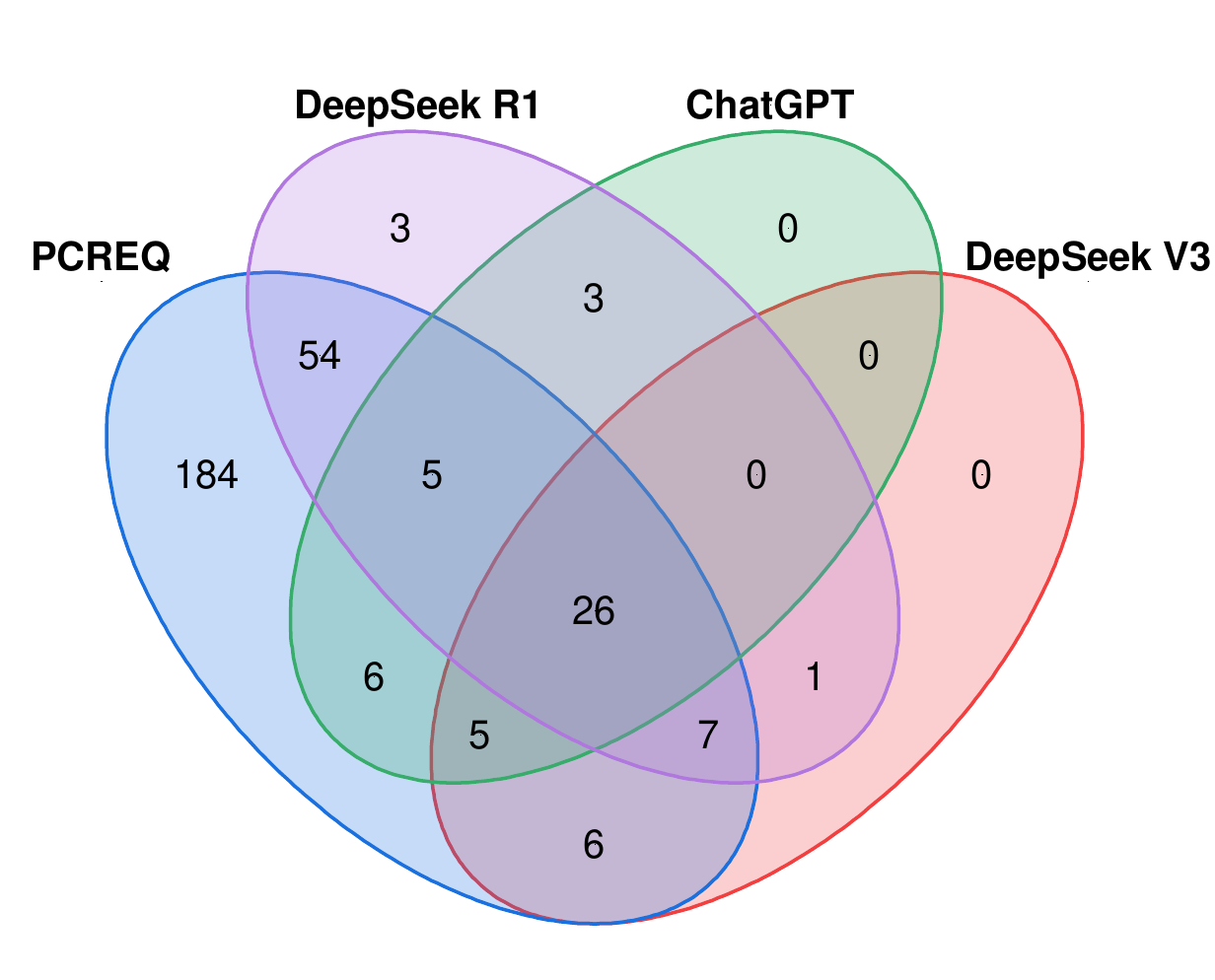}%
        \caption{Pip Unsolved}
        \label{fig:venn_llm:2}
    \end{subfigure}
\Description{Two Venn diagrams showing the overlap of successfully solved upgrade scenarios among \tool and three LLM-based methods (DeepSeek V3, DeepSeek R1, and ChatGPT) on \benchmark, separated into Pip Solved and Pip Unsolved categories.}
\vspace{-4mm}
    \caption{Venn diagram among \tool, DeepSeek V3, DeepSeek R1 and ChatGPT on \benchmark.}
    \label{fig:venn_llm}
    \vspace{-4mm}
\end{figure}

Moreover, Figure~\ref{fig:venn_llm} shows the intersection of compatible requirements inference among \tool, DeepSeek V3, DeepSeek R1, and ChatGPT on \benchmark. We can observe for both the pip solved and pip unsolved categories, \tool covers most of the scenarios solved by LLM-based methods. 
As shown in Figure~\ref{fig:venn_llm:1}, there are 11 scenarios solved by the LLM-based methods, but failed to be solved by \tool. 
As mentioned in 
Section~\ref{sec:rq1}. \tool makes misjudgments that prevent it from solving these issues. However, LLMs only need to upgrade the target TPL to the target version and then run the project. If no error occurs, our experimental setup considers that LLM-based methods make a correct inference. 

For the pip unsolved category, Figure~\ref{fig:venn_llm:2} shows that DeepSeek R1 and DeepSeek V3 can solve one scenario that \tool cannot solve. The project related to the scenario is \texttt{PedalNetRT}, with a target TPL of pytorch\_lightning. In \benchmark, the target version of pytorch\_lightning is 1.2.9, which 
uses packaging in its code. However, the dependency information of pytorch\_lightning 1.2.9 does not indicate that packaging is required, which causes the project to crash. ChatGPT and DeepSeek V3 use feedback error messages, thereby inferring that packaging is required for the project to run, thus solving the issue. 

For the remaining six scenarios that are successfully resolved by LLM-based methods, further analysis shows that all of them involve Project-TPL compatibility issues. In these cases, LLM-based methods restore normal project execution by downgrading the target dependency library. 
Taking the \texttt{BERT-NER} project as an example, the target dependency library is seqeval, with an initial version of 0.0.5 and a target upgrade version of 0.0.15. Due to Project-TPL API body incompatibilities between the project code and seqeval 0.0.15, the project encounters a runtime error:
\textit{ValueError: Invalid tag is found: [SEP]}.
After identifying and analyzing the error, the LLM (ChatGPT and DeepSeek R1) downgrades seqeval to version 0.0.10, which is compatible with the project, thereby successfully resolving the issue. \tool fails to resolve this issue mainly because it does not identify the Project-TPL code compatibility issue. 

\begin{figure}[!t]
\centering
\includegraphics[width=\linewidth]{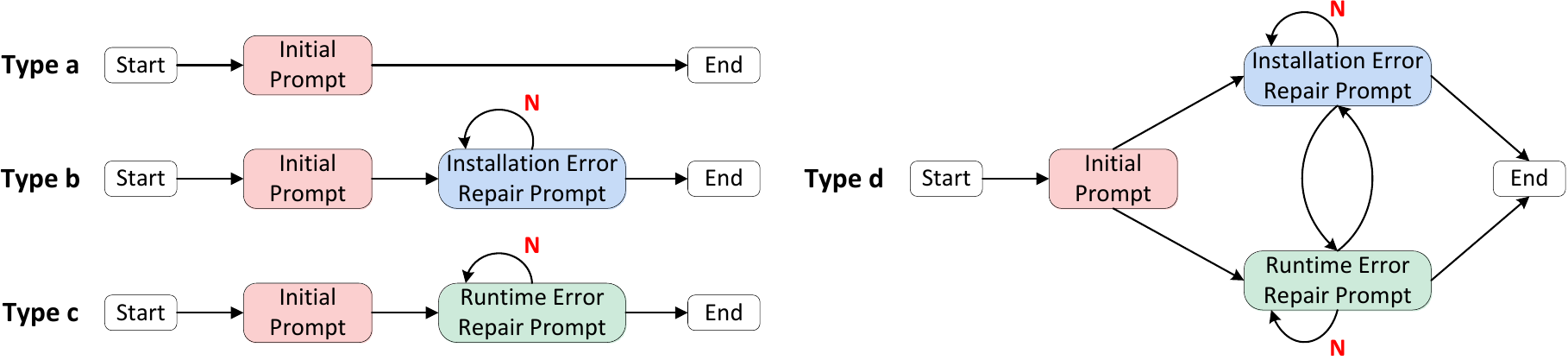}%
\Description{A schematic enumerating the possible state-transition workflows in the iterative LLM-based inference process, including sequences of initial prompting, installation-error repair iterations, runtime-error repair iterations, and termination on success or reaching the iteration limit.}
\vspace{-4mm}
\caption{All possible dynamic workflows in our LLM-based approach.}
\vspace{-4mm}
\label{llmtype}
\end{figure}

\begin{table}[!t]
\centering
\caption{Distribution of different dynamic workflows in the LLM-based approach across different LLMs and \benchmark's categories}
 \vspace{-4mm}
\label{Table:LLM}
\scalebox{0.75}{
\begin{tabular}{|l|c|c|c|c|c|c|c|c|c|c|}
\hline
 \multirow{2}{*}{\textbf{Type}} & \multicolumn{3}{c|}{\textbf{ChatGPT}} & \multicolumn{3}{c|}{\textbf{DeepSeek V3}} & \multicolumn{3}{c|}{\textbf{DeepSeek R1}} \\ \cline{2-10}
 & \textbf{Pip Solved} & \textbf{Pip Unsolved} & \textbf{Total} & \textbf{Pip Solved} & \textbf{Pip Unsolved} & \textbf{Total} & \textbf{Pip Solved} & \textbf{Pip Unsolved} & \textbf{Total} \\ \hline
a & 1,331 & 25 & 1,356 & 1,359 & 22 & 1,381 & 1,333 & 36 & 1,369 \\ \hline
    b & 112   & 1  & 113   & 78   & 1 & 79   & 99   & 6  & 105 \\  \hline
    c & 46    & 12  & 58    & 22    & 21 & 43    & 38    & 26 & 64 \\ \hline
    d & 49    & 7  & 56    & 4    & 1 & 5    & 30    & 31 & 61 \\ \hline
\end{tabular}
}
\vspace{-4mm}
\end{table}

\textbf{Analysis of the Dynamic Workflow in Our LLM-based Approach.} 
In the following, we comprehensively analyze the dynamic workflow in our LLM-based approach, to investigate how it works on the compatible requirements inference task. Figure~\ref{llm} illustrates the static inference process of our LLM-based approach, while Figure~\ref{llmtype} summarizes all possible dynamic workflows of the process. There are four categories of dynamic workflows: 

\begin{itemize}
    \item \textbf{Type a:} The LLM generates requirements based on the initial prompt and runs successfully without any installation or runtime errors. 
    
    \item \textbf{Type b:} Installation errors occur, but the LLM fixes the dependencies based on feedback and runs successfully. 
    
    \item \textbf{Type c:} Installation is successful but runtime errors occur, requiring the LLM to fix the code compatibility issues based on error prompts before running successfully. 
    
    \item \textbf{Type d:} Encountered a multi-round feedback repair process involving installation errors and runtime errors, with the LLM repeatedly processing until successful running. 

\end{itemize}

The distribution of different LLMs (ChatGPT, DeepSeek V3, and DeepSeek R1) across different dynamic workflows 
is presented in Table~\ref{Table:LLM}. 
We can observe that most scenarios fall under Type a, indicating successful execution without errors: ChatGPT (1,356), DeepSeek V3 (1,381), and DeepSeek R1 (1,369). Type b scenarios are fewer, i.e., ChatGPT (113), DeepSeek V3 (79), DeepSeek R1 (105), while Type c and Type d are the least common across all models, with DeepSeek R1 handling slightly more of these complex scenarios than the others.

Table ~\ref{Table:llm:promptcount} reports the number of prompt rounds required to resolve scenarios through Type b and Type c workflows successfully. 
For each scenario, the number of prompts corresponds to the number of iterations required before compatible requirements are found. The majority of successful resolutions for both Type b and Type c occur within a single prompt round, especially for ChatGPT (91 Type b, 22 Type c), DeepSeek V3 (78 Type b, 23 Type c), and DeepSeek R1 (91 Type b, 38 Type c). As the number of required prompts increases, the successful inference sharply declines, indicating that most models resolve compatibility issues efficiently, usually within one or two rounds. 

\begin{table}[!t]
\centering
\caption{Prompt rounds in Type b and Type c workflows scenarios across different LLMs}
 \vspace{-4mm}
\label{Table:llm:promptcount}
\scalebox{0.75}{
\begin{tabular}{|l|c|c|c|c|c|c|}
\hline
 \multirow{2}{*}{\textbf{Rounds}} & \multicolumn{2}{c|}{\textbf{ChatGPT}} & \multicolumn{2}{c|}{\textbf{DeepSeek V3}} & \multicolumn{2}{c|}{\textbf{DeepSeek R1}} \\ \cline{2-7}
     & \textbf{Type b} & \textbf{Type c}  & \textbf{Type b} & \textbf{Type c} & \textbf{Type b} & \textbf{Type c}\\ \hline
    1 & 91 & 22 & 78 & 23 & 91 & 38 \\ \hline
    2 & 17 & 8 & 1 & 12 & 8 & 14 \\  \hline
    3 & 4 & 3 & 0 & 0 & 4 & 6 \\  \hline
    4 & 1 & 1 & 0 & 2 & 1 & 4 \\ \hline
    5 & 0 & 3 & 0 & 3 & 1 & 1 \\ \hline
    6 & 0 & 0 & 0 & 3 & 0 & 1 \\ \hline
    7 & 0 & 21 & 0 & 0 & 0 & 0 \\ \hline
\end{tabular}
}
\vspace{-4mm}
\end{table}

\begin{figure}[!t]
    \centering

    \begin{subfigure}[b]{0.4\textwidth}
        \centering
        \includegraphics[width=\linewidth]{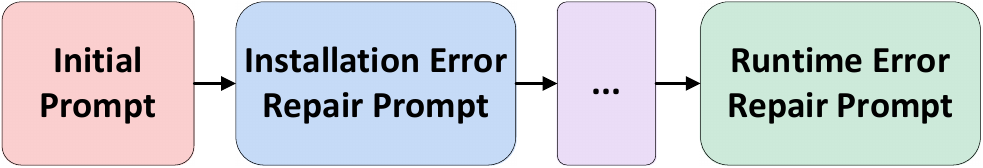}%
        \caption{}
        \label{fig:llmpath:1}
    \end{subfigure}
    \hfill
    \begin{subfigure}[b]{0.4\textwidth}
        \centering
        \includegraphics[width=\linewidth]{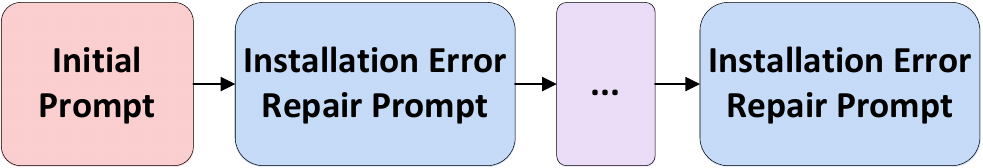}%
        \caption{}
        \label{fig:llmpath:2}
    \end{subfigure}


    \begin{subfigure}[b]{0.4\textwidth}
        \centering
        \includegraphics[width=\linewidth]{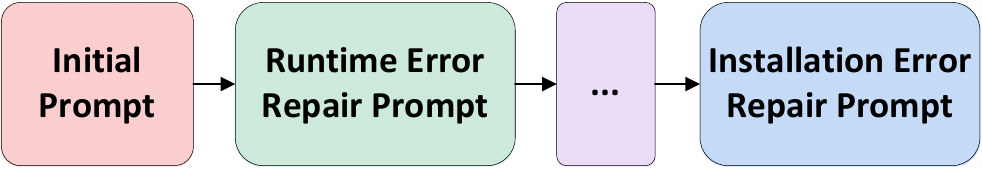}%
        \caption{}
        \label{fig:llmpath:3}
    \end{subfigure}
    \hfill
    \begin{subfigure}[b]{0.4\textwidth}
        \centering
        \includegraphics[width=\linewidth]{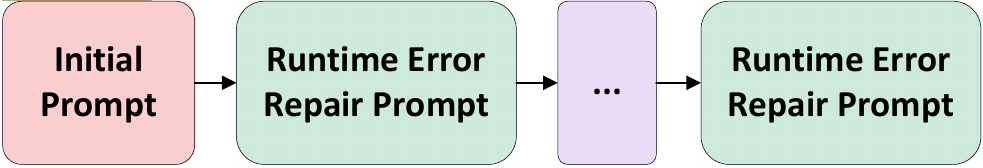}%
        \caption{}
        \label{fig:llmpath:4}
    \end{subfigure}
\Description{Four subfigures depicting the distinct sub-path variants that constitute Type d (the most complex iterative workflow) in the LLM-based approach, corresponding to different sequences of installation-error repairs and runtime-error repairs before reaching success or termination.}
\vspace{-4mm}
    \caption{All possible paths of Type d.}
    \label{llmpath}
    \vspace{-4mm}
\end{figure}

\begin{figure}[!t]
    \centering

    \begin{subfigure}[b]{0.45\textwidth}
        \centering
        \includegraphics[width=\linewidth]{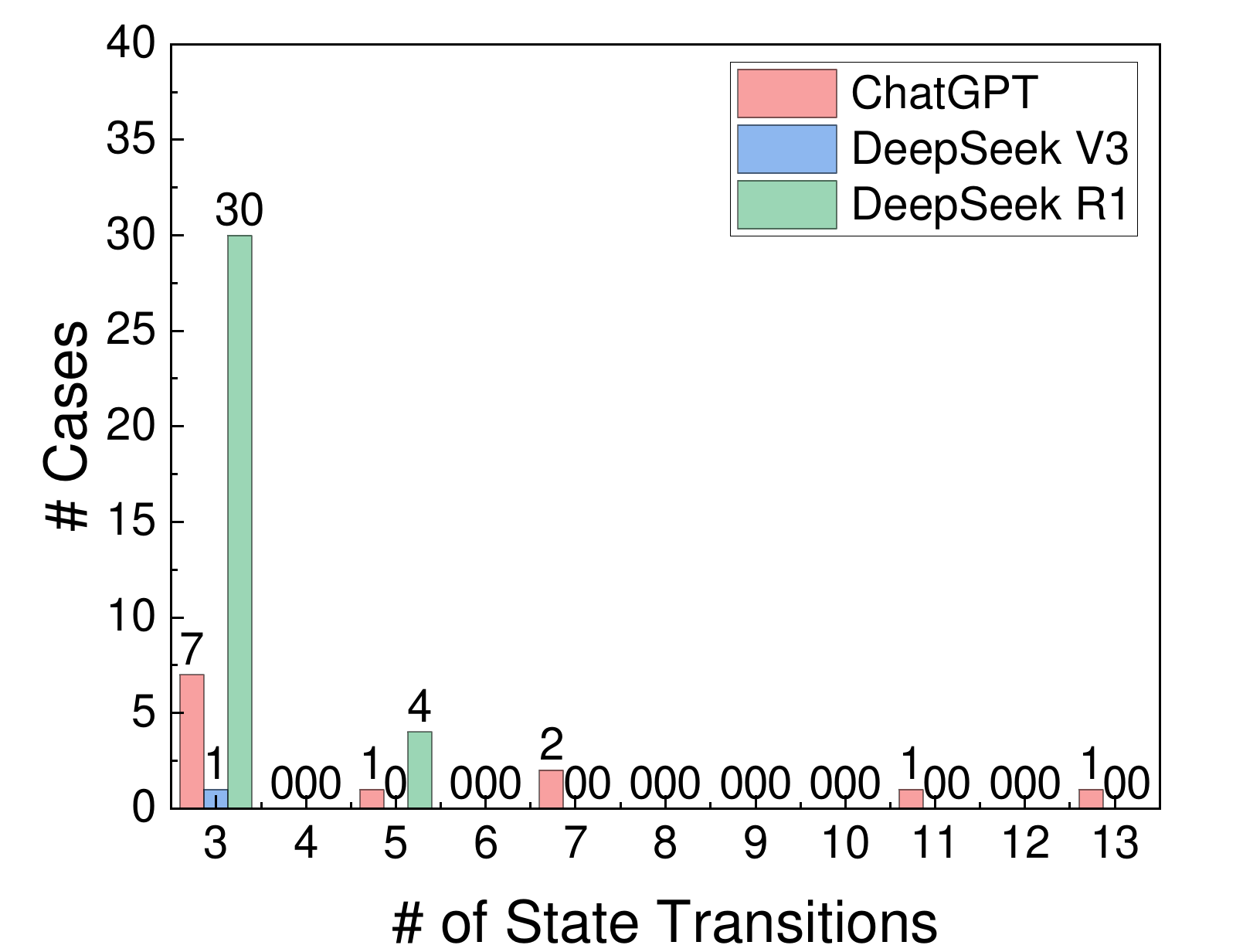}%
        \caption{Path a}
        \label{fig:path:1}
    \end{subfigure}
    \hfill
    \begin{subfigure}[b]{0.45\textwidth}
        \centering
        \includegraphics[width=\linewidth]{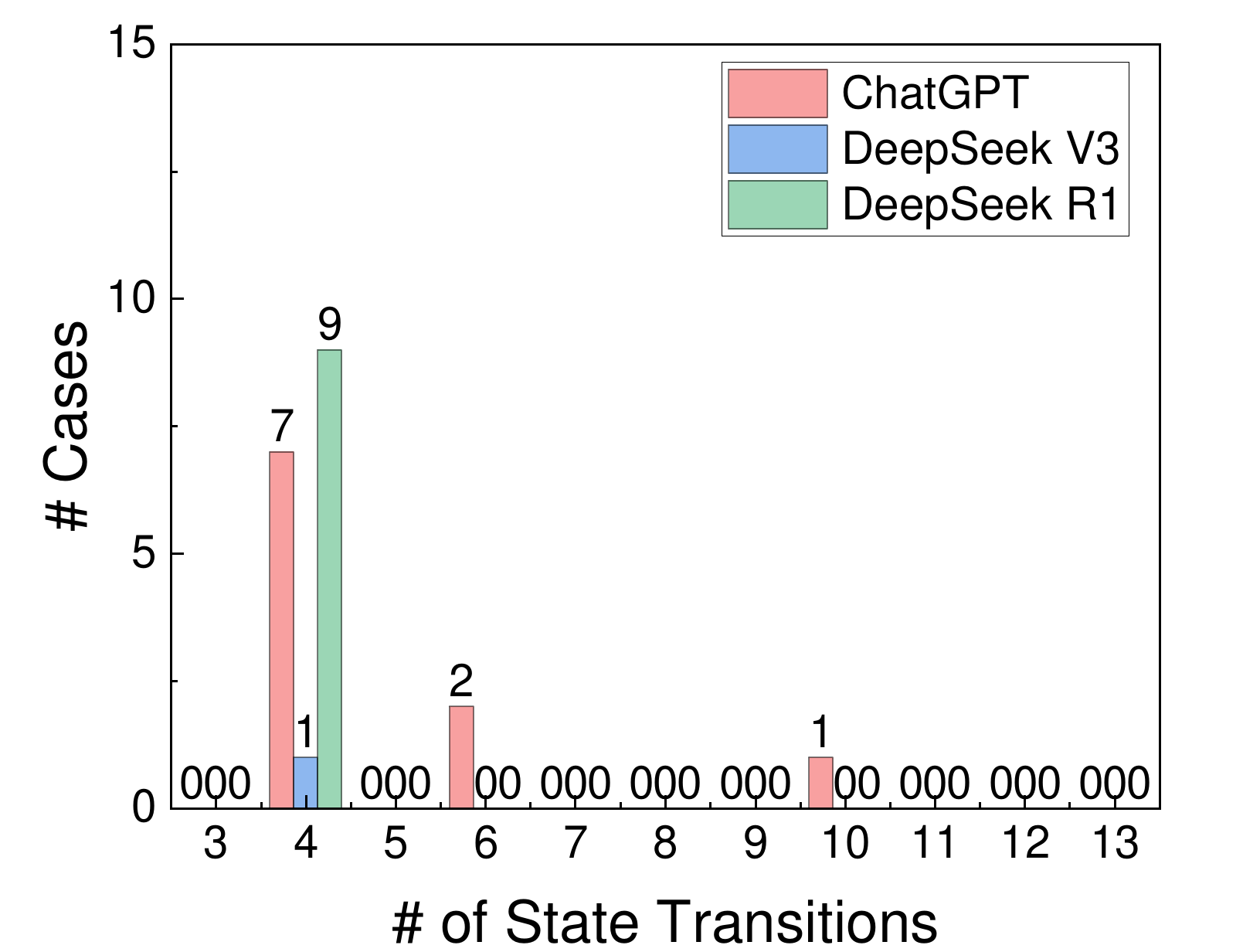}%
        \caption{Path b}
        \label{fig:path:2}
    \end{subfigure}


    \begin{subfigure}[b]{0.45\textwidth}
        \centering
        \includegraphics[width=\linewidth]{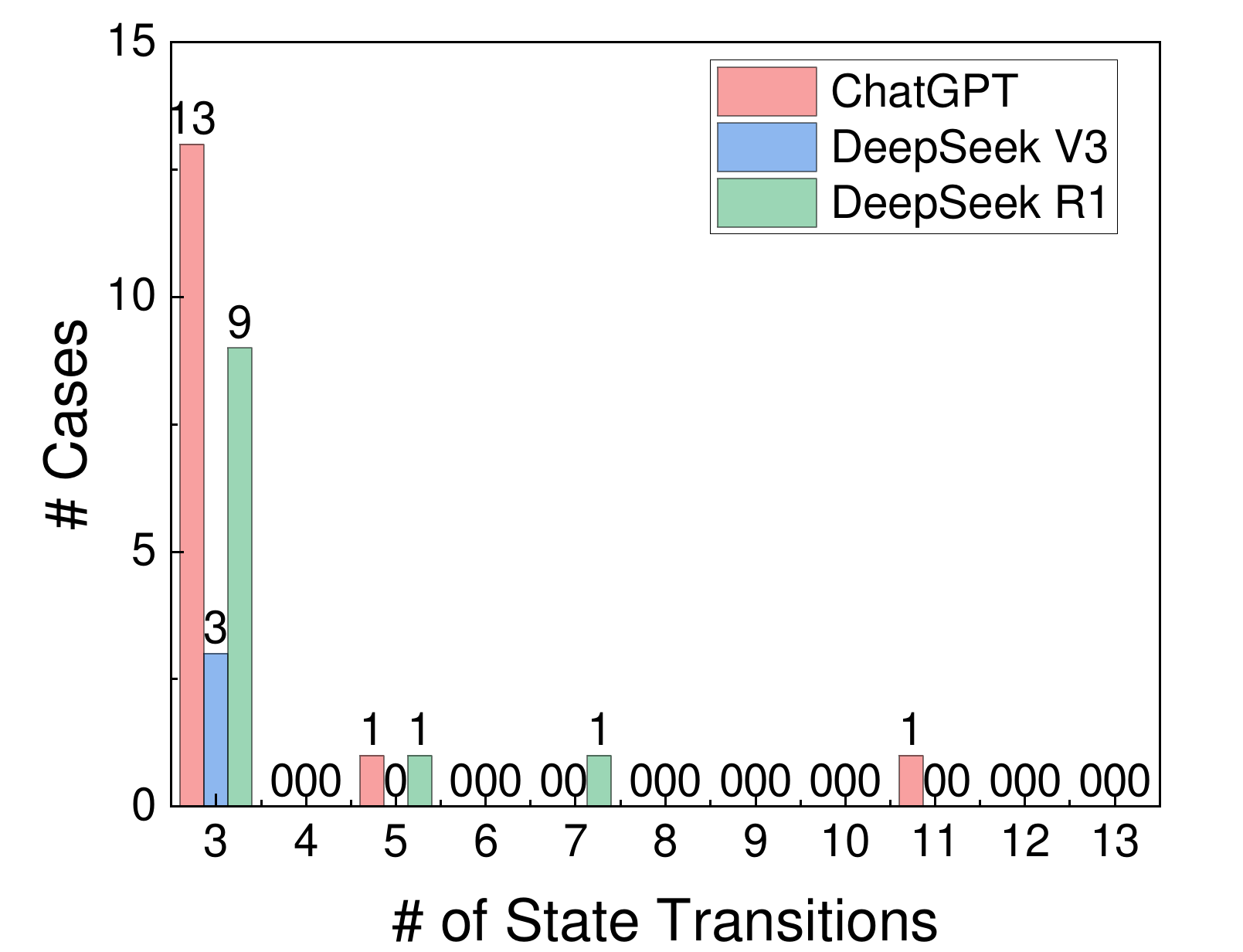}%
        \caption{Path c}
        \label{fig:path:3}
    \end{subfigure}
    \hfill
    \begin{subfigure}[b]{0.45\textwidth}
        \centering
        \includegraphics[width=\linewidth]{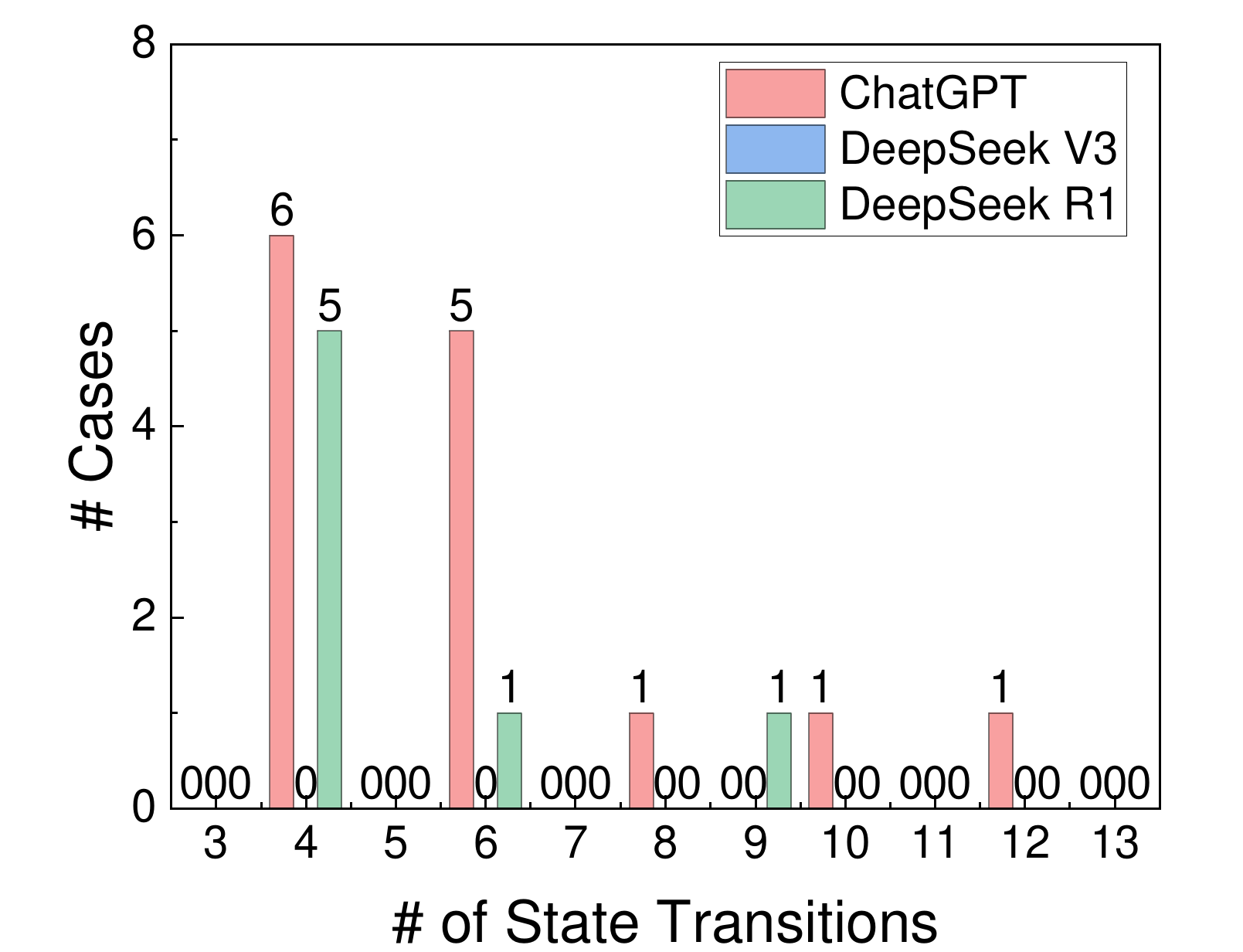}%
        \caption{Path d}
        \label{fig:path:4}
    \end{subfigure}
\Description{Four subfigures (Paths a--d) showing, for the complex Type d workflow, how many upgrade scenarios each LLM successfully resolves at each number of state transitions/repair rounds, highlighting where successes concentrate across the different paths.}
\vspace{-4mm}
    \caption{Distribution of solved scenarios in Type d across different inference paths.}
    \label{fig:path}
    \vspace{-4mm}
\end{figure}

Moreover, as depicted in Figure~\ref{llmpath}, we conduct a detailed analysis of Type d, starting from the initial prompt and then dividing it into four possible paths based on the installation error repair prompt and runtime errors repair prompt at the beginning and end, respectively. 
Figure~\ref{fig:path} shows the distribution of the number of scenarios solved by different LLMs under four sub-paths (Path a, b, c, d) in the complex iterative process Type d. The horizontal axis represents the state transitions (i.e., the number of rounds of repair prompts required), and the vertical axis represents the corresponding number of successful inference scenarios. 

We can observe that most solved scenarios are concentrated in shorter paths, particularly at 3-4 state transitions. For example, in Path a, DeepSeek R1 dominates with 30 scenarios solved in 3 transitions, whereas ChatGPT and DeepSeek V3 show fewer scenarios overall. In Path b, ChatGPT leads with 10 scenarios, of which 7 are solved in 4 transitions, followed by DeepSeek R1 with 9. Path c is mainly handled by ChatGPT, with 13 scenarios resolved at 3 transitions. In the most complex Path d, only a few scenarios are solved, mostly requiring 4-6 transitions. 
Overall, ChatGPT handles a broader distribution across paths, while DeepSeek R1 shows higher efficiency in Path a.
 
\begin{mdframed}[hidealllines=false,backgroundcolor=gray!10,roundcorner=3pt,skipabove=2pt]

\textbf{Answer to RQ3:} \tool significantly outperforms GPT-4o (75.56\%), DeepSeek V3 (71.98\%), and DeepSeek R1 (76.32\%) in terms of compatible requirements inference on \benchmark, with improvements of inference success rate by 18.47\%, 22.05\% and 17.71\%, respectively.

\end{mdframed}

\subsection{RQ4: What is the Time Cost of \tool in Compatible Requirements Inference? }

To answer RQ4, we measure the runtime of each upgrade scenario in \benchmark-PyEGo, \benchmark-ReadPyE, and \benchmark across different inference methods. In addition, we analyze the factors impacting the efficiency of \tool. 
Note that \tool, similar to ReadPyE and PyEGo, is a \mintinline{python}{requirements.txt} inference tool rather than a package installation tool. After \tool infers a compatible \mintinline{python}{requirements.txt} file, the actual package installation is performed separately using \mintinline{bash}{pip install -r requirements.txt}. Therefore, the runtime reported in RQ4 measures only the requirement inference process and is not directly comparable to the installation time of pip.

\begin{figure}[!t]
    \centering

   \begin{subfigure}[b]{0.32\textwidth}
        \centering
        \includegraphics[width=\linewidth]{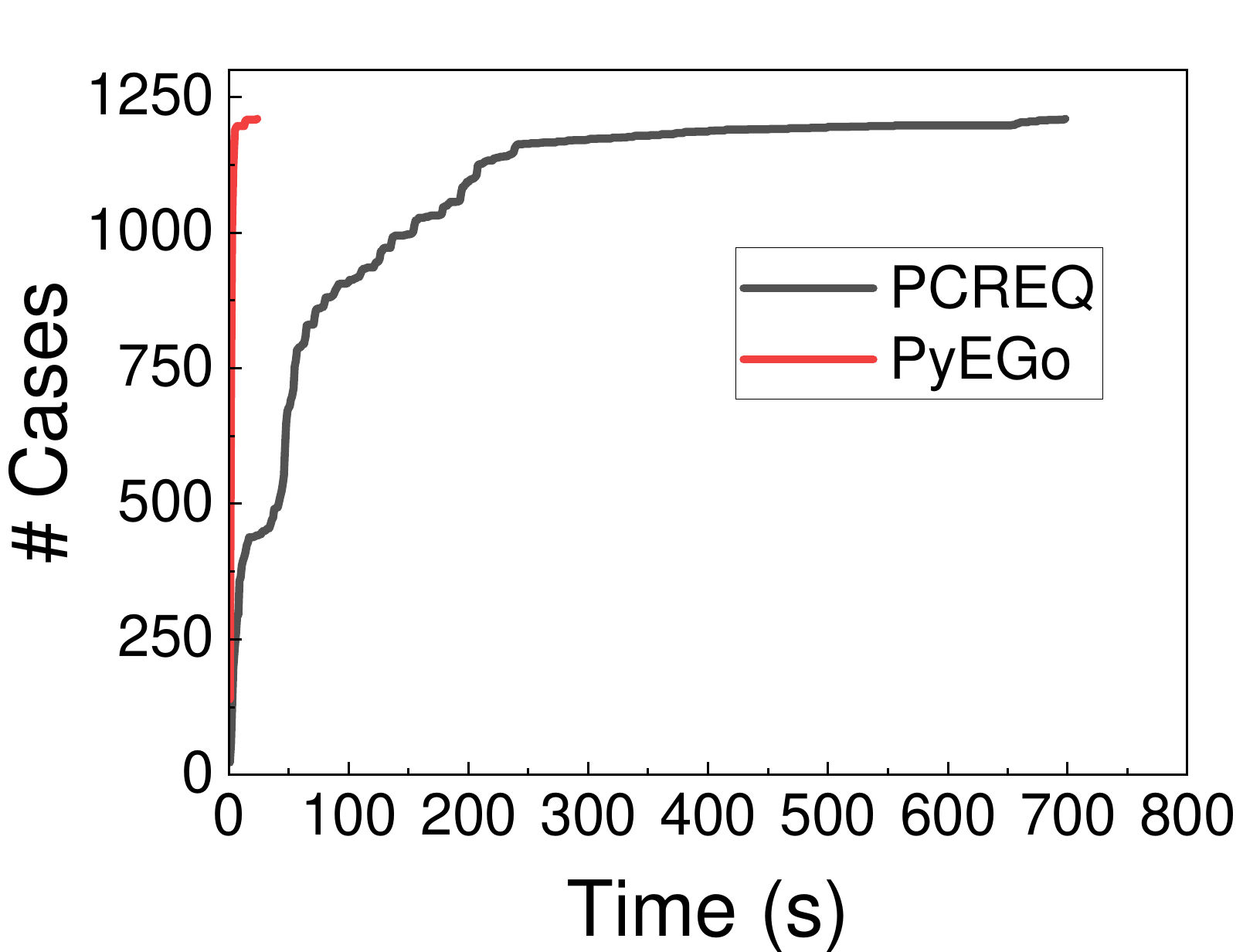}%
        \caption{\tool and PyEGo}
        \label{fig:time:1}
    \end{subfigure}
    \hfill
    \begin{subfigure}[b]{0.32\textwidth}
        \centering
        \includegraphics[width=\linewidth]{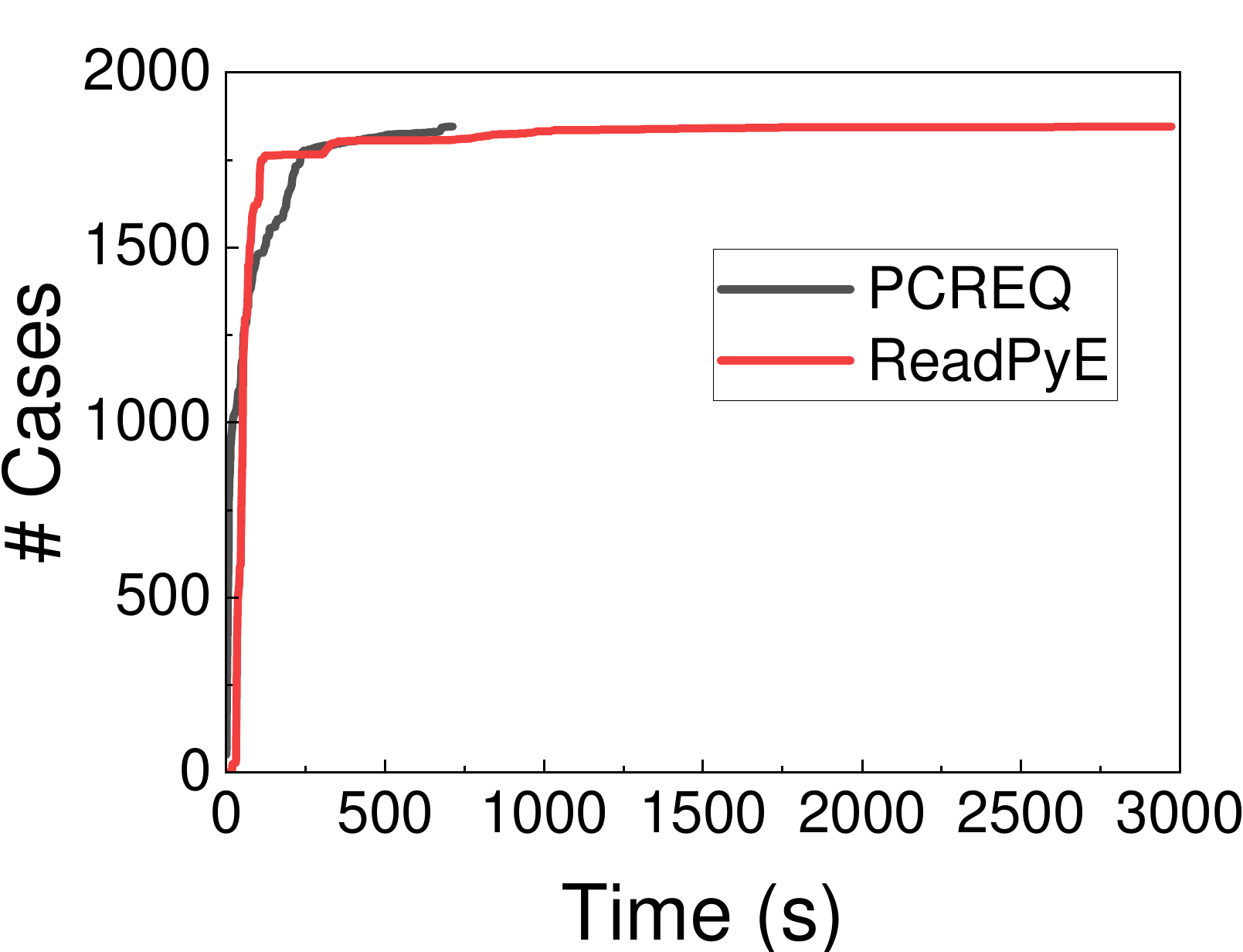}%
        \caption{\tool and ReadPyE}
        \label{fig:time:2}
    \end{subfigure}
    \hfill
     \begin{subfigure}[b]{0.32\textwidth}
        \centering
        \includegraphics[width=\linewidth]{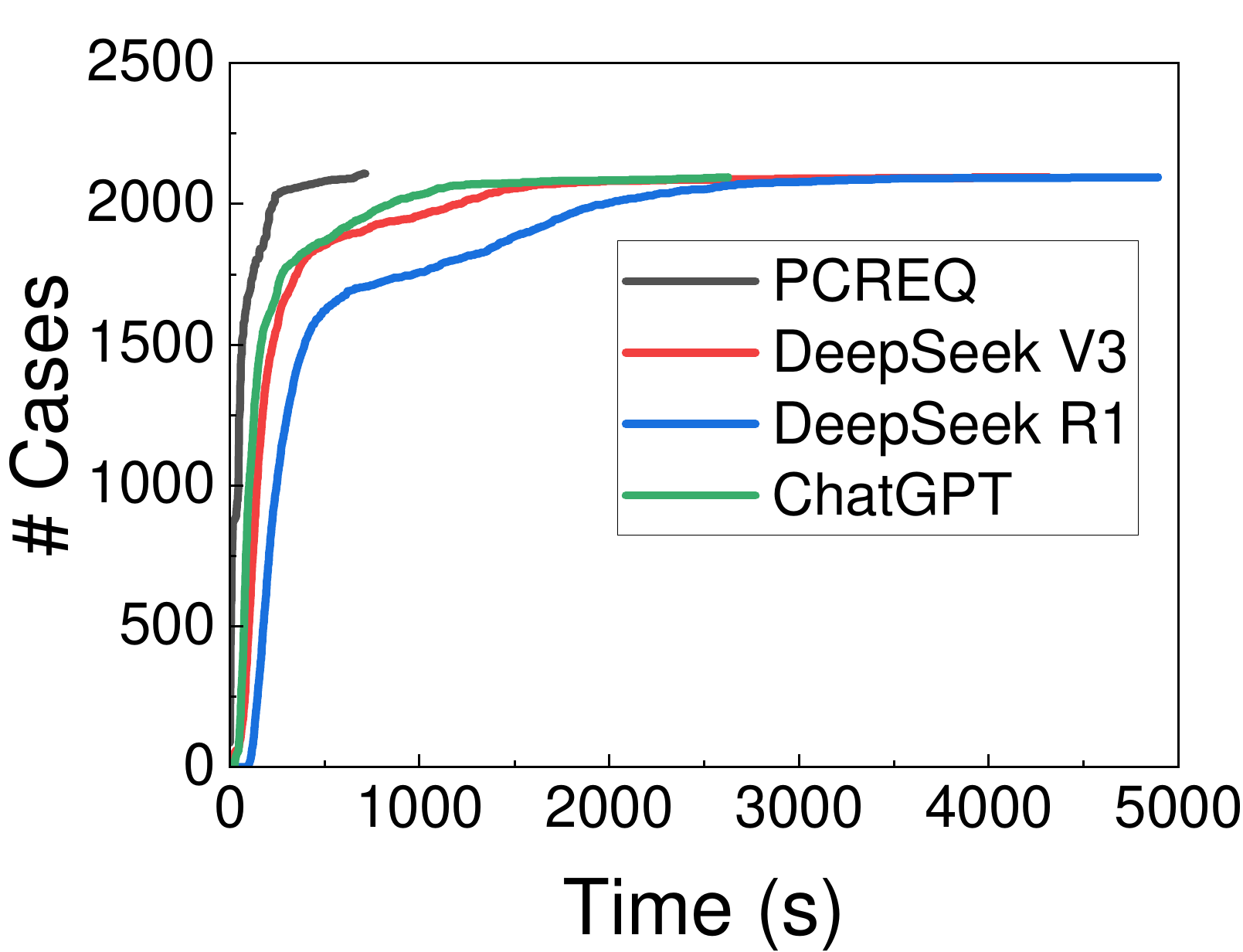}%
        \caption{\tool and LLMs}
        \label{fig:time:3}
    \end{subfigure}
\Description{Three cumulative count plots (CDF-style) of per-scenario inference runtime, comparing \tool against PyEGo, ReadPyE, and three LLM-based methods, respectively; the x-axis is time and the y-axis is the cumulative number of upgrade scenarios completed within that time.}
\vspace{-4mm}
    \caption{Cumulative count plots of inference times for \tool and PyEGo on \benchmark-PyEGo, \tool and ReadPyE on \benchmark-ReadPyE, \tool, DeepSeek (V3, R1), and ChatGPT on \benchmark.}
    \label{fig:time}
    \vspace{-4mm}
\end{figure}

\textbf{Comparison of Time Cost in Compatible Requirements Inference.} 
Figure~\ref{fig:time} collectively demonstrates that \tool achieves competitive time efficiency across different benchmark scenarios when compared with PyEGo, ReadPyE, and LLM-based methods. Specifically, Figure~\ref{fig:time:1} shows that \tool requires more processing time than PyEGo.  This is expected since \tool performs both version and code compatibility analysis, while PyEGo only resolves version constraints. By contrast, Figure~\ref{fig:time:2} shows that \tool significantly outperforms ReadPyE in inference time, as its cumulative distribution curve rises more steeply, indicating that most upgrade scenarios complete faster. Furthermore, Figure~\ref{fig:time:3} shows that in 90\% of the upgrade scenarios, \tool finishes processing within 193 s, confirming that it is efficient in inferring compatible requirements for the majority of upgrade scenarios. Compared with DeepSeek V3, DeepSeek R1, and ChatGPT, \tool demonstrates a clearer advantage by completing a greater proportion of upgrade scenarios within shorter time limits. 
Note that the runtime reported for \tool in RQ4 excludes the knowledge acquisition step. This is because the knowledge base is constructed offline as a one-time process and can be reused for multiple projects. Therefore, we only measure the online runtime of compatibility inference and resolution. For a fair comparison, the runtime of PyEGo and ReadPyE also excludes the construction time of their knowledge graphs.

\begin{figure}[!t]
    \centering

   \begin{subfigure}[b]{0.32\textwidth}
        \centering
        \includegraphics[width=\linewidth]{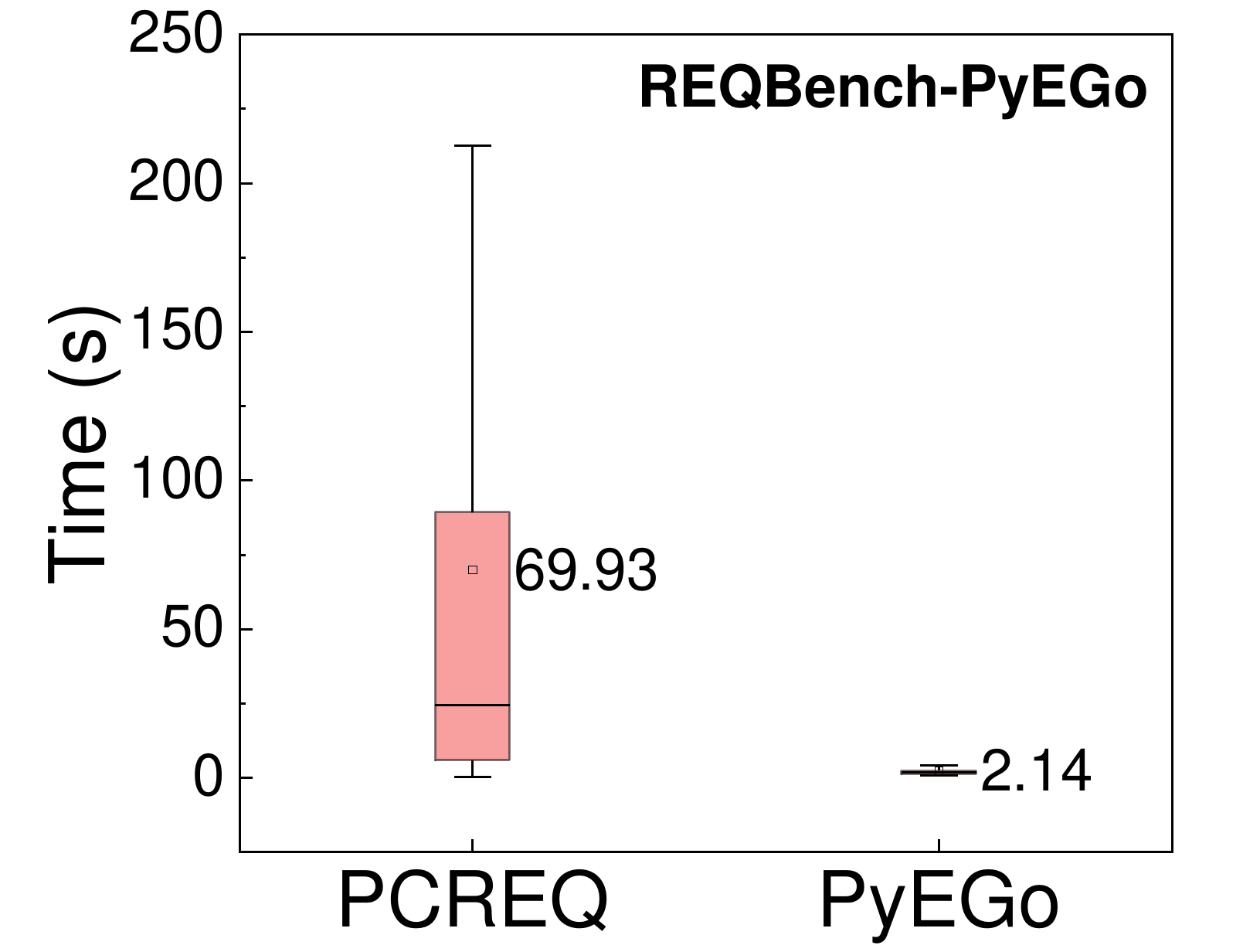}%
        \caption{\tool and PyEGo}
        \label{fig:time1:1}
    \end{subfigure}
    \hfill
    \begin{subfigure}[b]{0.32\textwidth}
        \centering
        \includegraphics[width=\linewidth]{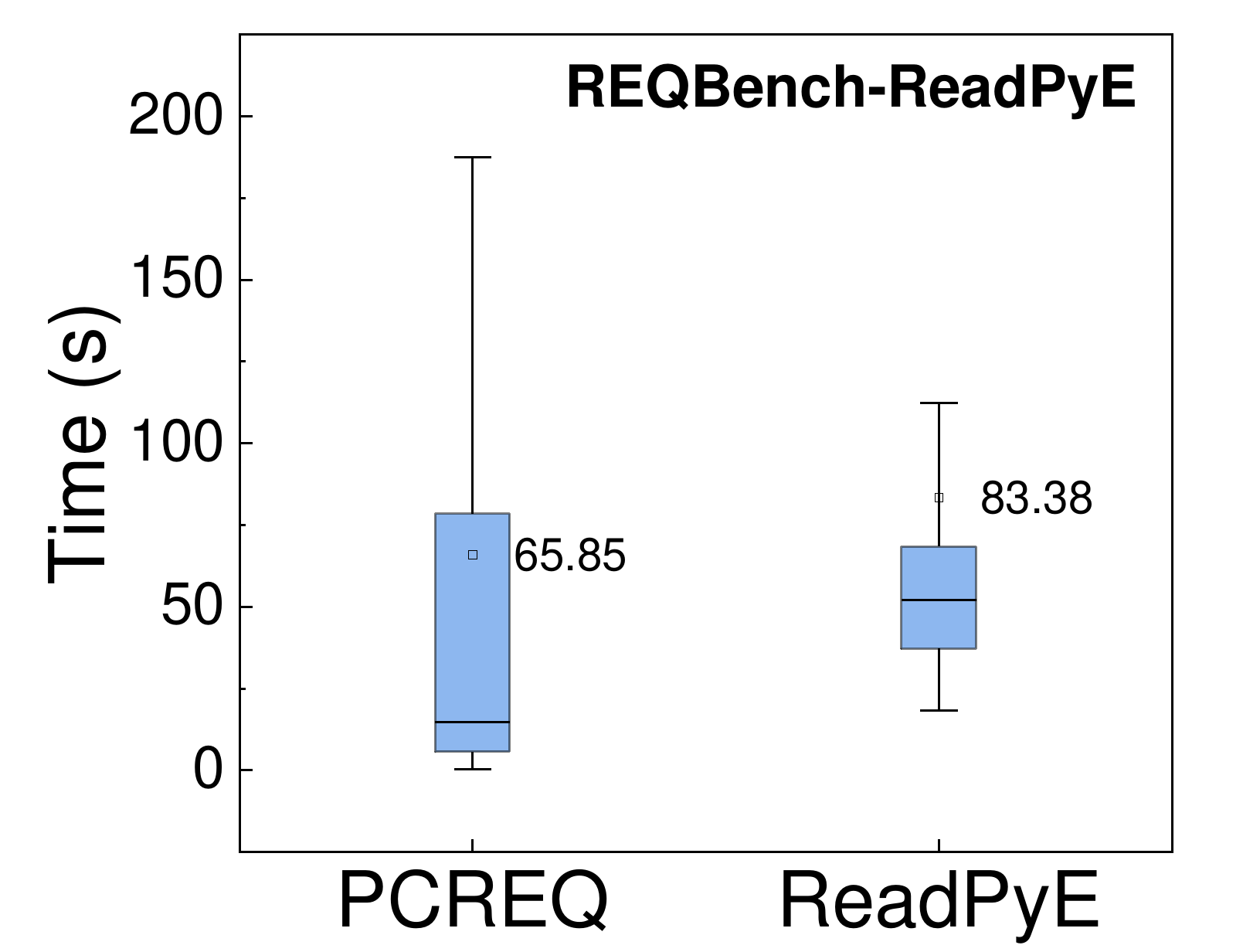}%
        \caption{\tool and ReadPyE}
        \label{fig:time1:2}
    \end{subfigure}
    \hfill
     \begin{subfigure}[b]{0.32\textwidth}
        \centering
        \includegraphics[width=\linewidth]{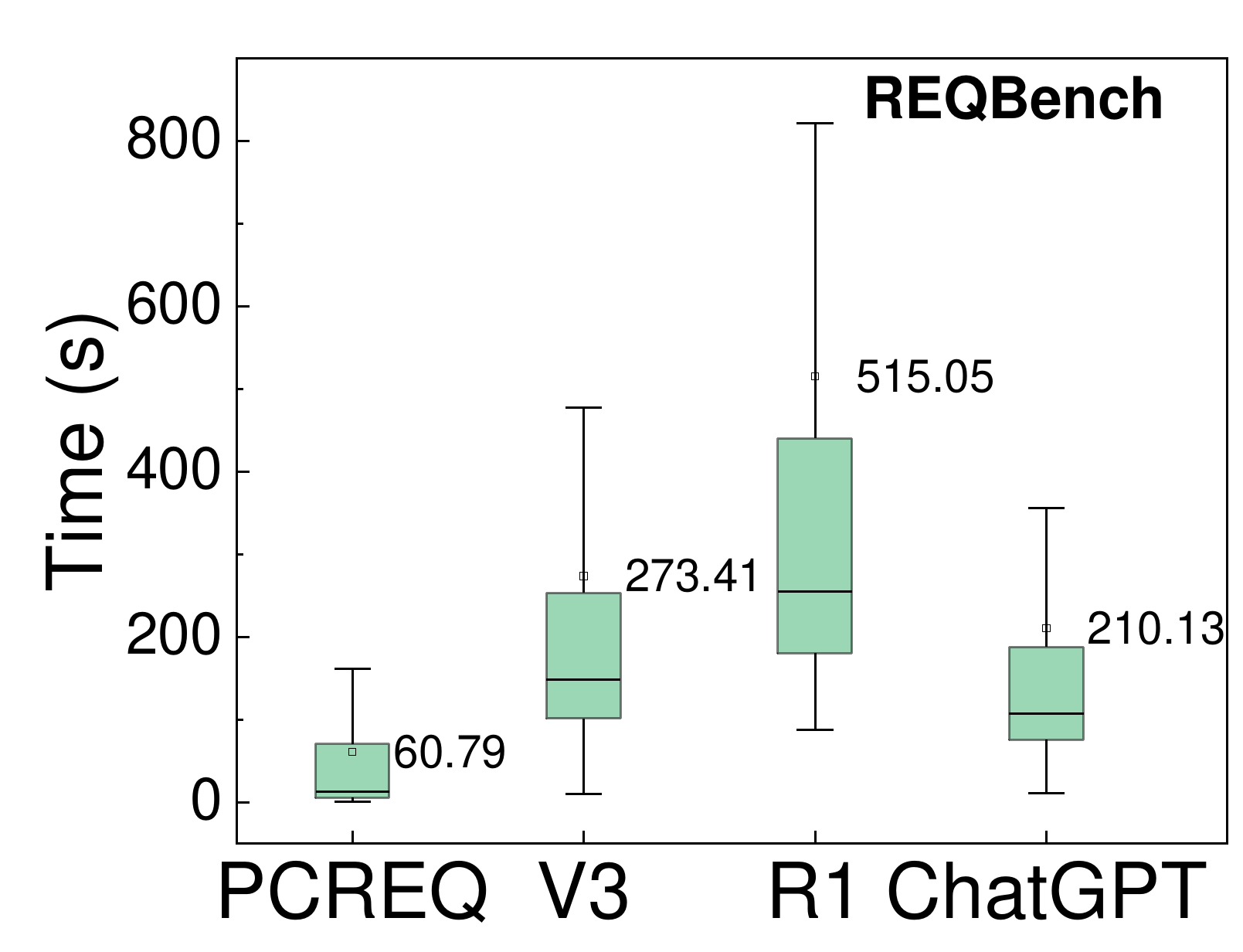}%
        \caption{\tool and LLMs}
        \label{fig:time1:3}
    \end{subfigure}
\Description{Three box plots summarizing the distribution of per-scenario inference runtime (median, quartiles, and outliers) for \tool compared with PyEGo, ReadPyE, and the LLM-based methods on their respective benchmark datasets.}
\vspace{-4mm}
    \caption{Box plots of inference times for \tool and PyEGo on \benchmark-PyEGo, \tool and ReadPyE on \benchmark-ReadPyE, and \tool, DeepSeek (V3, R1), and ChatGPT on \benchmark.}
    \label{fig:time1}
    \vspace{-4mm}
\end{figure}

In addition, after removing outliers, as depicted in Figure~\ref{fig:time1:1}, 
the average inference time of \tool in the \benchmark-PyEGo upgrade scenarios is 69.93 s, while PyEGo is 2.14 s. The average running time of \tool in the \benchmark-ReadPyE upgrade scenarios is 65.85 s, while ReadPyE is 83.38 s, as depicted in Figure~\ref{fig:time1:2}. The comparison with LLMs also demonstrates that \tool has good efficiency. As shown in Figure~\ref{fig:time1:3}, 
The average inference time of upgrade scenarios in the \benchmark is 60.79 s, 273.41 s, 515.05 s, and 210.13 s, respectively for \tool, DeepSeek V3, R1, and ChatGPT.  

\begin{table}[!t]
    \centering
    \caption{Statistical test results for \tool vs. different baseline methods in inferring time}
    \vspace{-4mm}
    \scalebox{0.85}{
    \begin{tabular}{|l|c|c|c|}
        \hline
        \textbf{Compared Methods} & \textbf{p-value} & \textbf{Effect Size $\delta$} & \textbf{Interpretation} \\ \hline
        \tool vs. PyEGo        & $<0.001$ & 0.857 & Large \\ \hline
        \tool vs. ReadPyE     & $<0.001$ & -0.375 & Medium \\ \hline
        \tool vs. DeepSeek V3 & $<0.001$ & -0.731 & Large \\ \hline
        \tool vs. DeepSeek R1 & $<0.001$ & -0.878 & Large \\ \hline
        \tool vs. ChatGPT     & $<0.001$ & -0.659 & Large \\ \hline
    \end{tabular}}
    \label{Table:stat:efficiency}
    \vspace{-4mm}
\end{table}

\begin{table}[!t]
\centering
\caption{Overview of analyzed projects and TPLs}
 \vspace{-4mm}
\label{Table:rq4}
\scalebox{0.75}{
\begin{tabular}{|l|c|c|c|c|c|c|}
\hline
\textbf{Project}                             & \textbf{TPL}               & \textbf{Start}    & \textbf{End}     &\textbf{\#Versions} & \textbf{\#TPL Call Chains} \\ \hline
\multirow{4}{*}{PyTorch-ENet}       & pillow            & 6.2.0    & 9.5.0   & 25        & 2  \\ \cline{2-6} 
                                    & torch             & 1.1.0    & 1.13.1  & 21        & 2  \\ \cline{2-6}
                                    & torchvision       & 0.3.0    & 0.14.1  & 22        & 3\\ \cline{2-6}        
                                    & numpy             & 1.16.0   & 1.21.6  & 35        & 4 \\ \hline
\multirow{4}{*}{MASTER-pytorch}     & pillow            & 7.2.0    & 8.4.0   & 10        & 2 \\ \cline{2-6} 
                                    & torch             & 1.5.1    & 1.10.2  & 10        & 3\\ \cline{2-6}
                                    & torchvision       & 0.6.1    & 0.11.2  & 10        & 3\\ \cline{2-6}        
                                    & numpy             & 1.16.4   & 1.19.5  & 20        & 4\\ \hline
\multirow{4}{*}{GLCIC-PyTorch}      & pillow            & 8.2.0    & 9.5.0   & 12        & 2\\ \cline{2-6} 
                                    & torch             & 1.9.0    & 1.13.1  & 9         & 3\\ \cline{2-6}
                                    & torchvision       & 0.10.0    & 0.14.1 & 9         & 3\\ \cline{2-6}        
                                    & numpy             & 1.19.2   & 1.21.6  & 14        & 3\\ \hline
\multirow{4}{*}{RetinaFace\_Pytorch}& pillow            & 6.1.0    & 9.5.0   & 26        & 3\\ \cline{2-6} 
                                    & torch             & 1.1.0    & 1.13.1  & 21        & 2\\ \cline{2-6}
                                    & torchvision       & 0.3.0    & 0.14.1  & 22        & 3\\ \cline{2-6}        
                                    & numpy             & 1.16.4   & 1.21.6  & 31        & 7\\ \hline
\multirow{4}{*}{siamese-pytorch}    & pillow            & 5.4.1    & 9.5.0   & 28        & 2\\ \cline{2-6} 
                                    & torch             & 1.0.1    & 1.13.1  & 22        & 2\\ \cline{2-6}
                                    & torchvision       & 0.2.1    & 0.14.1  & 24        & 3\\ \cline{2-6}        
                                    & numpy             & 1.16.1   & 1.21.6  & 34        & 3\\ \hline
\end{tabular}
}
\vspace{-4mm}
\end{table}

As shown in Table~\ref{Table:stat:efficiency}, the results of the Mann-Whitney U tests indicate that the inference time differences between \tool and PyEGo, ReadPyE, as well as between \tool and all LLM-based baselines, are statistically significant ($p < 0.001$). Furthermore, we introduce Cliff's Delta as the effect size measure and interpret it using commonly adopted thresholds (i.e., negligible effect when $|\delta| < 0.147$, small effect when $0.147 \leq |\delta| < 0.33$, medium effect when $0.33 \leq |\delta| < 0.474$, and large effect when $|\delta| \geq 0.474$). 
In our setting, $\delta > 0$ indicates that \tool tends to have higher runtime than the baseline, whereas $\delta < 0$ indicates that \tool tends to have lower runtime than the baseline. The effect size magnitude is interpreted based on $|\delta|$. 
The results show that the inference time difference between PCREQ and ReadPyE corresponds to a medium effect size (0.375), while the differences between PCREQ and the LLM-based baselines exhibit large effect sizes (0.731, 0.878, and 0.659, respectively). 
In summary, \tool exhibits moderate time overhead compared to lightweight tools like PyEGo, but remains significantly more efficient than ReadPyE and LLM-based baselines. \tool is able to complete over 90\% of upgrade scenarios in \benchmark under 193 s, with a post-outlier average time of 60.79 s. 
This demonstrates that \tool has practical applicability and strong scalability for real-world Python TPL upgrade scenarios.

\begin{figure}[!t]
    \centering

   \begin{subfigure}[b]{0.45\textwidth}
        \centering
        \includegraphics[width=\linewidth]{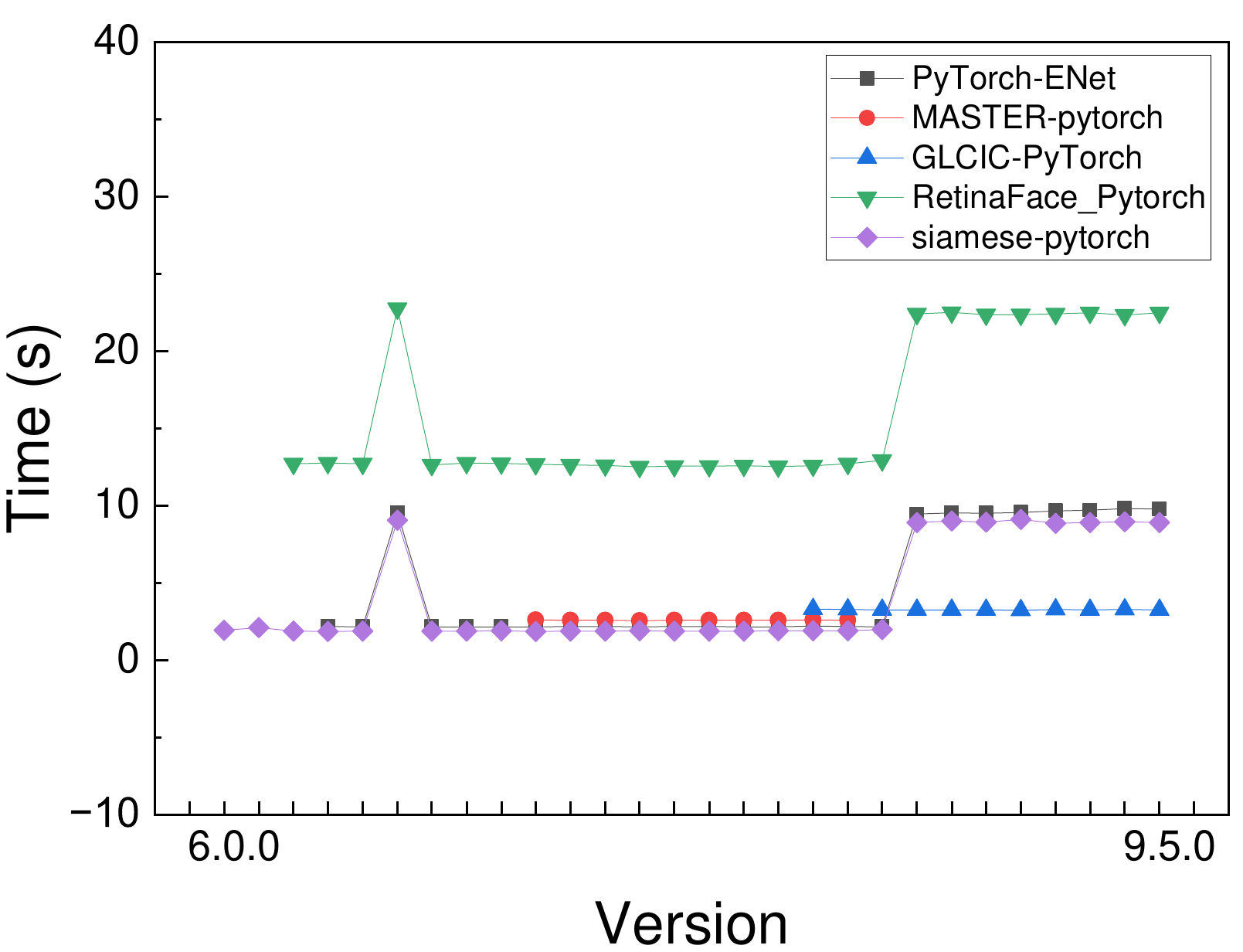}%
        \caption{pillow}
        \label{fig:time3:1}
    \end{subfigure}
    \hfill
    \begin{subfigure}[b]{0.45\textwidth}
        \centering
        \includegraphics[width=\linewidth]{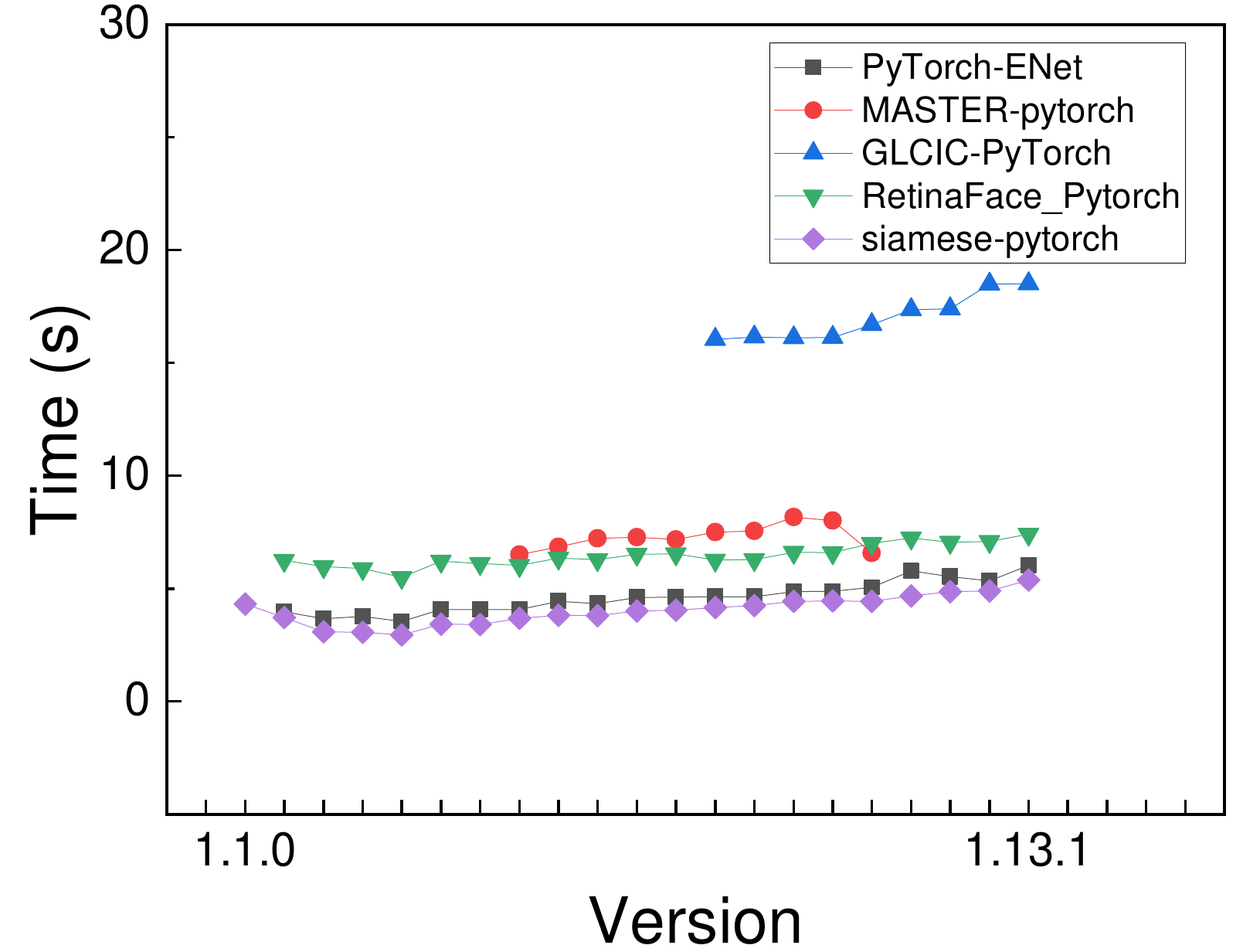}%
        \caption{torch}
        \label{fig:time3:2}
    \end{subfigure}


     \begin{subfigure}[b]{0.45\textwidth}
        \centering
        \includegraphics[width=\linewidth]{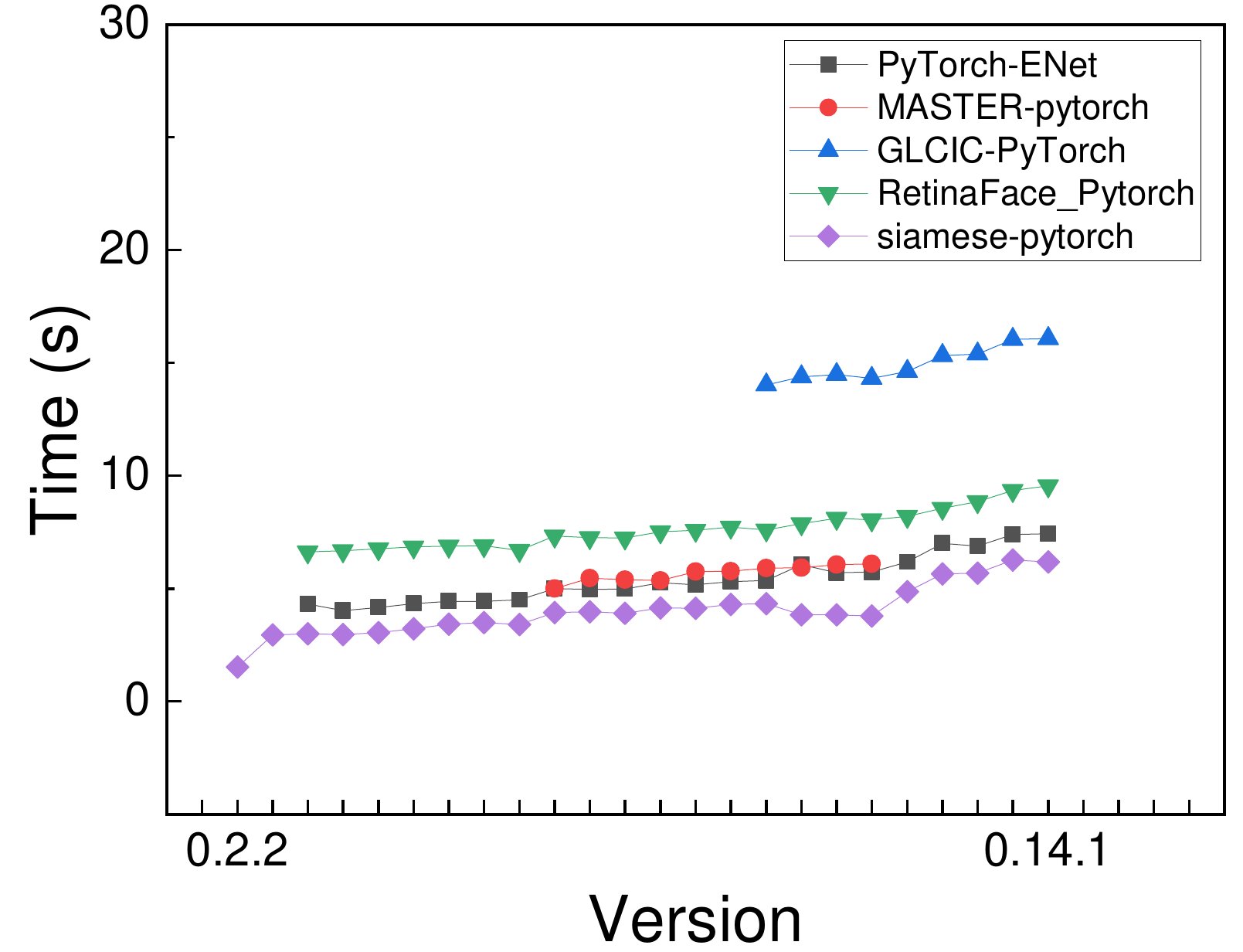}%
        \caption{torchvision}
        \label{fig:time3:3}
    \end{subfigure}
    \hfill
    \begin{subfigure}[b]{0.45\textwidth}
        \centering
        \includegraphics[width=\linewidth]{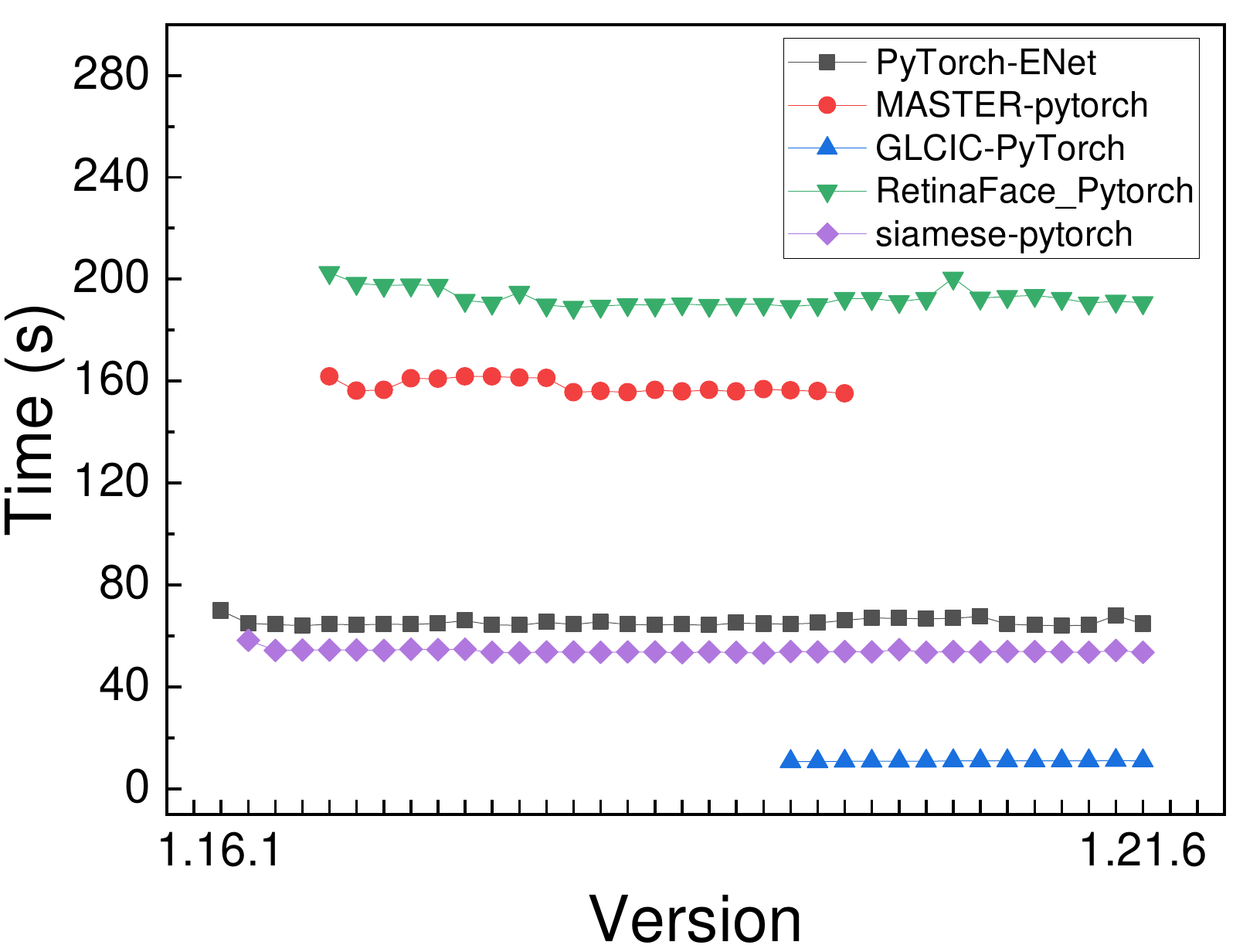}%
        \caption{numpy}
        \label{fig:time3:4}
    \end{subfigure}
\Description{Four subfigures reporting \tool's inference time across different target-library upgrade versions for pillow, torch, torchvision, and numpy; in each plot, multiple curves correspond to different projects, illustrating how runtime varies with target version and project.}
    \vspace{-4mm}
    \caption{Time cost of \tool across different versions of target TPLs: (a) pillow, (b) torch, (c) torchvision, and (d) numpy. Each curve represents a different project.}
    \label{fig:time3}
    \vspace{-4mm}
\end{figure}

\textbf{Impact Factors on Time Cost of \tool.} 
We conduct case studies to investigate the impact factors on \tool's inference time. 
As shown in Table~\ref{Table:rq4}, we select five representative projects from \benchmark, all of which depend on the same four TPLs, i.e., pillow, torch, torchvision, and numpy. 
Figure~\ref{fig:time3} presents the trend of inference time of different projects across different target TPLs along with the changes of target versions.

As shown in Figure~\ref{fig:time3:1}, the \texttt{PyTorch-ENet}, \texttt{siamese-pytorch}, and \texttt{RetinaFace\_Pytorch} projects all observe a significant increase in \tool inference time after upgrading pillow to versions 7.0.0 and 9.0.0-9.5.0, while the inference time of \texttt{GLCIC-PyTorch} remains largely unchanged. The reason is that the version of torchvision in \texttt{GLCIC-PyTorch}'s requirements is 0.10.0, which does not use the API removed in pillow versions 7.0.0 and 9.0.0-9.5.0, namely \mintinline{python}{PIL.PILLOW_VERSION}. However, \texttt{PyTorch-ENet}, \texttt{siamese-pytorch}, and \texttt{RetinaFace\_Pytorch} all use this removed API. To resolve these code compatibility issues, \tool takes a significantly longer time for inference.

Moving to version-specific effects, Figure~\ref{fig:time3:2} reveals that when \texttt{MASTER-pytorch} uses torch as the target TPL and 1.10.2 as the target version, the inference time decreases compared to previous versions. The reason is that \tool infers the version of torchvision to be 0.3.0, whereas the previous version, such as torch 1.10.0, is inferred to be torchvision 0.11.1. When torchvision undergoes a version change, \tool checks the compatibility between torchvision and the project. The later the version of torchvision, the more inference time it takes. 

When comparing inference time across projects, Figure~\ref{fig:time3:2} shows that the \texttt{MASTER-pytorch} curve is above \texttt{RetinaFace\_Pytorch}, while as shown in Figure~\ref{fig:time3:3}, the \texttt{MASTER-pytorch} curve is below \texttt{RetinaFace\_Pytorch}. The main reason is that when \texttt{RetinaFace\_Pytorch} targets torch as the TPL, it needs to handle two TPL call chains, whereas when targeting torchvision, it needs to handle three TPL call chains. By contrast, \texttt{MASTER-pytorch} requires handling three TPL call chains regardless of whether the target TPL is torch or torchvision. 

Furthermore, as shown in Figures~\ref{fig:time3:2} and~\ref{fig:time3:3}, the inference time of \texttt{GLCIC-PyTorch} is significantly higher than that of other projects. The main reason is that the starting version of \texttt{GLCIC-PyTorch} is later, meaning that the starting version has a larger amount of code. When \tool converts the APIs directly called by the project into fully qualified names, it takes more processing time.

Moreover, as shown in Figure~\ref{fig:time3:4}, the inference of \texttt{GLCIC-PyTorch} is significantly lower than that of \texttt{siamese-pytorch}. Both \texttt{siamese-pytorch} and \texttt{GLCIC-PyTorch} need to process three TPL call chains, and two of these call chains are identical. One of them is different, and the chain in \texttt{GLCIC-PyTorch} is $project \rightarrow opencv-python \rightarrow numpy$. Since \texttt{GLCIC-PyTorch} does not directly call the opencv-python API, the processing time for this chain is very short, resulting in a short overall inference time for \texttt{GLCIC-PyTorch}. 
In addition, 
Figure~\ref{fig:time3:4} shows that \texttt{RetinaFace\_Pytorch} takes significantly longer when taking numpy as the target TPL compared to the other three TPLs. The reason is that the TPL call chains ending with numpy are seven in number, which is more than the TPL call chains ending with the other TPLs.

Finally, for the two projects, nlp\_classification and \texttt{MASTER-pytorch}, we select torch as the TPL. The starting version of torch in nlp\_classification requirements is 1.5.0, and that of \texttt{MASTER-pytorch} is 1.5.1. In both scenarios, the target version is 1.5.1. When nlp\_classification is used as the target project, the inference time of  \tool is 37.63 s, whereas when \texttt{MASTER-pytorch} is used as the target project, the inference time is only 6.5 s. 
The primary reason for this discrepancy is that the \mintinline{python}{requirements.txt} file of nlp\_classification includes the grpcio package as a dependency. grpcio has over 150 candidate versions available, which significantly increases the time required for \tool to resolve a compatible requirements file using SMT. This results in an additional overhead of more than 30 s when nlp\_classification is the target project, thereby leading to a much longer inference time compared to \texttt{MASTER-pytorch}.

In summary, the inference time of \tool is mainly impacted by the following factors: (1) the use of deprecated or removed APIs, which triggers additional compatibility handling; (2) the number and depth of TPL call chains to be analyzed; (3) the inferred versions of related TPLs, with newer versions requiring more checks; and (4) the complexity of dependency resolution, especially when dependencies have a large number of available versions, which slows down SMT-based inference.

\begin{mdframed}[hidealllines=false,backgroundcolor=gray!10,roundcorner=3pt,skipabove=2pt]

\textbf{Answer to RQ4:} Compared with PyEGo, ReadPyE, and LLMs, \tool provides end-to-end automated reasoning that is more practical and efficient, while better ensuring compatibility. On average, it takes only 60.79 s per scenario on \benchmark.
\end{mdframed}

\section{Threats to Validity}\label{sec:threats}
In the following, we discuss the internal, external, and construct threats to the validity of our paper. 

\textbf{Threats to Internal Validity.} 
The main threat to internal validity lies in potential flaws in \tool's implementation. To reduce this threat, we thoroughly tested each module and cross-checked outputs against manually validated cases. We also ensured correctness through code reviews and intermediate result validation. Another threat comes from possible bias in failure classification within \benchmark. To mitigate this threat, we combined automated analysis with manual inspection and had multiple authors independently verify results. All data and tools are publicly available to support reproducibility.

\textbf{Threats to External Validity.} 
The main threat to external validity lies in the generalizability of our results beyond the evaluated dataset, including the representativeness of the selected TPLs. In addition, \benchmark is derived from our prior public dataset~\cite{lei2023deep}, which was authored largely by the same research group, and this may introduce potential selection bias in dataset construction. To reduce this threat, we applied explicit project and TPL selection criteria and further validated the selected dependencies using PCART before constructing \benchmark. The resulting \benchmark contains 2,095 upgrade scenarios from 34 real-world scientific/ML Python projects and 20 widely-used TPLs, covering both version and code compatibility issues. We believe this reflects common dependency-intensive upgrade scenarios in practice. Nevertheless, since the selected projects and TPLs mainly come from scientific and machine-learning domains, the current results may not fully generalize to Python projects from other domains, such as web development, automation, and system tools. Future work will include more TPLs and projects from additional domains to further evaluate generalizability. 
Additionally, we compared \tool with multiple representative baselines, including PyEGo, ReadPyE, GPT-4o, DeepSeek V3, and DeepSeek R1, ensuring broad coverage of traditional and LLM-based approaches. To further reduce bias, we 
used consistent evaluation procedures across all tools. The experimental results demonstrated that \tool performs best in compatible requirements inference on Python TPL upgrade scenarios.

\textbf{Threats to Construct Validity.} 
The main threat to construct validity lies in whether our evaluation metrics fully capture \tool's effectiveness. To mitigate this threat, we measured inference success rate as the primary metric, explicitly checking whether the generated \mintinline{python}{requirements.txt} enables successful execution without version and code compatibility issues. We also recorded inference time to reflect practical efficiency. Furthermore, we compared \tool's results against ground truth from real upgrade outcomes, and used consistent definitions of success across all tools to ensure fairness and completeness in evaluation.

\section{Related Work}\label{sec:relatedwork}

\subsection{Automated Dependency Inference for Python Programs}
Automatically inferring environment dependencies is crucial for ensuring Python software portability and reproducibility. Several approaches have been proposed to infer environment 
dependencies for Python programs.

DockerizeMe~\cite{horton2019dockerizeme} pioneers automatic dependency inference by constructing an inter-dependency graph to determine the required environment for Python code snippets and generating a corresponding Dockerfile. However, its effectiveness is limited by an incomplete knowledge base and version inference issues. PyEGo~\cite{ye2022knowledge} enhances this approach by leveraging a knowledge-based method that extracts dependencies based on syntax and module analysis, but it still faces challenges in handling OS-specific dependencies and finer-grained API-level differences. PyCRE~\cite{cheng2022conflict} introduces a conflict-aware inference technique using a domain knowledge graph, improving compatibility reasoning and reducing dependency mismatches. Similarly, ReadPyE~\cite{cheng2023revisiting} builds a more comprehensive knowledge graph to iteratively refine runtime environment predictions and adapt to real-world scenarios. V2 ~\cite{horton2019v2} addresses dependency inference from a different perspective, focusing on detecting and mitigating configuration drift caused by dependency updates. It employs a feedback-directed search and version upgrade matrices to explore viable environment configurations, ensuring stable execution over time. Additionally, SnifferDog~\cite{wang2021restoring} focuses on restoring execution environments for Jupyter notebooks by analyzing code and mapping dependencies to compatible package versions. While effective for many cases, SnifferDog faces limitations in API coverage and handling advanced Jupyter features. Peng \textit{et al.}~\cite{peng2024less} extended dependency inference research by conducting a large-scale empirical study on configuration issues in Python's PyPI ecosystem. The study provides valuable insights into dependency inconsistencies across thousands of TPLs, highlighting the need for more robust dependency inference mechanisms. 
Bartlett \textit{et al.}~\cite{bartlett2025raiders} proposed PLLM, an LLM-based approach for resolving Python dependency conflicts using retrieval-augmented generation and runtime feedback. PLLM infers dependencies and Python versions from source code and iteratively repairs configurations based on execution errors, enabling flexible dependency inference and automated environment construction. 

As discussed in Section~\ref{sec:survey}, these tools face several limitations in compatible requirements inference of Python TPL upgrade scenarios, such as outdated knowledge, high maintenance costs, and limited scope (e.g., file-level analysis or lack of transitive dependency handling). Our work, \tool, addresses these challenges by leveraging real-time knowledge acquisition, fine-grained static code analysis, project-level and transitive dependency analysis, to better support practical Python TPL upgrade scenarios.

\subsection{Dependency Conflict Detection and Resolution for Python Programs}
Dependency conflicts pose significant challenges in software development, often leading to installation failures and runtime errors. Several techniques have been developed to detect and resolve such issues for Python programs.

Watchman~\cite{wang2020watchman} monitors Python package ecosystem changes by constructing a full dependency graph and identifying potential version conflicts caused by package updates. SmartPip~\cite{wang2022smartpip} addresses dependency resolution by formulating the problem as an SMT constraint-solving task, significantly improving efficiency over traditional dependency resolution strategies. LooCo~\cite{wang2023automatically} introduces a behavior-consistent approach to relaxing version constraints while ensuring functional correctness, providing a novel way to mitigate dependency conflicts. HELP~\cite{cao2024diagnosis} applies SMT-based modeling to diagnose package installation failures due to dependency mismatches, outperforming pip's backtracking strategy in detecting and resolving issues. For Python build reproducibility, PyDFix~\cite{mukherjee2021fixing} focuses on identifying and fixing dependency-related build errors by iteratively resolving conflicting package versions. Xie \textit{et al.} developed Hera~\cite{xie2024pet}, a tool designed to automatically detect and fix cross-repository compatibility (CC) issues by building a cross-repository compatibility database offline and leveraging a system-level package dependency graph at runtime. Hera identifies incompatible API changes between packages installed via different managers (e.g., apt and pip) and provides advice to fix import errors in Python environments. Huang \textit{et al.} conducted a study on 446 dependency bugs (DBs) in deep learning stacks~\cite{huang2023demystifying}, identifying common symptoms, root causes, and fix patterns. They found that most DBs stem from inter-dependency constraints and are often introduced during environment setup but exposed later, highlighting the complexity of DL stack management.

Compared to existing dependency conflict resolution approaches that focus mainly on version constraint satisfaction, \tool addresses both version and code compatibility issues. It supports transitive dependency analysis, real-time knowledge acquisition, and fine-grained static code analysis, enabling more accurate and reliable inference of compatible Python environments.


\subsection{API Evolution and Compatibility Issues of Python Third-party Libraries}
A substantial body of research has been dedicated to understanding API evolution in Python TPLs, offering insights to assist developers and maintainers \cite{zhang2020python, zhang2021unveiling, wang2020exploring}. For example, Zhang \textit{et al.} \cite{zhang2020python} carried out one of the earliest in-depth investigations into how APIs change over time in Python frameworks. By examining six widely used Python TPLs and analyzing over 5,500 open-source projects built on top of them, they uncovered five unique API evolution patterns that are not typically found in Java ecosystems. In a follow-up study, Zhang \textit{et al.}~\cite{zhang2021unveiling} focused on the evolution of TensorFlow 2 APIs. By parsing documentation and classifying API modifications, they found that changes aimed at improving efficiency and maintaining compatibility accounted for more than half of the observed alterations.

Du \textit{et al.} \cite{du2022aexpy} proposed a systematic approach using an API-centric model to identify breaking changes in Python TPLs. They developed a prototype tool named AexPy to detect both declared and undeclared breaking changes in real-world TPLs. In related work, Montandon \textit{et al.} \cite{montandonunboxing} revealed that 79\% of breaking changes involving default parameter values in scikit-learn directly affect training and evaluation processes in machine learning workflows, potentially causing incorrect behavior in downstream applications.

Research on API deprecation has also gained attention. Wang \textit{et al.} \cite{wang2020exploring} explored the practices of deprecation handling in Python TPLs and concluded that poor documentation and vague deprecation notices hinder developers' ability to manage legacy code. To mitigate this, Vadlamani \textit{et al.}~\cite{vadlamani2021apiscanner} introduced APIScanner, a tool designed to alert developers when deprecated API calls are used.

To address the challenge of API compatibility, several automated solutions have been explored. Zhu \textit{et al.} developed Relancer \cite{zhu2021restoring}, an approach that leverages runtime error signals along with a hybrid search strategy guided by API usage patterns and documentation. This technique applies machine learning to identify appropriate fixes, enabling automatic updates to deprecated APIs in Jupyter Notebooks. Similarly, Haryono \textit{et al.} \cite{haryono2021characterization} conducted an empirical study on deprecated API updates and subsequently proposed MLCatchUp \cite{haryono2021mlcatchup}, which learns migration patterns from annotated API signatures to assist in automated updates. More recently, Navarro \textit{et al.}~\cite{navarro2023automated} released a closed-source tool that identifies deprecated APIs in Python code by extracting changelog data from TPLs via web crawling. This tool constructs a knowledge base and integrates with IDEs to suggest code-level corrections. Zhang \textit{et al.} proposed PCART~\cite{zhang2024pcart}, an end-to-end fully automated tool for detecting and repairing Python API parameter compatibility issues. It supports various parameter change types and combines dynamic and static analysis to accurately assess and fix API parameter compatibility issues. 

Compared to prior work on API evolution and compatibility issues, which primarily focuses on identifying, classifying, or repairing specific API changes (e.g., deprecation, parameter updates), \tool focuses on automatically inferring compatible runtime environments for TPL upgrades in Python programs. It not only detects fine-grained API-level incompatibilities 
but also integrates this with version constraint resolution and transitive dependency analysis to generate fully compatible requirements, addressing both installation and runtime compatibility in one unified process.

\section{Conclusion}\label{sec:conclusion}
In this paper, we introduced \tool, an open-source tool that combines version change analysis with code compatibility reasoning to achieve end-to-end automation in 
inferring compatible requirements for Python TPL upgrades, 
To comprehensively evaluate \tool's performance, we constructed a real-world benchmark (\benchmark) comprising 2,095 diverse cases from 34 scientific/ML Python projects (including 406 challenging issues unsolved by pip). Experimental results show that \tool achieves a 94.03\% success rate in inferring compatible requirements, significantly outperforming SOTA tools PyEGo (37.02\%) and ReadPyE (37.16\%), and also surpassing advanced LLM-based approaches (DeepSeek and ChatGPT) by 18--22\%. Furthermore, \tool demonstrates strong efficiency, processing each upgrade scenario in about 60.79 s on average. These results highlight the effectiveness and practicality of \tool in real-world Python TPL upgrade scenarios. 
In the future, we plan to address its current limitations and further improve its practicality and effectiveness by applying it to more projects and complex upgrade scenarios.

\section*{Data Availability}

All replication packages used in this study are publicly available. The implementation of \tool is hosted at \url{https://github.com/PCART-tools/PCREQ}
, the \benchmark is available at \url{https://github.com/PCART-tools/REQBench}
, and all evaluation artifacts are provided at \url{https://github.com/PCART-tools/PCREQ-evaluation}
. These repositories contain the complete source code, datasets, scripts, and experimental results used in this study, enabling full replication and follow-up research.

\begin{acks}
We would like to thank the anonymous reviewers for their insightful comments. 
This work was partially supported by the National Key R\&D Program of China under Grant No. 2024YFB3311503, the National Natural Science Foundation of China (Grant Nos. 62002163, 62372021, 62372219), and the Open Research Fund of State Key Laboratory of Novel Software Technology under Grant No. KFKT2025B13.
\end{acks}

\bibliographystyle{ACM-Reference-Format}
\bibliography{main}


\begin{thebibliography}{45}


\ifx \showCODEN    \undefined \def \showCODEN     #1{\unskip}     \fi
\ifx \showDOI      \undefined \def \showDOI       #1{#1}\fi
\ifx \showISBNx    \undefined \def \showISBNx     #1{\unskip}     \fi
\ifx \showISBNxiii \undefined \def \showISBNxiii  #1{\unskip}     \fi
\ifx \showISSN     \undefined \def \showISSN      #1{\unskip}     \fi
\ifx \showLCCN     \undefined \def \showLCCN      #1{\unskip}     \fi
\ifx \shownote     \undefined \def \shownote      #1{#1}          \fi
\ifx \showarticletitle \undefined \def \showarticletitle #1{#1}   \fi
\ifx \showURL      \undefined \def \showURL       {\relax}        \fi
\providecommand\bibfield[2]{#2}
\providecommand\bibinfo[2]{#2}
\providecommand\natexlab[1]{#1}
\providecommand\showeprint[2][]{arXiv:#2}

\bibitem[Abadi et~al\mbox{.}(2016)]%
        {abadi2016tensorflow}
\bibfield{author}{\bibinfo{person}{Mart{\'\i}n Abadi}, \bibinfo{person}{Paul Barham}, \bibinfo{person}{Jianmin Chen}, \bibinfo{person}{Zhifeng Chen}, \bibinfo{person}{Andy Davis}, \bibinfo{person}{Jeffrey Dean}, \bibinfo{person}{Matthieu Devin}, \bibinfo{person}{Sanjay Ghemawat}, \bibinfo{person}{Geoffrey Irving}, \bibinfo{person}{Michael Isard}, {et~al\mbox{.}}} \bibinfo{year}{2016}\natexlab{}.
\newblock \showarticletitle{$\{$TensorFlow$\}$: A System for $\{$Large-Scale$\}$ Machine Learning}. In \bibinfo{booktitle}{\emph{Proceedings of 12th USENIX Symposium on Operating Systems Design and Implementation (OSDI)}}. \bibinfo{pages}{265--283}.
\newblock


\bibitem[Anaconda(2025)]%
        {Anaconda}
\bibfield{author}{\bibinfo{person}{Anaconda}.} \bibinfo{year}{2025}\natexlab{}.
\newblock \bibinfo{title}{Anaconda}.
\newblock \bibinfo{howpublished}{Retrieved July 1, 2025 from \url{https://www.anaconda.com/}}.
\newblock


\bibitem[Bartlett et~al\mbox{.}(2025)]%
        {bartlett2025raiders}
\bibfield{author}{\bibinfo{person}{Antony Bartlett}, \bibinfo{person}{Cynthia Liem}, {and} \bibinfo{person}{Annibale Panichella}.} \bibinfo{year}{2025}\natexlab{}.
\newblock \showarticletitle{Raiders of the Lost Dependency: Fixing Dependency Conflicts in Python using LLMs}.
\newblock \bibinfo{journal}{\emph{arXiv preprint arXiv:2501.16191}} (\bibinfo{year}{2025}).
\newblock


\bibitem[Brown et~al\mbox{.}(2020)]%
        {brown2020language}
\bibfield{author}{\bibinfo{person}{Tom Brown}, \bibinfo{person}{Benjamin Mann}, \bibinfo{person}{Nick Ryder}, \bibinfo{person}{Melanie Subbiah}, \bibinfo{person}{Jared~D Kaplan}, \bibinfo{person}{Prafulla Dhariwal}, \bibinfo{person}{Arvind Neelakantan}, \bibinfo{person}{Pranav Shyam}, \bibinfo{person}{Girish Sastry}, \bibinfo{person}{Amanda Askell}, {et~al\mbox{.}}} \bibinfo{year}{2020}\natexlab{}.
\newblock \showarticletitle{Language Models are Few-shot Learners}.
\newblock \bibinfo{journal}{\emph{Advances in Neural Information Processing Systems}}  \bibinfo{volume}{33} (\bibinfo{year}{2020}), \bibinfo{pages}{1877--1901}.
\newblock


\bibitem[Cao et~al\mbox{.}(2024)]%
        {cao2024diagnosis}
\bibfield{author}{\bibinfo{person}{Yulu Cao}, \bibinfo{person}{Zhifei Chen}, \bibinfo{person}{Xiaowei Zhang}, \bibinfo{person}{Yanhui Li}, \bibinfo{person}{Lin Chen}, {and} \bibinfo{person}{Linzhang Wang}.} \bibinfo{year}{2024}\natexlab{}.
\newblock \showarticletitle{Diagnosis of Package Installation Incompatibility via Knowledge Base}.
\newblock \bibinfo{journal}{\emph{Science of Computer Programming}}  \bibinfo{volume}{235} (\bibinfo{year}{2024}), \bibinfo{pages}{103098}.
\newblock


\bibitem[Cheng et~al\mbox{.}(2023)]%
        {cheng2023revisiting}
\bibfield{author}{\bibinfo{person}{Wei Cheng}, \bibinfo{person}{Wei Hu}, {and} \bibinfo{person}{Xiaoxing Ma}.} \bibinfo{year}{2023}\natexlab{}.
\newblock \showarticletitle{Revisiting Knowledge-Based Inference of Python Runtime Environments: A Realistic and Adaptive Approach}.
\newblock \bibinfo{journal}{\emph{IEEE Transactions on Software Engineering}} \bibinfo{volume}{50}, \bibinfo{number}{2} (\bibinfo{year}{2023}), \bibinfo{pages}{258--279}.
\newblock


\bibitem[Cheng et~al\mbox{.}(2022)]%
        {cheng2022conflict}
\bibfield{author}{\bibinfo{person}{Wei Cheng}, \bibinfo{person}{Xiangrong Zhu}, {and} \bibinfo{person}{Wei Hu}.} \bibinfo{year}{2022}\natexlab{}.
\newblock \showarticletitle{Conflict-Aware Inference of Python Compatible Runtime Environments with Domain Knowledge Graph}. In \bibinfo{booktitle}{\emph{Proceedings of the 44th International Conference on Software Engineering (ICSE)}}. \bibinfo{pages}{451--461}.
\newblock


\bibitem[code2flow.com(2025)]%
        {code2flow}
\bibfield{author}{\bibinfo{person}{code2flow.com}.} \bibinfo{year}{2025}\natexlab{}.
\newblock \bibinfo{title}{Code2flow}.
\newblock \bibinfo{howpublished}{Retrieved May 15, 2025 from \url{https://code2flow.com/}}.
\newblock


\bibitem[DeepSeek(2025)]%
        {DeepSeek}
\bibfield{author}{\bibinfo{person}{DeepSeek}.} \bibinfo{year}{2025}\natexlab{}.
\newblock \bibinfo{title}{DeepSeek}.
\newblock \bibinfo{howpublished}{Retrieved July 1, 2025 from \url{https://chat.deepseek.com/}}.
\newblock


\bibitem[discuss.pytorch.org(2025)]%
        {cudnnerror}
\bibfield{author}{\bibinfo{person}{discuss.pytorch.org}.} \bibinfo{year}{2025}\natexlab{}.
\newblock \bibinfo{title}{RuntimeError: cuDNN error: CUDNN\_STATUS\_NOT\_INITIALIZED - PyTorch Forums}.
\newblock \bibinfo{howpublished}{Retrieved May 10, 2025 from \url{https://discuss.pytorch.org/t/runtimeerror-cudnn-error-cudnn-status-not-initialized}}.
\newblock


\bibitem[Du and Ma(2022)]%
        {du2022aexpy}
\bibfield{author}{\bibinfo{person}{Xingliang Du} {and} \bibinfo{person}{Jun Ma}.} \bibinfo{year}{2022}\natexlab{}.
\newblock \showarticletitle{AexPy: Detecting API Breaking Changes in Python Packages}. In \bibinfo{booktitle}{\emph{Proceedings of the IEEE 33rd International Symposium on Software Reliability Engineering (ISSRE)}}. \bibinfo{pages}{470--481}.
\newblock


\bibitem[GitHub.com(2025a)]%
        {torchvisionproblem}
\bibfield{author}{\bibinfo{person}{GitHub.com}.} \bibinfo{year}{2025}\natexlab{a}.
\newblock \bibinfo{title}{Issue \#4146 of torchvision}.
\newblock \bibinfo{howpublished}{Retrieved May 15, 2025 from \url{https://github.com/pytorch/vision/issues/4146}}.
\newblock


\bibitem[GitHub.com(2025b)]%
        {z3}
\bibfield{author}{\bibinfo{person}{GitHub.com}.} \bibinfo{year}{2025}\natexlab{b}.
\newblock \bibinfo{title}{Z3}.
\newblock \bibinfo{howpublished}{Retrieved May 15, 2025 from \url{https://github.com/Z3Prover/z3}}.
\newblock


\bibitem[Haryono et~al\mbox{.}(2021a)]%
        {haryono2021characterization}
\bibfield{author}{\bibinfo{person}{Stefanus~A Haryono}, \bibinfo{person}{Ferdian Thung}, \bibinfo{person}{David Lo}, \bibinfo{person}{Julia Lawall}, {and} \bibinfo{person}{Lingxiao Jiang}.} \bibinfo{year}{2021}\natexlab{a}.
\newblock \showarticletitle{Characterization and Automatic Updates of Deprecated Machine-Learning API Usages}. In \bibinfo{booktitle}{\emph{Proceedings of the IEEE International Conference on Software Maintenance and Evolution (ICSME)}}. \bibinfo{pages}{137--147}.
\newblock


\bibitem[Haryono et~al\mbox{.}(2021b)]%
        {haryono2021mlcatchup}
\bibfield{author}{\bibinfo{person}{Stefanus~A Haryono}, \bibinfo{person}{Ferdian Thung}, \bibinfo{person}{David Lo}, \bibinfo{person}{Julia Lawall}, {and} \bibinfo{person}{Lingxiao Jiang}.} \bibinfo{year}{2021}\natexlab{b}.
\newblock \showarticletitle{MLCatchUp: Automated Update of Deprecated Machine-Learning APIs in Python}. In \bibinfo{booktitle}{\emph{Proceedings of the IEEE International Conference on Software Maintenance and Evolution (ICSME)}}. \bibinfo{pages}{584--588}.
\newblock


\bibitem[Horton and Parnin(2018)]%
        {horton2018gistable}
\bibfield{author}{\bibinfo{person}{Eric Horton} {and} \bibinfo{person}{Chris Parnin}.} \bibinfo{year}{2018}\natexlab{}.
\newblock \showarticletitle{Gistable: Evaluating the Executability of Python Code Snippets on GitHub}. In \bibinfo{booktitle}{\emph{Proceedings of the 2018 IEEE International Conference on Software Maintenance and Evolution (ICSME)}}. \bibinfo{pages}{217--227}.
\newblock


\bibitem[Horton and Parnin(2019a)]%
        {horton2019dockerizeme}
\bibfield{author}{\bibinfo{person}{Eric Horton} {and} \bibinfo{person}{Chris Parnin}.} \bibinfo{year}{2019}\natexlab{a}.
\newblock \showarticletitle{Dockerizeme: Automatic Inference of Environment Dependencies for Python Code Snippets}. In \bibinfo{booktitle}{\emph{Proceedings of the IEEE/ACM 41st International Conference on Software Engineering (ICSE)}}. \bibinfo{pages}{328--338}.
\newblock


\bibitem[Horton and Parnin(2019b)]%
        {horton2019v2}
\bibfield{author}{\bibinfo{person}{Eric Horton} {and} \bibinfo{person}{Chris Parnin}.} \bibinfo{year}{2019}\natexlab{b}.
\newblock \showarticletitle{V2: Fast Detection of Configuration Drift in Python}. In \bibinfo{booktitle}{\emph{Proceedings of the 34th IEEE/ACM International Conference on Automated Software Engineering (ASE)}}. \bibinfo{pages}{477--488}.
\newblock


\bibitem[Huang et~al\mbox{.}(2023)]%
        {huang2023demystifying}
\bibfield{author}{\bibinfo{person}{Kaifeng Huang}, \bibinfo{person}{Bihuan Chen}, \bibinfo{person}{Susheng Wu}, \bibinfo{person}{Junming Cao}, \bibinfo{person}{Lei Ma}, {and} \bibinfo{person}{Xin Peng}.} \bibinfo{year}{2023}\natexlab{}.
\newblock \showarticletitle{Demystifying Dependency Bugs in Deep Learning Stack}. In \bibinfo{booktitle}{\emph{Proceedings of the 31st ACM Joint European Software Engineering Conference and Symposium on the Foundations of Software Engineering (ESEC/FSE)}}. \bibinfo{pages}{450--462}.
\newblock


\bibitem[Lei et~al\mbox{.}(2023)]%
        {lei2023deep}
\bibfield{author}{\bibinfo{person}{Huashan Lei}, \bibinfo{person}{Shuai Zhang}, \bibinfo{person}{Jun Wang}, \bibinfo{person}{Guanping Xiao}, \bibinfo{person}{Yepang Liu}, {and} \bibinfo{person}{Yulei Sui}.} \bibinfo{year}{2023}\natexlab{}.
\newblock \showarticletitle{Why Do Deep Learning Projects Differ in Compatible Framework Versions? An Exploratory Study}. In \bibinfo{booktitle}{\emph{Proceedings of the IEEE 34th International Symposium on Software Reliability Engineering (ISSRE)}}. \bibinfo{pages}{509--520}.
\newblock


\bibitem[Montandon et~al\mbox{.}(2023)]%
        {montandonunboxing}
\bibfield{author}{\bibinfo{person}{Jo{\~a}o~Eduardo Montandon}, \bibinfo{person}{Luciana~Lourdes Silva}, \bibinfo{person}{Cristiano Politowski}, \bibinfo{person}{Ghizlane El~Boussaidi}, {and} \bibinfo{person}{Marco~Tulio Valente}.} \bibinfo{year}{2023}\natexlab{}.
\newblock \showarticletitle{Unboxing Default Argument Breaking Changes in Scikit Learn}. In \bibinfo{booktitle}{\emph{Proceedings of the IEEE 23rd International Working Conference on Source Code Analysis and Manipulation (SCAM)}}. \bibinfo{pages}{209--219}.
\newblock


\bibitem[Mukherjee et~al\mbox{.}(2021)]%
        {mukherjee2021fixing}
\bibfield{author}{\bibinfo{person}{Suchita Mukherjee}, \bibinfo{person}{Abigail Almanza}, {and} \bibinfo{person}{Cindy Rubio-Gonz{\'a}lez}.} \bibinfo{year}{2021}\natexlab{}.
\newblock \showarticletitle{Fixing Dependency Errors for Python Build Reproducibility}. In \bibinfo{booktitle}{\emph{Proceedings of the 30th ACM SIGSOFT International Symposium on Software Testing and Analysis (ISSTA)}}. \bibinfo{pages}{439--451}.
\newblock


\bibitem[Navarro et~al\mbox{.}(2023)]%
        {navarro2023automated}
\bibfield{author}{\bibinfo{person}{Nacho Navarro}, \bibinfo{person}{Salwa Alamir}, \bibinfo{person}{Petr Babkin}, {and} \bibinfo{person}{Sameena Shah}.} \bibinfo{year}{2023}\natexlab{}.
\newblock \showarticletitle{An Automated Code Update Tool For Python Packages}. In \bibinfo{booktitle}{\emph{Proceedings of the IEEE International Conference on Software Maintenance and Evolution (ICSME)}}. \bibinfo{pages}{536--540}.
\newblock


\bibitem[OpenAI(2025)]%
        {ChatGPT}
\bibfield{author}{\bibinfo{person}{OpenAI}.} \bibinfo{year}{2025}\natexlab{}.
\newblock \bibinfo{title}{ChatGPT}.
\newblock \bibinfo{howpublished}{Retrieved July 1, 2025 from \url{https://chatgpt.com/}}.
\newblock


\bibitem[Paszke et~al\mbox{.}(2019)]%
        {paszke2019pytorch}
\bibfield{author}{\bibinfo{person}{Adam Paszke}, \bibinfo{person}{Sam Gross}, \bibinfo{person}{Francisco Massa}, \bibinfo{person}{Adam Lerer}, \bibinfo{person}{James Bradbury}, \bibinfo{person}{Gregory Chanan}, \bibinfo{person}{Trevor Killeen}, \bibinfo{person}{Zeming Lin}, \bibinfo{person}{Natalia Gimelshein}, \bibinfo{person}{Luca Antiga}, {et~al\mbox{.}}} \bibinfo{year}{2019}\natexlab{}.
\newblock \showarticletitle{PyTorch: An Imperative Style, High-Performance Deep Learning Library}.
\newblock \bibinfo{journal}{\emph{Advances in Neural Information Processing Systems}}  \bibinfo{volume}{32} (\bibinfo{year}{2019}).
\newblock


\bibitem[Peng et~al\mbox{.}(2024)]%
        {peng2024less}
\bibfield{author}{\bibinfo{person}{Yun Peng}, \bibinfo{person}{Ruida Hu}, \bibinfo{person}{Ruoke Wang}, \bibinfo{person}{Cuiyun Gao}, \bibinfo{person}{Shuqing Li}, {and} \bibinfo{person}{Michael~R Lyu}.} \bibinfo{year}{2024}\natexlab{}.
\newblock \showarticletitle{Less is More? An Empirical Study on Configuration Issues in Python PyPI Ecosystem}. In \bibinfo{booktitle}{\emph{Proceedings of the IEEE/ACM 46th International Conference on Software Engineering (ICSE)}}. \bibinfo{pages}{1--12}.
\newblock


\bibitem[pip.pypa.io(2025)]%
        {pip20.3}
\bibfield{author}{\bibinfo{person}{pip.pypa.io}.} \bibinfo{year}{2025}\natexlab{}.
\newblock \bibinfo{title}{Changes to the Pip Dependency Resolver in 20.3 (2020)}.
\newblock \bibinfo{howpublished}{Retrieved Dec 10, 2025 from \url{https://pip.pypa.io/en/latest/user_guide/\#changes-to-the-pip-dependency-resolver-in-20-3-2020}}.
\newblock


\bibitem[Proposals(2014)]%
        {pep440}
\bibfield{author}{\bibinfo{person}{Python~Enhancement Proposals}.} \bibinfo{year}{2014}\natexlab{}.
\newblock \bibinfo{title}{PEP 440 -- Version Identification and Dependency Specification}.
\newblock \bibinfo{howpublished}{Retrieved Mar 10, 2026 from \url{https://peps.python.org/pep-0440/}}.
\newblock


\bibitem[Proposals(2015)]%
        {pep508}
\bibfield{author}{\bibinfo{person}{Python~Enhancement Proposals}.} \bibinfo{year}{2015}\natexlab{}.
\newblock \bibinfo{title}{PEP 508 -- Dependency Specification for Python Software Packages}.
\newblock \bibinfo{howpublished}{Retrieved Mar 10, 2026 from \url{https://peps.python.org/pep-0508/}}.
\newblock


\bibitem[PyPI(2025)]%
        {PyPI}
\bibfield{author}{\bibinfo{person}{PyPI}.} \bibinfo{year}{2025}\natexlab{}.
\newblock \bibinfo{title}{PyPI · The Python Package Index}.
\newblock \bibinfo{howpublished}{Retrieved July 1, 2025 from \url{https://pypi.org/}}.
\newblock


\bibitem[pypi.org(2025)]%
        {coverage}
\bibfield{author}{\bibinfo{person}{pypi.org}.} \bibinfo{year}{2025}\natexlab{}.
\newblock \bibinfo{title}{Coverage}.
\newblock \bibinfo{howpublished}{Retrieved Dec 10, 2025 from \url{https://pypi.org/project/coverage/}}.
\newblock


\bibitem[TIOBE(2025)]%
        {TIOBE}
\bibfield{author}{\bibinfo{person}{TIOBE}.} \bibinfo{year}{2025}\natexlab{}.
\newblock \bibinfo{title}{TIOBE Index - TIOBE}.
\newblock \bibinfo{howpublished}{Retrieved Dec 10, 2025 from \url{https://www.tiobe.com/tiobe-index/}}.
\newblock


\bibitem[Vadlamani et~al\mbox{.}(2021)]%
        {vadlamani2021apiscanner}
\bibfield{author}{\bibinfo{person}{Aparna Vadlamani}, \bibinfo{person}{Rishitha Kalicheti}, {and} \bibinfo{person}{Sridhar Chimalakonda}.} \bibinfo{year}{2021}\natexlab{}.
\newblock \showarticletitle{APIScanner-Towards Automated Detection of Deprecated APIs in Python Libraries}. In \bibinfo{booktitle}{\emph{Proceedings of the IEEE/ACM 43rd International Conference on Software Engineering: Companion Proceedings (ICSE-Companion)}}. \bibinfo{pages}{5--8}.
\newblock


\bibitem[Wang et~al\mbox{.}(2022)]%
        {wang2022smartpip}
\bibfield{author}{\bibinfo{person}{Chao Wang}, \bibinfo{person}{Rongxin Wu}, \bibinfo{person}{Haohao Song}, \bibinfo{person}{Jiwu Shu}, {and} \bibinfo{person}{Guoqing Li}.} \bibinfo{year}{2022}\natexlab{}.
\newblock \showarticletitle{Smartpip: A Smart Approach to Resolving Python Dependency Conflict Issues}. In \bibinfo{booktitle}{\emph{Proceedings of the 37th IEEE/ACM International Conference on Automated Software Engineering (ASE)}}. \bibinfo{pages}{1--12}.
\newblock


\bibitem[Wang et~al\mbox{.}(2023)]%
        {wang2023automatically}
\bibfield{author}{\bibinfo{person}{Huiyan Wang}, \bibinfo{person}{Shuguan Liu}, \bibinfo{person}{Lingyu Zhang}, {and} \bibinfo{person}{Chang Xu}.} \bibinfo{year}{2023}\natexlab{}.
\newblock \showarticletitle{Automatically Resolving Dependency-Conflict Building Failures via Behavior-Consistent Loosening of Library Version Constraints}. In \bibinfo{booktitle}{\emph{Proceedings of the 31st ACM Joint European Software Engineering Conference and Symposium on the Foundations of Software Engineering (ESEC/FSE)}}. \bibinfo{pages}{198--210}.
\newblock


\bibitem[Wang et~al\mbox{.}(2020a)]%
        {wang2020exploring}
\bibfield{author}{\bibinfo{person}{Jiawei Wang}, \bibinfo{person}{Li Li}, \bibinfo{person}{Kui Liu}, {and} \bibinfo{person}{Haipeng Cai}.} \bibinfo{year}{2020}\natexlab{a}.
\newblock \showarticletitle{Exploring How Deprecated Python Library APIs Are (Not) Handled}. In \bibinfo{booktitle}{\emph{Proceedings of the 28th ACM Joint Meeting on European Software Engineering Conference and Symposium on the Foundations of Software Engineering (ESEC/FSE)}}. \bibinfo{pages}{233--244}.
\newblock


\bibitem[Wang et~al\mbox{.}(2021)]%
        {wang2021restoring}
\bibfield{author}{\bibinfo{person}{Jiawei Wang}, \bibinfo{person}{Li Li}, {and} \bibinfo{person}{Andreas Zeller}.} \bibinfo{year}{2021}\natexlab{}.
\newblock \showarticletitle{Restoring Execution Environments of Jupyter Notebooks}. In \bibinfo{booktitle}{\emph{Proceedings of IEEE/ACM 43rd International Conference on Software Engineering (ICSE)}}. \bibinfo{pages}{1622--1633}.
\newblock


\bibitem[Wang et~al\mbox{.}(2020b)]%
        {wang2020watchman}
\bibfield{author}{\bibinfo{person}{Ying Wang}, \bibinfo{person}{Ming Wen}, \bibinfo{person}{Yepang Liu}, \bibinfo{person}{Yibo Wang}, \bibinfo{person}{Zhenming Li}, \bibinfo{person}{Chao Wang}, \bibinfo{person}{Hai Yu}, \bibinfo{person}{Shing-Chi Cheung}, \bibinfo{person}{Chang Xu}, {and} \bibinfo{person}{Zhiliang Zhu}.} \bibinfo{year}{2020}\natexlab{b}.
\newblock \showarticletitle{Watchman: Monitoring Dependency Conflicts for Python Library Ecosystem}. In \bibinfo{booktitle}{\emph{Proceedings of the ACM/IEEE 42nd International Conference on Software Engineering (ICSE)}}. \bibinfo{pages}{125--135}.
\newblock


\bibitem[Xie et~al\mbox{.}(2024)]%
        {xie2024pet}
\bibfield{author}{\bibinfo{person}{Yifan Xie}, \bibinfo{person}{Zhouyang Jia}, \bibinfo{person}{Shanshan Li}, \bibinfo{person}{Ying Wang}, \bibinfo{person}{Jun Ma}, \bibinfo{person}{Xiaoling Li}, \bibinfo{person}{Haoran Liu}, \bibinfo{person}{Ying Fu}, {and} \bibinfo{person}{Xiangke Liao}.} \bibinfo{year}{2024}\natexlab{}.
\newblock \showarticletitle{How to Pet a Two-Headed Snake? Solving Cross-Repository Compatibility Issues with Hera}. In \bibinfo{booktitle}{\emph{Proceedings of the 39th IEEE/ACM International Conference on Automated Software Engineering (ASE)}}. \bibinfo{pages}{694--705}.
\newblock


\bibitem[Ye et~al\mbox{.}(2022)]%
        {ye2022knowledge}
\bibfield{author}{\bibinfo{person}{Hongjie Ye}, \bibinfo{person}{Wei Chen}, \bibinfo{person}{Wensheng Dou}, \bibinfo{person}{Guoquan Wu}, {and} \bibinfo{person}{Jun Wei}.} \bibinfo{year}{2022}\natexlab{}.
\newblock \showarticletitle{Knowledge-Based Environment Dependency Inference for Python Programs}. In \bibinfo{booktitle}{\emph{Proceedings of the 44th International Conference on Software Engineering (ICSE)}}. \bibinfo{pages}{1245--1256}.
\newblock


\bibitem[Zhang et~al\mbox{.}(2026)]%
        {zhang2024pcart}
\bibfield{author}{\bibinfo{person}{Shuai Zhang}, \bibinfo{person}{Guanping Xiao}, \bibinfo{person}{Jun Wang}, \bibinfo{person}{Huashan Lei}, \bibinfo{person}{Gangqiang He}, \bibinfo{person}{Yepang Liu}, {and} \bibinfo{person}{Zheng Zheng}.} \bibinfo{year}{2026}\natexlab{}.
\newblock \showarticletitle{PCART: Automated Repair of Python API Parameter Compatibility Issues}.
\newblock \bibinfo{journal}{\emph{IEEE Transactions on Software Engineering}} \bibinfo{volume}{52}, \bibinfo{number}{3} (\bibinfo{year}{2026}), \bibinfo{pages}{723--753}.
\newblock


\bibitem[Zhang et~al\mbox{.}(2021)]%
        {zhang2021unveiling}
\bibfield{author}{\bibinfo{person}{Zejun Zhang}, \bibinfo{person}{Yanming Yang}, \bibinfo{person}{Xin Xia}, \bibinfo{person}{David Lo}, \bibinfo{person}{Xiaoxue Ren}, {and} \bibinfo{person}{John Grundy}.} \bibinfo{year}{2021}\natexlab{}.
\newblock \showarticletitle{Unveiling the Mystery of API Evolution in Deep Learning Frameworks: A Case Study of TensorFlow 2}. In \bibinfo{booktitle}{\emph{Proceedings of the IEEE/ACM 43rd International Conference on Software Engineering: Software Engineering in Practice (ICSE-SEIP)}}. \bibinfo{pages}{238--247}.
\newblock


\bibitem[Zhang et~al\mbox{.}(2020)]%
        {zhang2020python}
\bibfield{author}{\bibinfo{person}{Zhaoxu Zhang}, \bibinfo{person}{Hengcheng Zhu}, \bibinfo{person}{Ming Wen}, \bibinfo{person}{Yida Tao}, \bibinfo{person}{Yepang Liu}, {and} \bibinfo{person}{Yingfei Xiong}.} \bibinfo{year}{2020}\natexlab{}.
\newblock \showarticletitle{How Do Python Framework APIs Evolve? An Exploratory Study}. In \bibinfo{booktitle}{\emph{Proceedings of the IEEE 27th International Conference on Software Analysis, Evolution and Reengineering (SANER)}}. \bibinfo{pages}{81--92}.
\newblock


\bibitem[Zhao et~al\mbox{.}(2023)]%
        {zhao2023knowledge}
\bibfield{author}{\bibinfo{person}{Zhongkai Zhao}, \bibinfo{person}{Bonan Kou}, \bibinfo{person}{Mohamed~Yilmaz Ibrahim}, \bibinfo{person}{Muhao Chen}, {and} \bibinfo{person}{Tianyi Zhang}.} \bibinfo{year}{2023}\natexlab{}.
\newblock \showarticletitle{Knowledge-Based Version Incompatibility Detection for Deep Learning}. In \bibinfo{booktitle}{\emph{Proceedings of the 31st ACM Joint European Software Engineering Conference and Symposium on the Foundations of Software Engineering (ESEC/FSE)}}. \bibinfo{pages}{708--719}.
\newblock


\bibitem[Zhu et~al\mbox{.}(2021)]%
        {zhu2021restoring}
\bibfield{author}{\bibinfo{person}{Chenguang Zhu}, \bibinfo{person}{Ripon~K Saha}, \bibinfo{person}{Mukul~R Prasad}, {and} \bibinfo{person}{Sarfraz Khurshid}.} \bibinfo{year}{2021}\natexlab{}.
\newblock \showarticletitle{Restoring the Executability of Jupyter Notebooks by Automatic Upgrade of Deprecated APIs}. In \bibinfo{booktitle}{\emph{Proceedings of the 36th IEEE/ACM International Conference on Automated Software Engineering (ASE)}}. \bibinfo{pages}{240--252}.
\newblock


\end{thebibliography}

\end{document}